\newif\ifShowKeys
\ShowKeystrue
\documentclass[11pt]{article}
	\pdfoutput=1
\usepackage[T1]{fontenc}

\usepackage{listings}
\lstnewenvironment{arkady}  % usage \begin{arkady} ... \end{arkady}
{\lstset{language=C,frame=trbl,basicstyle = \footnotesize \ttfamily , breaklines = true,showstringspaces=false}}{}

\usepackage[usenames,dvipsnames]{xcolor}
\usepackage[setpagesize=false,pagebackref=false, 
linktocpage, bookmarksopen=true, colorlinks=true, 
linkcolor=Maroon,citecolor=Maroon,urlcolor=Maroon]{hyperref}

\usepackage[parsep]{collref}				% collect references in groups
\usepackage{amsmath, amssymb,amsthm}
\usepackage{stackrel}
\numberwithin{equation}{section}
\usepackage{bm,environ,mathrsfs,array,arydshln}
\usepackage{booktabs,float,slashed}
\usepackage{appendix}
\usepackage[mathcal]{euscript}
\usepackage{tensor} 						% Ratcliffe package to write tensors
\usepackage{mathabx}
\usepackage[vcentermath]{youngtab}

\usepackage{graphicx,epsfig,epic}
\usepackage{tikz} 
\usetikzlibrary{arrows,decorations.pathreplacing,decorations.markings,snakes}
\usetikzlibrary{cd}

\usepackage{datetime}
\usepackage{comment}					% to comment large parts of text

\allowdisplaybreaks

\usepackage{framed}						% for shaded equations \begin{shaded} + \end{shaded} or \bs + \es
\definecolor{shadecolor}{rgb}{0.9996078, 0.984314, 0.960784}
\definecolor{framecolor}{rgb}{0,0,0}
\definecolor{TFTitleColor}{RGB}{1,1,1}
\definecolor{TFFrameColor}{RGB}{249	218	181}		
\definecolor{TFFrameColor}{RGB}{230 230 230 }

\newenvironment{frshaded}{%
    \MakeFramed {\FrameRestore}}%
    {\endMakeFramed}
\newcommand{\blue}[1]{\textcolor{blue}{#1}}

\definecolor{myred}{RGB}{233, 33, 45}

\newcommand{\bs}{\begin{frshaded}}			% framed with background in shadecolor 
\newcommand{\es}{\end{frshaded}\noindent}

\def\ba#1\ea{\begin{align}#1\end{align}}		        %  clever way to bypass the known problem...
\newcommand{\be}{\begin{equation}}
\newcommand{\ee}{\end{equation}}
\newcommand{\bea}{\begin{equation} \begin{aligned}} 
\newcommand{\eea}{\end{aligned} \end{equation}}
\newcommand{\mc}{\mathcal }

\newcommand{\mk}{\mathfrak}
\newcommand{\la}{\label}
\newcommand{\eps}{\varepsilon}

\newcommand{\lp}{\notag \\ & }

\DeclareMathOperator{\tr}{\text{tr}}

\newcommand{\ie}{\textit{i.e.} }

\newcommand{\N}{\mathcal N}

\newcommand{\sql}{\sqrt\l}
\renewcommand{\l}{\lambda}
\newcommand{\lb}{\bar\lambda}

\newcommand{\ads}{\text{AdS}}
\newcommand{\vp}{\varphi}

\newcommand{\cp}{\text{CP}^{3}}
\newcommand{\J}{{\mc J}}
\renewcommand{\S}{{\mc S}}
\newcommand{\E}{{\mc E}}
\renewcommand{\L}{{\rm L}}
\newcommand{\T}{{\rm T}}
\newcommand{\re}{{\rm e}}
\newcommand{\rn}{{\rm n}}
\newcommand{\ww}{\wedge}

\newcommand{\foot}{\footnote}
\newcommand{\ci}{\cite}
\def\ov{\over}
\newcommand{\rf}[1]{(\ref{#1})}
\def\no{\nonumber}
\def\OO{\mc O}
\def \adss {$\text{AdS}_{5}\times S^{5}$\ }

\def \ha {{1 \ov 2}}
\def \te {\textstyle}
\def \l {\lambda} 
\def \iffa {\iffalse}

\def \la {\label}
\def \l {\lambda} 
\def \gs  {g_{\rm s}}

\def \ha {{1\ov 2}}

\def \te {\textstyle}

\def \r {\rho}  \def \ka {\kappa} \def \s {\sigma} \def \vp {\varphi}
\def \del {\partial} \def \td {\tilde}
\def \ha {{1 \ov 2}}

\def \sql {{\sqrt{\l}}\ }

\def \ci {\cite}

\def \foot {\footnote}

\def \adss {AdS$_5 \times S^5\ $}
\def \a {\alpha } 
\def \eps {\epsilon}

\def \T {{\rm T}}
\def \ha {\tfrac{1}{2}}
 
 \def \k {{\rm p}}

\def \ed {\small \bibliography{BT-Biblio}
\bibliographystyle{JHEP-v2.9}  \end{document} }

\def \a  {\alpha}
\def \td {\tilde}
 
 \def \abjm {_{\rm ABJM}}
\def \bl {{\bar \l}}

\def \adsc {AdS$_4\times {\rm CP}^3$}

  \def \N  {{\cal N}}

 \def \adsss   {$\ads_{4}\times S^{7}/\mathbb{Z}_{k}$} %SK2

\def \tr {{\rm tr}}
\def \T  {{\rm T}} \def \four {\tfrac{1}{4}}

\def \EE {{\cal Q}}

 \def \E {{\cal E}}
\def \sql {\sqrt{\l}}
 \def \g {\gamma} 
 \def \J {{\cal J}}   \def \E {{\cal E}}

\def \k {\ka}

 \def \g {\gamma}

\def \om {\omega}

\def \rn {{\rm n}}

 \def \abjm {_{\rm ABJM}}
\def \ads {{\rm AdS}}
\def \cp {{\rm CP^3}} 
 \def \EE {{\rm E}}
  \def \Q  {{\cal Q}}
\def \sbl {\sqrt{\bar \l}}
 \def \const  {{\rm const}}
\def \ho {\hat \omega}  \def \om {\omega}
\def \vv {{\rm v}}
\def \aA  {{\rm v}}
\newcommand{\pauli}[5]{\sigma^{#1}\otimes \sigma^{#2}\otimes \sigma^{#3}\otimes \sigma^{#4}\otimes \sigma^{#5}}

\def \KK {{\cal K}}

\def \bag {{\bar g}}
\begin{document}
%%%%%%%%%%%%%%%%%%%%%%%%%%%%%%%%%%%%%%%%%%%%%%%%%%%%%%%%%
%%%%% NOTICE the title page is commented out with \begin{comment}...\end{comment}   but is ready to be used

\begin{titlepage}
\begin{tabbing}
\hspace*{11.5cm} \=  \kill % set the tabbings
%\>  %\small\today\ -- \currenttime \\
\> %none
\end{tabbing}

\vspace*{15mm}
\begin{center}
{\Large\sc \bf   On non-planar  ABJM anomalous dimensions  %in $(S,J)$ sector\\
\vskip 0.2cm
from  M2 branes   in  AdS$_{4}\times S^{7}/\mathbb{Z}_{k}$
 % of ABJM  theory  from  quantum  M2 branes
}
%\vskip 9pt
%{\Large\sc      Notes}
\vspace*{10mm}

{\large Matteo Beccaria${}^{a,}$\footnote{E-mail: \texttt{matteo.beccaria@le.infn.it,  s.kurlyand23@imperial.ac.uk, tseytlin@imperial.ac.uk}}, \ \ Stefan A. Kurlyand${}^{b}$, \ \ Arkady A. Tseytlin${}^{b,}$\footnote{Also at the Institute for Theoretical and Mathematical Physics (ITMP) of MSU   and Lebedev Institute.}} 
\vspace*{4mm}
	
${}^a$ Universit\`a del Salento, Dipartimento di Matematica e Fisica \textit{Ennio De Giorgi},\\ 
		and INFN - sezione di Lecce, Via Arnesano, I-73100 Lecce, Italy
			\vskip 0.1cm
${}^b$ Blackett Laboratory, Imperial College London SW7 2AZ, U.K. 
			\vskip 0.3cm
%\vskip 0.2cm {\small E-mail: \texttt{matteo.beccaria@le.infn.it,  s.kurlyand23@imperial.ac.uk, tseytlin@imperial.ac.uk}}
\vspace*{0.8cm}
\end{center}

\begin{abstract} 
Planar  parts of  conformal  dimensions of primary operators in $U_k(N) \times U_{-k}(N)$    ABJM theory 
 are controlled  by integrability.  Strong coupling asymptotics    of  planar dimensions of operators
 with large spins   can be found  from   the  energy  of  semiclassical  strings in AdS$_{4}\times$CP$^3$  but 
 computing non-planar corrections requires  understanding higher genus  string  corrections.
 As was pointed out  in  arXiv:2408.10070,  there 
 is an alternative way  to find the  non-planar  corrections 
  by  quantizing  M2 branes   in AdS$_{4}\times S^7/\mathbb{Z}_{k}$
  which are wrapped around the 11d circle of radius $1/k= \lambda/N$
  and  generalize spinning  strings in  AdS$_4\times$CP$^3$. 
   Computing  the 1-loop correction to the  energy of 
 M2 brane  that  corresponds to  the long folded string with large spin $S$ 
  in AdS$_4$   allowed   to obtain a prediction 
  for  the  large $\lambda$ limit  of non-planar corrections  to the  cusp anomalous dimension. 
  Similar  predictions were   found for non-planar    dimensions of operators 
  dual to    M2 branes that generalize the ``short'' and ``long'' circular strings with two equal  spins  $J_1=J_2$ 
   in    CP$^3$. 
Here we   consider two 
 more non-trivial examples  of  1-loop M2 brane  computations 
   that   correspond to: 
   (i)  long folded string with large spin $S$  in  AdS$_4$  and orbital momentum $J$ in CP$^3$ 
   whose energy determines  the  generalized cusp  anomalous dimension, and  (ii)
   circular string with     spin $S$  in   AdS$_4$    and  spin $J$  in  CP$^3$. 
   We find   the leading terms of the expansion of the corresponding 1-loop M2 brane energies in $1/k$. 
   We also  discuss   similar semiclassical  1-loop M2 brane  computation in flat 11d background
   and comment on possible relation to  higher genus   corrections to energies in 10d  string theory.

\end{abstract}
\vskip 0.5cm
	{
		%Keywords: {\sc insert here keywords}
	}
\end{titlepage}
%%%%%%%%%%%%%%%%%%%%%%%%%%%%%%%%%%%%%%%%%%%%%%%%%%%%%%%%%
%\makeatletter
%\newcommand*{\toccontents}{\@starttoc{toc}}
%\makeatother
%\toccontents

{\small 
\tableofcontents
}
%\vspace{1cm}
 \setcounter{equation}{0}
\setcounter{footnote}{0}
%\newpage 

\section{Introduction} % and summary}

An important problem in the study of superconformal quantum field theories admitting a large $N$ expansion  (such as $\N= 4$ SYM and ABJM \cite{Aharony:2008ug})
is to compute  the conformal dimensions $\Delta$ of primary operators as functions of the 't Hooft coupling $\l$ at each order in  $1/N$, 
%As is well known, the planar part $\Delta_{0}(\l)$ appearing at leading order in the large $N$ expansion
\be\la{1}
\Delta(\l,N) = \Delta_{0}(\l)+\frac{1}{N^2}\Delta_{1}(\l)+\dots \ . 
\ee
The planar part $\Delta_{0}(\l)$ is determined in principle  by  integrability
 with  its large $\l$ expansion  matched  by  the large tension expansion of  string 
 energies in the dual string theory \cite{Beisert:2010jr}. This 
   does not in general apply to non-planar  $\Delta_n$ corrections. 
 In particular, determining   large $\l$  behaviour   of $\Delta_n$  is challenging 
  on string theory even  for the states with large  quantum numbers 
 as this requires computing 
 higher genus corrections  to semiclassical string energies.  
  
While computing non-planar corrections  at large $\l$ is an  open problem in the $\N=4$ SYM theory 
it was  recently realised  \cite{Giombi:2024itd}  that 
it is possible to find  the leading  strong-coupling  terms in 
   non-planar corrections $\Delta_{n}(\l)$
in the $U(N)_{k}\times U(N)_{-k}$ 
ABJM theory by using its duality to M-theory, i.e.  quantum M2 brane theory on $\ads_{4}\times S^{7}/\mathbb{Z}_{k}$.\foot{This was  demonstrated   earlier for  other  observables   for which M2  brane  predictions 
can be  matched with  localization results on the gauge theory side  
   \cite{Giombi:2023vzu,Beccaria:2023ujc,Beccaria:2025vdj}.}
%(see  also \cite{Drukker:2020swu,Beccaria:2023sph,Drukker:2023jxp,Drukker:2023bip}).}

The idea  is to consider an  M2 brane  counterpart  of a semiclassical   string solution with  large quantum numbers
in $\ads_4 \times \cp$  and  find   the 
 1-loop   correction to its energy as a function of the 11d radius $1/k$  or  string coupling (see Appendix \ref{apa}). 
%SK3
The M2  brane  action $S=  S_{\rm B} + S_{\rm F} $   \cite{Bergshoeff:1987cm,deWit:1998yu} has the following bosonic part 
%The bosonic part of the M2 brane action is 
\ba
\la{a9}
S_{\rm B} &= S_{\rm V}+S_{\rm WZ}, \qquad  S_{\rm V} = -T_{2}\int d^{3}\xi\sqrt{-g},\qquad \ \ 
g_{ij}= \del_i X^M \del_j X^N G_{MN}(X) ,  \\
S_{\rm WZ} &= -T_{2}\int d^{3}\xi\ \tfrac{1}{3!}\eps^{ijk}C_{MNK}(X)\, \partial_{i}X^{M}\partial_{j}X^{N}\partial_{k}X^{K} \ , \qquad 
\qquad  T_2= {1\ov (2\pi)^2 \ell_P^3} \ .
\la{a10}\ea
%Av2 
The explicit form of the  quadratic term in the   fermionic part $S_{\rm F} $  that we use in this paper 
  can be found,  e.g.,  in \ci{Giombi:2024itd}. 

The large $N$, fixed $k$   expansion corresponds to the expansion in  large  effective dimensionless M2  brane tension ($\L$  is the 11d  scale, see \rf{A.1})
\be\la{2}
\T_{2} = \L^{3}\, T_{2} = \frac{1}{\pi}\sqrt{2Nk}, \qquad \qquad T_{2} = \frac{1}{(2\pi)^{2}\ell_{P}^{3}} \ . 
\ee
In general, an   observable  (like the energy  of a  spinning membrane  corresponding 
to  the  dimension of an operator with large quantum numbers) 
 computed in the semiclassical  expansion will  then  be  %have a large $\T_{2}$ structure
\be
\la{1.4}
E = \T_{2}\, \EE_{0}(k)+\EE_{1}(k)+\T_{2}^{-1}\, \EE_{2}(k)+\dots,\qquad  \qquad \T_{2}\gg 1.
\ee
Expanding  further  in large $k$   with  fixed $\l \equiv \frac{N}{k}$ 
 corresponds to the 't Hooft expansion in the 3d gauge theory which  is dual to  perturbative expansion 
 in  type IIA string theory in 
$\ads_{4}\times \cp$ with  string coupling $\gs$ (\ref{A.8}) 
and the effective dimensionless string tension   given  by  (cf. \rf{a8},\rf{a15})
\be \la{3} 
\T= {\pi \ov 2k} \T_2 = \sqrt{ \l\ov2} =\frac{\sqrt{\bar\l}}{2\pi}\ , \qquad 
\qquad \bar\l \equiv  2\pi^{2}\l 
\ . \ee
Representing  $\EE_{n}(k)$  in \rf{1.4}  as a  series in   powers of %$1\ov k^2$   given by 
\be\la{4}  \frac{1}{k^{2}} = \frac{\l^{2}}{N^{2}} = \frac{\gs^{2}}{8\pi\T}\ ,   \ee
and expressing it as an expansion in  $\gs^2$ and then in  $\T^{-1}$ 
  determines  the    strong coupling (large $\l$) corrections  at each order in $1/N^2$  to the corresponding
   anomalous dimension  in  gauge theory.

In particular, the 1-loop term  $\EE_{1}(k)$   encodes  the  leading 
inverse string tension   corrections  at each order in $\gs^2$ in type IIA theory on \adsc. 
Ref.  \cite{Giombi:2024itd}  computed   $\EE_{1}(k)$ for the two  types of  M2   brane   solutions  generalizing 
  (i)  long folded string with spin $S$  in $\ads_4$   related to cusp  anomalous dimension, and  
  (ii)  ``long'' and ``short'' circular strings  with two equal  spins $J_1=J_2$  in $\cp$. 
  We shall review the resulting expressions   in section 1.1. 
  
The aim of the present  paper is  to  consider two 
 more non-trivial examples in the  $(S,J)$  %( or ``$SL(2)$'') 
 sector   that generalize
   (i)  long folded string with spin $S$  in $\ads_4$  and orbital momentum $J$ in $\cp$ 
      related to the  generalized cusp  anomalous dimension, and  
  (ii)  circular string with     spin $S$  in $\ads_4$    and  spin $J$  in $\cp$. 
We shall  summarize  our   conclusions  in section 1.2. 

%%%%%%%%%%%%%%%%%%%%%%%%%%%%%%%%%%%%%%%%%%%%%%%%%

\subsection{Review}

Starting with a  classical  string solution in \adsc\   with large   quantum numbers  $Q = { \sqrt{\lb}}\,  \Q$
(e.g.   spins in  both $\ads_4$ and $\cp$) 
its  $\ads_4$ energy computed in  large tension expansion may be written as 
\be\la{5}
E = \Big[\sqrt{\lb}\, \E_{0}(\Q)+\E_{1}(\Q)+\frac{1}{\sqrt{\lb}}\E_{2}(\Q)+\OO\big(\frac{1}{(\sqrt{\lb})^2}\big)\Big] + \OO(\gs^2) \ . 
\ee
1-loop  string corrections  $\E_{1}$ to the energies of   spinning string solutions in $\ads_{4}\times \cp$  were  computed, e.g., in   \cite{McLoughlin:2008ms,Alday:2008ut,McLoughlin:2008he,Bandres:2009kw,Beccaria:2012qd}.
Such solutions have  direct M2 brane counterparts in \adsss   --  the membrane is wrapped on 11d circle of radius 
 $1/k$  and reduces to the   string solution on the $\ads_{4}\times \cp$. 
The classical M2  brane   energy, i.e. the first term in \rf{1.4} then coincides with the first term in \rf{5}
 with the tensions related as in \rf{3},  i.e. 
 $\EE_0= {\pi^2\ov k} \E_0$.  Also, $ \EE_1 =\E_{1} + \OO({1\ov k^2}) ,  $
 with subleading terms in $1\ov k^2$ determining $\gs^2$ corrections, etc.  

\subsubsection*{Cusp anomalous dimension}

The first example considered  in   \cite{Giombi:2024itd}  was the M2  brane analog of the  long  folded string  with
 large spin  $\S = {S\ov \sbl}\gg 1 $. Its  energy determines the  null  cusp anomalous  dimension
 as in  \ci{Gubser:2002tv,Frolov:2002av}, i.e.  the  leading coefficient   $f(\l, N)$  in the large spin expansion 
of the conformal dimension of an  operator like $O_S=\tr\big[D^{S}(Y^{1}Y_{4}^{\dagger})\big]$\foot{Here  $Y^i$ are   bi-fundamental scalars  of ABJM theory  and $D$ is covariant derivative in a  null  direction.}
\ba
\la{1.5}
& E= \Delta(\l, S) \stackrel{S\gg 1}{=} S+f(\l, N)\, \log S + \dots,  \qquad f(\l, N) = f_{0}(\l)+\frac{1}{N^{2}}{\rm f}_{1}(\l)+\frac{1}{N^{4}}{\rm f}_{2}(\l)+\dots, \\
\la{1.6}
&\qquad \ \ \  f_{0}(\l) \stackrel{\l\gg 1} = \sqrt{2\l}+{\rm f_{0}}(\l), \qquad\qquad  {\rm f}_{r}(\l) \stackrel{\l\gg 1}{=} \l^{2r}\big(a_{1r}+\tfrac{1}{\sql}a_{2r}+\dots\big) , \ \ \ \   r=0,1,2, ...\ .
\ea
The leading coefficients in the  planar part $f_{0}(\l)$  found by quantizing the long folded spinning string in 
$\ads_{4}\times \cp$ \cite{McLoughlin:2008ms,Alday:2008ut,Krishnan:2008zs,Bianchi:2014ada}  are 
\be\la{9}
f_{0}(\l) =\te  \sqrt{2\l}-\frac{5\log 2}{2\pi}-\big(\frac{K}{4\pi^{2}}+\frac{1}{24}\big)\frac{1}{\sqrt{2\l}}+\mc O(\frac{1}{(\sql)^{2}})
= 2h(\l)-\frac{3\log 2}{2\pi}-\frac{K}{8\pi^{2}}h^{-1}(\l)+\mc O\big(h(\l)^{-2}\big),
\ee
where
% $K$ is Catalan's constant and, in the second equality, we expressed the result in terms of the
$h(\l)$  is a  ``renormalized tension''    (containing 
a $\log 2$ correction \cite{McLoughlin:2008he} with the exact  form suggested in  \cite{Gromov:2014eha})
\be
\la{1.8}
h(\l) =\te  \frac{1}{\sqrt 2}\sqrt{\l-\frac{1}{24}}-\frac{\log 2}{2\pi}+\mc O(e^{-2\pi \sqrt{2\l}}) \ . 
\ee
%SK4 The relation not to 2\pi but 4\pi with extra 1/2 factor see below 1.39 in GKT
The   expression \rf{9}  is consistent with integrability   and is related to the \adss  or $\N=4$ SYM result 
by the replacement  ${\sqrt{ \l_{_{\rm  SYM}}} \ov 4\pi}  \to h(\l_{_{\abjm}})$ and factor $\tfrac{1}{2}$: $f_{0}(\l) = \tfrac{1}{2}f_{0_{\rm SYM}}(\l_{\rm SYM})$ %SK4
\ci{Gromov:2008qe}.\foot{This is related 
to the fact that  planar part of  the cusp anomalous   dimension  is controlled  just by the asymptotic Bethe  ansatz.  
There is no reason to expect that a similar simple relation should  hold at a non-planar level.}

\def \rmk  {{\rm n}}

The  semiclassical quantization of the corresponding  M2 brane  solution  leads to the  following representation  for $f$ in \rf{1.5}  \cite{Giombi:2024itd} (cf. \rf{1.4}) 
%generalisation of the long spinning folded string in $\ads_{4}$ has the form
\ba
\la{1.9}
& f(k, \T_{2}) = \frac{\pi}{k}\T_{2}+q_{0}(k)+\T_{2}^{-1}\, q_{1}(k)+\T_{2}^{-2}\, q_{2}(k)+\dots, \\
\la{1.10}
& q_{s}(k) = k^{s}\big(p_{s}^{(0)}+\frac{1}{k^{2}}p_{s}^{(1)}+\frac{1}{k^{4}}p_{r}^{(2)}+\dots\big), \ \ \ \ \  s=0,1,2, ...
\ea
Comparing   \rf{1.5},(\ref{1.6}) and (\ref{1.9}) one finds that  at any non-planar order  $1/N^{2r}$
 the leading large $\l$ contribution to ${\rm f}_{r}(\l)$, \ie the coefficient $a_{1r}$,
is fixed in terms of the  coefficients   appearing in the large $k$ expansion \rf{1.10} of $q_{0}(k)$, i.e. 
%\be
%\la{1.11}
$a_{1r} = p_{0}^{(r)}, \  r = 0,1, 2, \dots$.
%\ee
The expression for   the 1-loop M2  brane correction $q_{0}(k)$  obtained  in 
%In particular, from the analysis in 
\cite{Giombi:2024itd} is 
%, one has the following prediction for the whole set of coefficients $\{p_{0}^{(s)}\}$ from 
%the large $k$ expansion
\be
\la{1.12}
q_{0}(k)  = {1\ov \pi} \Big[
 -\frac{5}{2} \log 2  +{4\ov k^{2}}\zeta(2)  +{4\ov k^{4}}\zeta(4)  - {1616\ov 15 k^6} \zeta(6) + ...\Big] 
 =  -\frac{5\log 2}{2\pi}+\frac{2\pi}{3}{1\ov k^2} +\frac{2\pi^{3}}{45}{1\ov k^4} -\frac{1616\pi^{5}}{14175}{1\ov k^6} +\dots \  ,
\ee
where  the expression for  $1\ov k^2$  is given in  \rf{4}. 
%where all terms in the expansion were determined in terms of a generating function. 
This  determines the   leading strong coupling  asymptotics of the  non-planar $1\ov N^{2r}$ terms in the 
 cusp anomalous dimension.

\subsubsection*{Circular $J_1=J_2$  solution}

The  second example discussed in \cite{Giombi:2024itd}   was the M2   brane generalization
 of the  string  solutions \ci{Frolov:2003qc,Bandres:2009kw}  with two spins 
$J_1=J_2\equiv J $  in $\cp$. 
The dual ABJM operator should  have the structure % be built out of the 4 bi-fundamental scalars as 
$O_{J_1,J_2} = \tr[(Y^{1}Y_{2}^{\dagger})^{J_{1}}(Y^{3}Y_{4}^{\dagger})^{J_{2}}]+\dots$.
 %SK
 For the M2 brane solution  which admits the ``short'' (or ``slow'' $\J\equiv  {J\ov \sbl}\ll 1$) limit, the 
 corresponding energy  contains  the following 1-loop  M2 brane corrections
  \cite{Giombi:2024itd}  %is the $1/k^{2}$ correction in 
 \ba
 \la{1.16}
 E =& 2\sqrt{\sqrt{\lb}\, J}+\tfrac{1}{2}+\tfrac{1}{2}\lb^{-1/4}J^{1/2}-\tfrac{9}{4}\zeta(3)\lb^{-3/4}J^{3/2}+\mc O(\lb^{-1}J^{2})
  \lp
 +\frac{1}{k^{2}}  \Big[\zeta(2)\big(-4\lb^{3/4}J^{-3/2}+8\lb^{1/4}J^{-1/2}
\big) +\mc O(\lb^{-1/4}J^{1/2})\Big]
 +\mc O\big({1\ov k^{4}}\big) \ , 
 \ea
 where $\bar \l$ was defined in \rf{3}. Here 
 the first line is the 1-loop string correction that  represents the strong-coupling expansion of  the  conformal dimension 
 $\Delta(J, \l)$ of the dual ``short'' operator  in the planar limit. 
 % a prediction for the next-to-leading strong coupling corrections to the 
% conformal dimension $\Delta(J)$ of the dual ``short'' operator with flat-space leading order scaling $\Delta\sim \lb^{1/4}\sqrt{J}$ \cite{Gubser:2002tv}.
The second line is the  large $k $  expansion  of  the  leading   non-planar correction (cf. \rf{1.9},\rf{1.12}), 
i.e.  the leading large tension  asymptotics  of  the  quantum string  (torus) contribution.
%%%%%%%%%%
Equivalently, \rf{1.16}  may be written as  % {\bf [check]} 
\be
 \la{166}
 (E-\ha)^2 = 4\sqrt{\lb}\, J\,  \big(1 + \tfrac{1}{ 2 \sbl} + ... \big)  - \tfrac{9}{\sbl} \,  \zeta(3) J^{2}+...
 %\mc O(\lb^{-1}J^{2})
+ \frac{1}{k^{2}} \,16 \zeta(2)\big(-\bl\, J^{-1}+2  \sbl + ...\big) %+\mc O(\lb^{-1/4}J^{1/2})\Big]
 +\mc O\big({ {1\ov k^{4}}}\big) \ .
 \ee
 %%%%%%%%%%%%%%%%%%%%%
For the M2 brane solution which admits  the  ``long'' (or ``fast'' $\J\gg 1$) limit, one finds 
 \cite{Giombi:2024itd}\foot{Here $\zeta(2)$ is the standard Riemann zeta-function value, i.e.   $\pi^2\ov 6$.  Like in \rf{1.12},  
 below we will keep the  $\zeta(2n)$   factors  in the $1\ov k^{2n}$  terms in their  implicit form 
 to emphasize their common origin  from the summation over the  M2 brane  modes in the  second  circular  direction
 (identified with the 11d direction).}
\ba %SK3
\la{1.18}
%Av2
E =& 2J+\tfrac{1}{4}\lb J^{-1}\big(1-2\log 2\, \lb^{-1/2}+\dots\big)+\tfrac{1}{2}c_{1}\lb J^{-2}+\mc O(\lb^{2}J^{-3})  \lp
\quad +\frac{1}{k^{2}}\zeta(2)\big(-8\lb^{-1/2}J-2\lb^{1/2}J^{-1}+\tfrac{3}{16}\lb^{3/2}J^{-3}+\mc O(\lb^{5/2} J^{-5}) \big)+\mc O\big({1\ov k^{4}}\big) 
\ .
\ea
The  first line is the sum of the classical energy and string 1-loop contribution expanded in the ``long'' limit \cite{Bandres:2009kw}
with $c_{1} \simeq -0.336$. The second line represents the membrane contribution, or the leading non-planar correction, to the anomalous dimension of the dual ABJM operator with large spin $J$.

Let us recall  that   the string corrections in  \rf{1.18}  that scale as odd powers of $1/J$, i.e.  $E^{\rm odd}(J, \sqrt{\bar \l })$, 
do not receive wrapping corrections, \ie they are  controlled by  the asymptotic  Bethe Ansatz  %on gauge side 
\cite{Gromov:2008qe}  and  are %happen to be   %. As a result, these corrections are 
  directly related
to their counterparts  $E^{\rm odd}_{\ads_{5}}(J, \sqrt\l )$   in the $\ads_{5}\times S^{5}$ case  as  \cite{McLoughlin:2008he}
\be
\la{1.19}
E^{\rm odd}_{\ads_{4}}(J, \sqrt{\lb}) = \tfrac{1}{2}E^{\rm odd}_{\ads_{5}}(2J, 2\bar h(\lb))\ , 
\qquad \qquad  \bar h(\lb) \equiv  2\pi h(\l) = \sqrt{\lb}- \log 2+ \dots \ , 
\ee
where $ h(\l) $   was  given in \rf{1.8}. Thus the $\log 2$ term in the first line of \rf{1.18}   may be absorbed into 
the redefinition of the $\bar \l$ factor. This relation is also consistent with \rf{1.5},\rf{9}.\foot{Note that  the  large $\J$ 
expansion of the  classical  string energy $E_0=\sqrt \l \E_0$  has the form 
%Note  that  the overall factor of  $1/2$ and the scaling of $J$ by $2$ are, in a sense, trivial. For instance, in $\ads_{5}$ and for 
%large $\J$, the energy takes the form 
%$\E= \J+\sum_{n=0}^{\infty}c_{n}/\J^{2n+1}$ 
 $E_0(J,\sql)= J+\sum_{n=0}^{\infty}c_{n}(\sql)^{2n+2} J^{-2n-1}$, which implies
that $E_0(J, \sql) = \frac{1}{2}E_0(2J, 2\sql)$. This relation extends to odd terms 
in the  string  quantum (inverse tension)   corrections  with extra  replacements of $\sqrt \l$ in the coefficients. 
The  factor of 2 in front of $\bar h$
in (\ref{1.19})  stems from a 
%is $\sql\to 2\bar h = 2\sqrt{\lb}+\dots$, where  the  factor of 2 in front of $\bar h$ arises from 
 comparison of 
the BMN dispersion relation in  the $\ads_{4}\times \cp $ and \adss  cases.
}

%%%%%%%%%%%%%%%%%%%
\subsection{Summary} 
%Non-planar corrections in $(S,J)$ sector of ABJM from semiclassical M2 branes}

\subsubsection*{Generalized cusp  anomalous dimension}

Below  we will  first  consider  a generalization of the  1-loop M2 brane  computation of non-planar corrections to the cusp anomaly 
in \cite{Giombi:2024itd} to the case of the ``generalized'' cusp anomaly  that includes also dependence on the 
angular  momentum $J$  in $\cp$.
In the $\ads_{5}\times S^{5}$  case  the corresponding   long folded string $(S,J)$ 
 solution  and  the  associated   cusp anomaly  function  was  studied in \cite{Belitsky:2006en,Frolov:2006qe,Giombi:2010fa}.  
For the string in $\ads_{4}\times \cp$,  the 1-loop correction to the energy of  such 
 $(S,J)$ solution was  computed in
\cite{McLoughlin:2008ms,Alday:2008ut}. 
While  for generic   values  of the  spins   the form   of the folded   string solution is complicated, it simplifies 
in the limit 
\be
\la{1.20}
\S \gg \J \gg 1, \quad\qquad  x \equiv  \frac{\log\S}{\pi\, \J}  %= \frac{\sqrt{\lb}}{\pi\, J}\log\frac{S}{\sqrt{\lb}} 
= \text{fixed}, \qquad \qquad \S= {S\ov \sbl}, \ \ \    \J= {J \ov \sbl } \ . 
\ee
In this case  one finds a  generalization of \rf{1.5}   where  $f(\l,N) \to f(\l,N, x)$. 
%%%%%%%%%%%%%%%
\iffa 
The structure of the expansion (\ref{1.5}) is similar, but now all functions ${\rm f}_{s}(\l)$ depend also on $J$. \footnote{
\blue{Remark to be improved} In $\ads_{4}$ relation between conserved charge (spin) and winding $w$ is 
in general $S=2\T\, w$, while in $\cp$ the charge (angular momentum) is 
$J=\T\, w$. Here $\T$ is defined in (\ref{A.8}).
This is different from $\ads_{5}$ case where $\T=L^{2}/\alpha'$ is same 
in AdS and sphere parts.
}
\fi
%%%%%%%%%%%%%%%%%%%
The  tree-level string (planar) term in the expansion of $f$ in (\ref{1.5}) is   then   the following 
 generalization of \rf{1.6}  %AT4
\be\la{21}
f_{0}(\l, x) = {\sqrt{\bar\l}\ov \pi} \,{\sqrt{1+x^{2}}\ov x } +  {\rm f}_{0}(\l,x) \ , \qquad \ \ \ 
{\rm f}_{0}(\l,x)= {\rm f}_{01}(x) + \OO ({1\ov \sqrt \l}) \ . 
%= \sqrt{2\l}\, \sqrt{1+\frac{J^{2}}{2\l\log^{2}S}}+{\rm f}_{0}(\l,x).
\ee
%The leading term in ${\rm f}_{0}(\l,x) = \frac{E_{1}(J, S, \sqrt{\lb})}{\log S}+\mc O(1/\sqrt{\lb})$ 
%comes from the 1-loop energy correction computed in \cite{McLoughlin:2008ms} that reads
%AT4
Here ${\rm f}_{01}(x)$  (that generalizes $- {5\log 2\ov 2 \pi} $ in \rf{9})
  is the coefficient of $\log S$ in the 
  1-loop   $\ads_{4}\times \cp$ string  correction 
\cite{McLoughlin:2008ms} 
that   may be written as 
\ba %SK4 %AT4
\la{1.22}
%E_{1}(J, S, \sqrt{\bl}) =& \, \mathrm{f}_{0}(\l, x)\log S + ... \ , \\
E_{1}(J, S, \sqrt{\lb}) =    -   {x \log \S \over \pi \sqrt{1+x^{2}}}\log 2+  \ha  E_{1}^{\ads_{5}}(2J, 2S, 2\sqrt{\lb}) \equiv \mathrm{f}_{01}( x)\log \S\ ,  
\ea
where $E_{1}^{\ads_{5}}(J, S, \sql)$ is the  corresponding expression in the \adss  string   case  \cite{Frolov:2006qe}
\ba
\la{1.23}
E_{1}^{\ads_{5}}(J, S, \sql) =&  \frac{\log \S}{\pi x \sqrt{1+x^{2}}}
\Big[x (\sqrt{1+x^{2}}-x) 
\lp \qquad  +2(1+x^{2})\log(1+x^{2})-(1+2x^{2})\log\big[\sqrt{1+2x^{2}}(x+\sqrt{1+x^{2}})\big]
\Big].
\ea
Combining \rf{1.22}   with the 
classical string term in \rf{21}, i.e. $E_0=  S +  \sqrt{2\l}\,\sqrt{1+\frac{1}{x^{2}}} \log S + ...$
one observes that 
the relation (\ref{1.22}) is consistent with the direct  generalization of  (\ref{1.19})  that 
includes also the  spin $S$ argument.\foot{Let us  note also that the 2-loop 
corrections to the corresponding string energy or generalized cusp anomaly  were computed  in  the 
%SK4
\adss case in   \cite{Giombi:2010fa}. The 2-loop 
corrections to the null cusp anomaly in the \adsss \, case  were computed in \cite{Bianchi:2014ada}.  }

%Two-loop string correction to energy in the considered limit
 %was found in  and similar result for the two-loop contribution to  $h(\l)$ in  \cite{Bianchi:2014ada}.

% In particular, the $\log 2$ term in (\ref{1.22}) comes from the coupling redefinition in (\ref{1.19}).

To find non-planar corrections to  this generalized  scaling function  we shall   consider 
 the M2 brane generalization  of the  long folded $(S,J)$ 
  string in the limit \rf{1.20}. 
The resulting    analog of    (\ref{1.9})  will   have $q_r$ now depending also on $x$ 
with  the 1-loop   term  $q_{0}(x,k)$  having the following large $k$ expansion %generalizing \rf{1.12} 
\ba %SK4
\la{1.24}
q_{0}(x, k) =& \, \mathrm{f}_{01}(x)+\bar q_{0}(x, k), \\
\bar q_{0}(x, k)=& {\pi \ov x}\Big(\tfrac{2}{3}{\sqrt{1+{x^{2}}}}+\frac{1}{\sqrt{1+x^{2}}}\Big)\frac{1}{k^{2}} 
\lp
+\frac{\pi^{3}}{x^3}\Big[\tfrac{2}{45}{(\sqrt{1+x^{2}})^3 }+\tfrac{19}{45}{\sqrt{1+x^{2}}}{}
-\tfrac{1}{60}\frac{1}{\sqrt{1+x^{2}}}\Big]\frac{1}{k^{4}}  +\dots. \la{1.24a}
\ea
This  generalizes  (\ref{1.12}) to the  non-zero $J$ case  (\rf{1.24} reduces to \rf{1.12}   for $x\to \infty$). Explicitly, 
the large $x$  and small $x$ expansions of $\bar q_0$    may be written as (cf. \rf{3})
% In particular, the first term in (\ref{1.24})
%is the $x\to \infty$ limit of $E_{1}/\log S$ as can be checked using $\J  = \frac{1}{\pi x}\log \S$, \cf (\ref{1.20}). 
\ba\la{22} %SK3 %SK4
\bar q_{0}(x, k) \Big|_{x\gg 1} 
=& \frac{2\pi}{3}{\l^2\ov N^{2}}\Big(1+ \frac{J^2}{\l\, \log^{2}S}+   \mc O\big(\l^{-2}J^4 \log^{-4} S\big) \Big)+\mc O\big( { 1\ov N^4}\big)\ , \\
\bar q_{0}(x, k) \Big|_{x\ll 1} =& 
\frac{5\pi}{3\sqrt 2}\frac{\l^2}{N^2}\Big(\frac{J }{\sqrt{\l}\, \log S}+ \mc O\big(\l^{1/2}J^{-1}\log S \big) \Big)+\mc O\big( { 1\ov N^4}\big) \ . \la{23}
\ea
These expressions   represent the predictions for the   non-planar corrections to the generalized ABJM cusp anomaly  function 
in the strong coupling limit.

\subsubsection*{Circular $(S,J)$  solution}

%%%%%%%%%%%%%%%%%%%%%%%%%%%%%%%%%%%%%%%%
Another  1-loop computation  we will consider  in this paper   is for an M2 brane generalization 
of the circular  $(S,J)$ string solution, in which the string  wraps  
a circle in $\ads_{4}$ and  a circle in $\cp$. % (in this case $S=mJ$ where $m$ is a wrapping number).  
The 1-loop   correction to  such    solution in $\ads_{5}\times S^{5}$ was   computed  in \cite{Arutyunov:2003za,Park:2005ji,Beisert:2005mq}  while  for its 
 $\ads_{4}\times \cp$  analog   this was done in  \cite{McLoughlin:2008he}.
We will consider string energy as a function of $J=\sbl\, \J$ and\footnote{In general, the classical solution depends on the  two integer 
 winding numbers ${\rmk, \rm m}$  so that  ${\rmk}\,S+{\rm m}\,J=0$. 
 We will  fix 
 ${\rmk}=1$  (so that $u= - {\rm m}$)  but  the dependence on $\rmk$   can be  always  reinstated by an appropriate  rescaling  of $J$.
}
\be\la{25}
u \equiv   \frac{\S}{\J} = \frac{S}{J} \ . 
\ee
%and again the two regimes of ``short'' ($\J\ll 1$) and ``long'' ($\J\gg 1$) string. In ``short'' limit, we obtain 
In general, 
the M2  brane energy   has the following structure 
\be\la{26}
E(S, J, k, \sqrt{\lb}) = E_0  
+E_{1}(u,\J)+E_{1}^{\rm M2}(u, \J, k)+ ... \ , \ee
where $E_0$ is the classical  string (or membrane)   energy, $E_1$ is the 1-loop string correction   
and  $E_{1}^{\rm M2}(u,\J, k)$ in \rf{26}  stands for the genuine  M2 brane 1-loop contribution,  i.e. a series  in 
   $1/k^{2}$. % (see  (\ref{1.17})).

In the ``short''   string limit  ($\J\ll 1$, $u$=fixed)   one finds that %we  find  that the M2  brane energy   has the following structure 
\ba
%E(S, J, k, \sqrt{\lb}) =& E_0  
%+E_{1}(u,\J)+E_{1}^{\rm M2}(u, \J, k)+ ... \ , \\
\la{24} E_0   = &\sqrt{\lb}\, \Big[u+\sqrt{1+u^{2}}\, \J+\frac{u}{2(1+u^{2})}\J^{2}+\dots\Big]\ , \\
\la{27}
E_{1}(u,\J) =& A_{0}(u)+A_{1}(u)\, \J+A_{2}(u)\, \J^2 +   \dots \ , 
\ea
%with  the functions $A_{i}(u)$  given  by convergent infinite sums.
where  $A_i$ are given  by convergent infinite sums (see Appendix  \ref{app:circular-short-oneloop}),     while  (cf. \rf{1.16},\rf{1.18}) 
\ba
E_{1}^{\rm M2}(u,\J, k) =& \frac{1}{k^{2}}\zeta(2) \Big(-6u+\frac{10}{\sqrt{1+u^{2}}}\J+\frac{u(3+13u^{2})}{(1+u^{2})^{2}}\J^{2}+\dots\Big)\lp
+\frac{1}{k^{4}}\zeta(4)\Big(-\frac{3}{2}u-\frac{35}{\sqrt{1+u^{2}}}\J+\frac{162+165u^{2}-137u^{4}}{4u(1+u^{2})^{2}}\J^{2}+\dots\Big)+\mc O\big( {1 \ov k^{6}}\big).\la{28}
\ea
In ``long''   string limit ($\J\gg 1$, $u$=fixed)  the classical string energy  is given by 
\be
E_0 = \sqrt{\lb}\, \Big[(1+u)\,\J+\frac{1}{2\J}u(1+u)-\frac{1}{8\J^{3}}u(1+u)(1+3u+u^{2})+\dots\Big], \la{29}
\ee
i.e.  has the familiar ``fast string''   expansion ($E= S + J\,  +\  $terms with odd powers in $1/J)$. 
String 1-loop correction  $E_1$  expanded for $\J \gg 1$   contains  terms with both odd and even powers of $1/J$. 
 The former  combine with the classical $E_0$ term  %contribution 
to  satisfy the analog of the  relation \rf{1.19} (including also  the dependence  on $S$) 
to the  corresponding terms in the \adss   case 
 \cite{Park:2005ji,Beisert:2005cw} 
\ba
\la{1.33}
&E^{\rm odd}_{\ads_{5}}(J, S, \sql) = J+S+\frac{\l}{2J}\, u(1+u)-\frac{\l^{2}}{8J^{3}}\, u(1+u)(1+3u+u^{2})\no\\ &\quad \qquad \qquad \qquad \qquad 
+\frac{\l^{3}}{16 J^{5}}\, \Big[ u(1+u)(1+7u+13u^{2}+7u^{3}+u^{4})
+ {1\ov \sql} u^{3}(1+u)^{3}\Big] +... %\mc O(1/J^{7}).
\ea
The even  part of the $\ads_4$  string 1-loop term is 
\be
\la{1.34}
E_{1}^{\rm even}=\Big[ -\tfrac{3}{4}u^{2}(1+u)^{2}\zeta(2)+\tfrac{15}{8}u^{3}(1+u)^{3}\zeta(4)+\dots\Big]\frac{1}{\J^{2}}
+\mc O\big({1\ov \J^{4}}\big).
\ee
The coefficients of  $1/\J^{2n}$ terms in \rf{1.34} 
 are given  by  infinite sums  that converge for sufficiently small $u$ \cite{Beisert:2005cw,McLoughlin:2008he}.
The membrane correction in \rf{26}  expanded   at large $\J$   is found to have the following form (cf. \rf{28})
%In this ``long'' limit, the membrane corrections take the form 
\ba %SK3
E_{1}^{\rm M2}(u,\J,k) =& \frac{1}{k^{2}}\zeta(2)\Big[10\, \J-\frac{u(6+11u)}{\J}+\frac{u(12+68u+88u^{2}+27u^{3})}{4\J^{3}}+\mc O(\J^{-5})   \Big]\lp
+\frac{1}{k^{4}}\zeta(4)\Big[\frac{81}{2u(1+u)}\, \J^{3}+\frac{81+103u-221u^{2}}{4u(1+u)}\, \J+\mc O(\J^{-1}) \Big]+\mc O\big( {1 \ov k^{6}}\big)  \ . \la{32}
\ea
Note  that the  leading $ \frac{1}{k^{2}} \J$ term in \rf{32} 
  combined  with the classical $J$ term in \rf{29}   gives 
$E = (1+\tfrac{10\zeta(2)}{k^2\bar{\lambda}^{1/2}}+\dots)J+\dots$
which is similar  to the  structure of the 
linear in $J$ term  in the energy \rf{1.18} for the $J_1=J_2$ solution found in \cite{Giombi:2024itd}.

%The leading term in the above expansion for small $u$, \ie $\J\gg 
%\S$, can be combined with the classical part as $E = (1+\tfrac{10\zeta(2)}{k^2\bar{\lambda}^{1/2}}+\dots)J+\dots$. Although the meaning of this term is not clear and may suggest renormalisation of the dispersion relation, we remark that a similar behaviour appears in the long $\J_{1}=\J_{2}$   case  in Eq.~(3.79) in \cite{Giombi:2024itd}.

\subsubsection*{Flat space case}

The matching  of the leading  strong-coupling asymptotics of non-planar corrections  
 found from localization to the  1-loop M2 brane corrections in \ci{Giombi:2023vzu,Beccaria:2025vdj} 
 implies  that  the latter should be  reproducing the leading in  large  string tension term  in  the quantum string higher genus
 contribution  in \adsc  (cf.  \rf{4}).   In particular, the 1-loop   
   energy  of a semiclassical  M2  brane state  should be 
   reproducing   certain part of the  quantum string 1-loop  (torus)   correction  to  the energy  of the corresponding   string 
  state. 
  
\def \rE  {{\rm E}}  
\def \bae {\bar \rE_1}  
\def \La  {\Lambda}
\def \ls {\ell_{\rm s}} 
\def \lpp {\ell_P} 
\def \m {\mu}

 As   was  noted in \ci{Giombi:2024itd} the  same should then apply  also in the flat space 
 limit. Namely, starting with a  semiclassical string state in 10d  IIA  string theory,  considering its M-theory  analog represented  by an  M2  brane wrapped on the  11d   circle  and computing the  1-loop   correction to its  energy 
   should   capture certain    leading  terms  in  the  corresponding  torus, etc., corrections 
  \ci{Sundborg:1988ai,Amano:1988ht,Chialva:2003hg,Chialva:2004xm,Sen:2016gqt}
   to the string energy or  the   mass shift in 10d string theory.
   
  For example, starting with 
   the  $\mathbb R^{1,9} \times  S^1$ 
  counterpart   of the (``short'') circular $J_1=J_2=J$   string  rotating in 2 orthogonal planes in \adsc 
   and  taking the flat-space limit   \ci{Giombi:2024itd} of the M2 brane 1-loop   expression  in 
    \rf{1.16}  or \rf{166}  one finds\foot{Note that   \ci{Mezincescu:1987kj}   discussed 
    the   1-loop correction to the 
  energy   of a different $J_1=J_2$   solution  in   flat 11d space: there   
  the membrane   was rotating in 2 planes with the ``radii''    being periodic functions of  the two world-volume coordinates 
  but   was  not wrapped on $S^1$.}
   \ba \la{001}
   &E=  2 \sqrt{ \a'^{-1} J}\Big[1 -  \tfrac{1}{2}  \zeta(2) \, \gs^2 J^{-2}   -\tfrac{19}{60}\zeta(6)\, \gs^6J^{-4}+
    ... \Big]\ , \\
  & \a' E^2 =\a' E_0^2   - 4  \zeta(2)\,  \gs^2 J^{-1}    +  \zeta^2(2)\,  \gs^4 J^{-3}  -\tfrac{38}{15}\zeta(6)\, \gs^6J^{-3}+ ... \ , \qquad\ \ \  \a' E_0^2 =4 J \ .   \la{002}
\ea
Below  we will  reproduce \rf{001}  by  the 1-loop M2 brane    computation  directly  in flat space  and extend this expression 
 to  all orders in expansion  in $\gs$. 

Considering M2 brane   in flat  11d  background  we have ($\mu=0, ..., 9$, \ $\a'=\ls^2$) 
 \ba  &  ds^2_{11}=d x^\m dx_\m  +  dx_{10}^2, \ \ \  \qquad \qquad  x_{10}\equiv  x_{10} + 2 \pi  R_{11} \ , \la{008} \\
  T= {1\ov 2\pi \ls^2 } = 2 \pi R_{11}  T_2 \ , \ \ \ \  &
 \ \    T_2= {1\ov (2\pi)^2 \lpp^3}=   {1\ov (2\pi)^2\ls^3\   \gs }   \ , \ \ \ \ \ R_{11} = \gs\,  \ls\ ,\ \  \ \ \  \ \lpp = \gs^{1/3}\, \ls \ .
 \la{007}
 \ea
 Here $T$ is the  tension of the 10d string  related  by  the double  dimensional reduction to  the M2  brane action
 with tension $T_2$. 
 As our aim is to compare the semiclassical 
 M2 brane expansion  to string perturbation theory, we express  the M-theory parameters in terms of the string theory ones. 
 
 Specialising  to the   classical M2 brane solutions with topology $\Sigma \times S^1$, where 
$S^1$ wraps the M-theory circle of radius $R_{11}$ and $\Sigma$ is the  world surface of the associated  10d string solution 
the corresponding quantum  M2  brane corrections to its energy are  organized  as an expansion in $T_2 ^{-1}=
(2\pi)^2 \ls^3 \, \gs$
as in \rf{1.4}, 
\be \la{71}
E= E_0  + {\rm E_1}    + T_2^{-1} {\rE_2} +  T_2^{-2} {\rm E_3} +  ...\  .
 \ee
Here  $\rE_r$  depend on dimensionless parameters of the solution (spins) 
  and on $R_{11}/\ls = \gs$.  Note that in the flat space case there are no non-trivial  tree-level (genus 0) 
 $\a'$ corrections to  string energies (i.e.    $E_0$   is just the classical  energy)   so that   $\rE_n$  will  represent   the genuine  M2  brane   corrections  depending on $\gs$. 
 Expanding \rf{71}  in small $\gs$   we get  a series of $\gs^{n}$ contributions  that may be compared with  higher 
  genus   string-theory corrections to the mass shift of the  corresponding semiclassical string state.

In  the above example of the  $J_1=J_2$  circular string solution we find   the  following  
 structure of 
the 1-loop M2  brane correction to its  energy (cf. \rf{001})\foot{Note that   $J$, $\bae$ and $\La$ are dimensionless  while   $E$ has  canonical dimension of  inverse length. } 
\ba\la{003} 
& E=E_0 +  \rE_1 \ , \ \ \quad  \    \rE_1 =  {1\ov \a' E_0} \, \bae  \ , \qquad
\ E^2= E_0^2 + \Delta E^2, \ \quad \ \   \a' \Delta E^2 =  2 \bae +   {  (\bae)^2\ov \a' E_0^2}    \ ,     \\ 
&\qquad  \bae = \sum_{n=1}^\infty c_n \zeta(4n-2)   {1  \ov \La^{4n-2}}  =    -  {2}  \zeta(2) \, {\gs^2\ov J} 
   -\tfrac{19}{15}\zeta(6)\, {\gs^6\ov J^3} + ...
       \ , \qquad\qquad   \ \ \ \ \   {1\ov \La^2} \equiv {\gs^2\ov 2 J} \ . \la{004}
\ea
Here $c_n $ are  rational  coefficients that we will determine below  for all $n$. 
 We  will also  discuss the case of a  folded  string with spin $J$ where $\bae$   should have a similar expansion in powers of $ {\gs^2  J^{-1}}$.\foot{Note that since $J=2\pi T \J= \J/\a'$ where   $\J$ is a  parameter  of a  spinning  string solution,  
 the   expansion  in  ${1\ov \La^2} = {\gs^2\ov 4 \pi T \J}$ is  analogous to  the expansion in ${1\ov k^2} = {\gs^2\ov 2\pi {\rm T}}$ in \rf{4}
 discussed above (cf. \rf{1.10}).}
 
 %calculation in flat space and also verify the result numerically. However, within the expansion we consider, no imaginary part is observed. 

   One  may try to  compare the $\gs^2$ term in \rf{002} or \rf{004} with the   1-loop mass shift  for  the corresponding string state
   $(\Delta E^2)_{\rm str} \equiv \Delta M^2 $ 
   that can be found    from the torus 2-point amplitude
   \be \la{799}
  \a'  \Delta E^2 = \gs^2 \big[{\rm R} (J)  + i\, {\rm I} ( J) \big] \ . \ee 
   Here  the real  and  imaginary parts are non-trivial functions of  $J\sim E_0^2  $. The presence of an 
   imaginary part  
   reflects the possibility of a decay of a massive states into lighter states.\footnote{By optical theorem, 
   the imaginary part is related to  the decay rate 
    as   $\Gamma ={1\ov 2 E_0}  \text{Im}(\Delta M^2)$.}. 
           The imaginary part  has the form  \ci{Iengo:2006gm,Chialva:2003hg,Chialva:2004xm}
       %    has the form 
\begin{equation}\la{80}
 {\rm I} ( J)    \sim   J^\gamma\  ,
\end{equation}
with $\gamma$ depending on a specific string state (for the  folded string  $\gamma={1\ov 2} $ \ci{Iengo:2006gm,Chialva:2003hg},
while  for the  rotating circular string   $\gamma= -{ 2}$ \ci{Chialva:2003hg,Chialva:2004xm}).  
The real part $ {\rm R} (J) $  of the torus correction is given by a complicated modular integral 
and  appears to be explicitly  known only  in an 
  ``averaged''  (over states at a given mass level)   form  %closed string states 
   \ci{Chialva:2009pg}, e.g., 
%\foot{The averaging is performed over
% the states at a given mass level $E_0$   and we have included
 % the result only for a zero Neveu-Schwarz charge.}
\begin{equation}\la{81}
  {\rm R} (J)   \sim   J^{-3/4} \ . 
\end{equation}
We conjecture that the  leading   M2 brane  result  $\sim \gs^2  J^{-1}$ in \rf{002}, \rf{004} 
may be reproducing the large $J$ asymptotics of  the function  $ {\rm R} (J) $  corresponding to the  spinning 
circular string state. The terms 
 subleading in large $J$   may 
be  captured by   higher-loop  M2 brane corrections in \rf{71}. % that are non-trivial to compute. 
A resummation of   higher loop terms in \rf{71}   may also produce 
an imaginary part in the resulting $\Delta E^2$  that should be present  in the string theory result
 but  absent  in the M2  brane 1-loop correction in \rf{002}, \rf{004}.

\iffa 
%%%%%%%%%%%%%%%%%%%%%%%%%
naively   M2  brane loop corrections   are
Delta E(g_s, J) =  f1 (g^2_s  J^{-1})
 + 1/T2  f2 (g^2_s  J^{-1})   + 1/T2^2  f3 (g^2_s J^{-1})  + ...
1/T2  ~ g_s
and since  pure string  correction =0   and  masses of KK excitations
are  such that 1/m^2 ~ g^2_s J^{-1}  that enter each propagator
one could think that it is unlikely that may resum into a correction
to just g^2_s  term
 but point   is that starting with 2-loop order  diagrams  like OO
may contain  one massless and one massive propagator
and may be that may lead to "non-analytic"   contributions in g_s...
This is  of course  pure speculation.  But that may  provide hope to
get  Im terms from higher M2   loops  resummed
though  it does look far-fetched...
%%%%%%%%%%%%%%%%%%%%%
\fi

\medskip

The rest of this paper is organized as follows. 
In section 2  we  review the 1-loop   correction to energy of long folded $(S,J)$   string in $\ads_{4}\times \cp$. 
Then in section 3  we compute the leading $1/k^2$ 
M2  brane corrections  to its energy or generalized cusp anomaly.
In section 4    we  present a similar  discussion of  the circular $(S,J)$ string in $\ads_{4}\times \cp$
while  in section 5   we generalize the computation  of  the 1-loop  correction to its energy to the M2  brane case. 
The  flat  space case is discussed in section 6. 
There are also several appendices with    some  details  of   computations  
in the main part.

%%%%%%%%%%%%%%%%%%%%%%%%%%%%%%%%%%%%%%%%
\section{Long folded $(S,J)$ string in $\ads_{4}\times \cp$}

%Many spinning string solutions explored in \cite{Frolov:2002av,Frolov:2003tu,Frolov:2003qc} for $\ads_{5}\times S^{5}$
%also serve as solutions for strings in $\ads_{4}\times \cp$, sharing many features, including quantum corrections.
 In this section  we review the structure of the 1-loop  correction  \cite{McLoughlin:2008ms} to the   energy 
 of the folded string with one spin  $S$  in  $\ads_{4}$
 and one angular momentum  $J$  in $\cp$. 
 Using the  coordinates  defined   in (\ref{A.2}), (\ref{A.5}), the   folded  string solution  in the 
 long string limit \rf{1.20}  \cite{Frolov:2006qe,Roiban:2007ju}  takes a  simple form 
\ba\la{31} 
& \te t = \kappa\,\tau, \quad  \rho=\mu\,\sigma, \quad \alpha = 0, \quad \beta = \kappa\, \tau, \qquad 
 \eta = \nu\, \tau, \quad \gamma=\frac{\pi}{4}, \quad \theta_{1,2}=\frac{\pi}{2}, \quad \phi_{1,2}=0,\\
\la{2.3}
&\qquad \qquad \kappa^{2}=\mu^2 +\nu^{2}, \quad \qquad x \equiv  \frac{\mu}{\nu} = \text{fixed},\quad  \qquad  \kappa, \nu, \mu  \gg 1 \ . 
\ea
Here  $\rho(\sigma)$  periodic in $\s\in(0, 2 \pi)$  is actually  a combination of 4 segments: it is 
given   by  $\mu\,\sigma$ for  $\sigma \in (0, {\pi\ov2})$, 
 by  $\mu\, (\pi -\sigma)$ for  $\sigma \in ({\pi\ov2}, \pi)$,  etc. 
% should represent  4 segments of a  folded string: $\rho$ increases for $0<\sigma<\pi/2$, then decreases, etc. 
The corresponding energy and the  two spins  are, in general,  defined as\foot{The string tension  in \rf{3} is  given by $\sbl \ov 2\pi$.  Note that  while  the radii  of $\ads_4$ and $\cp$ in \rf{A.7} 
 differ by 2, the   $d \eta$ term in \rf{A.5} that determines  the  angular momentum $J$ 
  has the  prefactor $\cos^2 \g \, \sin^2 \g = {1\ov 4}$  that compensates for this difference  which explains 
 why  \rf{34},\rf{33}  have  the same form  as in the similar \adss  case.}
%\footnote{Notice that our definition of $\E, \S$ is slightly different from  that in  \cite{McLoughlin:2008ms}.}
\ba & (E_0,S,J) =\sbl\,  (\E_0,\S,\J) \ , \qquad \qquad 
\E_0 = \frac{\kappa}{2\pi}\int_{0}^{2\pi}d\sigma\cosh^{2}\rho= \S + \k , \quad \la{34}\\
&\S = \frac{\kappa}{2\pi}\int_{0}^{2\pi}d\sigma\sinh^{2}\rho, \qquad \qquad \ \ \ \ \ \ 
\J = \frac{1}{2\pi}\int_{0}^{2\pi}d\sigma\, \nu=\nu.  \la{33}
\ea
In the $\mu \gg 1$ limit 
\ba\la{345}
\qquad \S = \frac{\kappa}{4\pi\mu}e^{\pi \mu}+\dots, \qquad\qquad \mu= {1\ov \pi}  \log\S +\dots,
%\S = 4\times\frac{\kappa}{2\pi}\int_{0}^{\pi/2}d\sigma\sinh^{2}(\mu\,\sigma).
\ea
and thus from \rf{2.3} we get 
\be\la{35}
\E_0 =  \S  + \J \sqrt{ 1 + x^2}+... = 
\S + \frac{1}{\pi }{\sqrt{1+x^2}\ov x}\log \S  +... \ ,  
\ee
where the coefficient  of $\log \S$ corresponds to the  leading term in the  scaling function in \rf{21}. 
%The case $\J=0$ corresponds to $x\to\infty$ and gives same classical scaling function $f_{0}(x)\to \sqrt{2\lambda}$.
%\subsection{One-loop correction}

To compute the 1-loop string correction to the energy one needs to start with    %sum up the corresponding fluctuation frequencies
 the quadratic fluctuation action and  find  the fluctuation frequencies $p_0$   \cite{Frolov:2006qe,McLoughlin:2008ms}.
 Using the static gauge and 
denoting by $p_1\in\mathbb Z$  the mode number  corresponding to the  periodic 
$\sigma$-direction  we 
find    for the   bosonic  and fermionic fluctuation frequencies\footnote{Some frequencies have constant shifts 
compared to the ones in \cite{McLoughlin:2008ms} as we use a different parametrization 
of the fluctuation fields adapted to our  choice of coordinates.   These shifts  cancel in the  contribution  to the energy.}
\begin{alignat}{2}
\la{2.11}
{\rm B}:\ \ \ &(p_{0})_{1} = \sqrt{p_1^{2}+2\kappa^{2}-\nu^{2}}, \qquad \ \ \ \ \ \ \ \ \ \ (p_{0})_{2,3} = (p_{0})_{4,5} = \pm\tfrac{1}{2}\nu+\sqrt{p_1^{2}+\tfrac{1}{4}\nu^{2}},\\
\la{2.12}
&(p_{0})_{6} = \sqrt{p_1^{2}+\nu^{2}}, \qquad \qquad \ \ \ \ \ \ (p_{0})_{7,8} = \sqrt{p_1^{2}+2\kappa^{2}\pm 2\sqrt{\kappa^{4}+p_1^{2}\nu^{2}}}\ , \\
\la{2.13}
{\rm F}:\ \ \  & (p_{0})_{1,2} = \sqrt{p_1^{2}+\kappa^{2}}, \qquad\qquad\qquad \ \ \ 
(p_{0})_{3,4} = \sqrt{p_1^{2}+\kappa^{2}}\pm \nu, \\
\la{2.14}
& (p_{0})_{5,6} = (p_{0})_{7,8} = \tfrac{1}{\sqrt 2}\sqrt{2p_1^{2}+\kappa^{2}\pm\sqrt{\kappa^{4}+4p_1^{2}\nu^{2}}}.
\end{alignat}
Then the  1-loop  correction to the energy is  given by
\be
\la{2.15}
E_{1} = \frac{1}{2\kappa}\sum_{p_1=-\infty}^{\infty}\sum_{\{p_{0}\}}(-1)^{\rm F}\, p_{0}\ .
\ee
In the  limit  $\kappa\gg 1$  the sum over the spatial mode  number $p_1$ may be converted into an integral,
and then one obtains  the expression \rf{1.22}  quoted  in the Introduction. 
%%%%%%%%%%%%%%%%%%%%%%%%%%%%%%%%%%%%%%%%%%%%
\iffa 
 \footnote{
As we mentioned in the Introduction, the relation (\ref{1.32}) holds (here we have only odd terms at large $J$)
and implies that  result in $\ads_{4}$ has the expected structure predicted by the
asymptotic Bethe Ansatz. In particular, the $\log 2$ term in (\ref{1.22})
comes from
classical expression of $E-S = J\sqrt{1+x^{2}}$ after using (\ref{1.32}): 
$J \sqrt{1 +  \frac{1}{\pi^{2}}\frac{\l}{J^{2}}\log^{2} S}\to \frac{1}{2}\times (2J)\sqrt{1+\frac{1}{\pi^{2}}\frac{\bar h(\lb)^{2}}{J^{2}}\log^{2} S}$,
and using $\bar h(\lb) = \sqrt{\lb}-\log 2+\dots$ we get 
$\sqrt{1+x^{2}}-\frac{1}{\sqrt{\lb}}\log 2\frac{x^{2}}{\sqrt{1+x^{2}}}+\dots$, 
in agreement with (\ref{1.22}).}
\fi 
%%%%%%%%%%%%%%%%%%%%%%%%%%%%%%%%%%%%%%%%
Let us note  that the 
 small/large  $x$  expansions of  $E_{1}$ in  the $\ads_{4}$ and  the $\ads_{5}$ cases read
 (cf. \rf{1.22})
\ba
\no % \la{2.16}
&x \ll 1:  \ \ E_{1}^{\ads_{4}} = \J\, (-\log 2\ x^{2}-\tfrac{2}{3}x^{3}+\tfrac{1}{2}\log 2\, x^{4}+\tfrac{2}{5}x^{5}+\dots),\no \\ 
 &\qquad \qquad  
E_{1}^{\ads_{5}} = \J\,(-\tfrac{4}{3}x^{3}+\tfrac{4}{5}x^{5}+\dots), 
\\
&x\gg 1: \ \  E_{1}^{\ads_{4}} = \J\, (-\tfrac{5}{2}\log 2\ x+\tfrac{3+2\log 2+4\log x}{4x}+\dots), \no \\ &\qquad \qquad  
E_{1}^{\ads_{5}} = \J\,(-3\log 2\ x+\tfrac{3+4\log x}{2x}+\dots).\la{2.16}
\ea

%Two-loop string correction to energy in the considered limit
 %was found in \cite{Giombi:2010fa} and similar result for the two-loop contribution to  $h(\l)$ in  \cite{Bianchi:2014ada}.

\section{1-loop energy of long folded $(S,J)$ M2 brane  in $\ads_{4}\times S^{7}/\mathbb{Z}_{k} $}
%$\ads_{4}\times \cp$}
%\section{M2 brane generalization of long folded $(S,J)$ solution}

The above  folded   string  solution has  a straightforward 
uplift to  the M2 brane solution in  $\ads_{4}\times S^{7}/\mathbb{Z}_{k} $ 
with   the second spatial  direction of the  membrane   wrapped on the 11d circle $\vp\in (0,2\pi) $ in \rf{A.3}.
The values of the classical membrane action in \rf{a9} (given just by the volume term) 
 and of  the   conserved charges 
are then the same  as in  \rf{34},\rf{33}  (the M2 brane \rf{2}  and the string tensions  are related by 
\rf{3}). 

To  find the 1-loop  M2 brane correction to  the energy \rf{2.15}   we will use  as in 
\cite{Giombi:2024itd}  
the static gauge 
relating the world-volume coordinates $\xi^i=(\tau, \s, \s')$   to  the target space coordinates 
$t, \rho$ and $\vp$. % (that will not then have fluctuating parts). 
The  classical  value of the  induced 3d metric  corresponding to the solution \rf{31} and $\vp=\s'$ 
 is then (see  \rf{A.1},\rf{a9})
\be\la{02}
\bag_{ij} = \frac{\L^{2}}{4}
\begin{pmatrix}
-\mu^{2} & 0& 0 \\
0 & \mu^{2} & 0  \\
0 & 0 & \frac{4}{k^{2}}
\end{pmatrix}, \qquad \qquad 
\sqrt{-\bag } = \frac{\L^{3}}{4k}\, \mu^{2}.
\ee

\def \mk   {{\rm K}}

%%%%%%%%%%%%%%%%%%%%%%%%%%%%%%
\subsection*{Bosonic fluctuations}

The fluctuations of  the  8 ``transverse''  bosonic coordinates (denoted by a hat)
  will be defined  as  (cf. \rf{31})
\ba\la{03}
& \alpha =\frac{1}{\sinh(\mu\,\sigma)}\ \hat\alpha(\xi), \qquad\qquad 
\beta = \kappa\tau+\frac{1}{\sinh^{2}(\mu\,\sigma)}\hat\beta(\xi)\ , \\
&\eta = \nu \tau+\hat\eta(\xi), \qquad
\gamma = \tfrac{\pi}{4}+\hat\gamma(\xi), \qquad
\theta_{1,2} = \tfrac{\pi}{2}+\hat\theta_{1,2}(\xi), \qquad
\phi_{1,2} = \hat\phi_{1,2}(\xi).
\ea
Expanding the volume part of the  M2 brane action \rf{a9} to quadratic  order in fluctuations 
we get 
\be
\la{3.9}
S_{\rm V}^{(2)} = \int d^{3}\xi \sqrt{-\bag }\, L_{\rm V}^{(2)}, \qquad
\qquad  L_{\rm V}^{(2)} = \tfrac{1}{2}\hat\Phi_{p}\big[\mk_{\rm V}(\partial_{i})\big]_{pq}\hat\Phi_{q}, 
\ee
where $M_{\rm V}$ is a 2nd order differential operator with constant  coefficients in the $\mu \gg 1$ limit  (cf. in \rf{2.3}). 
% operator  with co
Similarly, expanding the  WZ  term in \rf{a10} gives 
(in the long string limit   $\coth(\mu\sigma)\to 1$)
\ba %SK2
S_{\rm WZ}^{(2)} =& -T_{2} (-\tfrac{3}{8}\L^{3})\int \cosh(\mu\sigma)\sinh^{2}(\mu\sigma)\frac{1}{\sinh(\mu\sigma)}\hat\alpha\, 
d(\kappa\tau)\ww (\mu d\sigma)\ww \frac{1}{\sinh^{2}(\mu\sigma)}\partial_{\s'}\hat\beta\, d\s'\lp
\to  \tfrac{3}{8}\T_{2}\kappa\mu\int d\tau d\sigma d\s'\ \hat\alpha\partial_{\s'}\hat\beta
= \int d^{3}\xi \sqrt{-\bag }\, L_{\rm WZ}^{(2)} ,\qquad \qquad
L_{\rm WZ}^{(2)} = \frac{3k \kappa}{2\mu}\, \hat\alpha\partial_{\s'}\hat\beta.\la{3.11}
\ea
Combining \rf{3.9}   and \rf{3.11}   we get the total  kinetic  operator 
$\mk(\del_i)  = \mk_{\rm V}+\mk_{\rm WZ}$  so that  the equation  for fluctuations reads 
%,  and the (bosonic) equations of motion  $\delta L_{2}=0$, given by
\be\la{36} %SK2
\big[\mk(\partial_{i})+\mk^{\text{\sc T}}(-\partial_{i})\big]_{pq}\, \hat\Phi_{q}=0 \ , \qquad \qquad 
\hat \Phi_{q} = \sum_{p_{1}, p_{2}\in\mathbb Z}\int {d p_0 \ov 2\pi} \ \td \Phi_{q}(p_0,p_1, p_{2}) \ e^{i (p_{0}\tau+p_{1}\sigma+p_{2}\s')} \ . 
\ee
The characteristic frequencies  $p_0=p_0 (p_1,p_2)$ can be found from the   equation 
\ba
\la{3.14}
& \qquad \qquad \mc D_{\rm B}(p_{0},p_{1},p_{2}) \equiv \det\big[\mk(ip_{i})+\mk^{\text{\sc T}}(-ip_{i})\big]=0 \ , \\
&
\mc D_{\rm B}(p_{0},p_{1},p_{2}) = P_{8}(p_{0},p_{1},p_{2})\prod_{s_{1},s_{2}\in\{-1,1\}}
\Big[p_{0}^{2}-p_{1}^{2}+s_{1}p_{0}\sqrt{\kappa^{2}-\mu^{2}}-\tfrac{1}{4}\mu^{2}kp_{2}(kp_{2}+2s_{2})\Big]\ .
\ea
Here  $P_{8}(p_{0},p_{1},p_{2})$ is a  complicated polynomial of degree 8 in  $p_{0}$  that we do not give explicitly here.
In the string theory  limit, i.e.  for $p_{2}=0$, one recovers the expressions for 
 the bosonic  fluctuation 
frequencies  given  in (\ref{2.11}),(\ref{2.12}).

%%%%%%%%%%%%%%%%%%%%%%%
\subsection*{Fermionic fluctuations}

To find the fermionic fluctuation frequencies we will follow the approach described in 
%The membrane can be viewed as a 3d surface in 11d spacetime and it is convenient to adopt the formalism of 
Appendix A   of \cite{Giombi:2024itd}. 
One defines
an orthonormal basis $\re_{i}=\re_{i}^{M}\partial_{M}$ on the membrane 3d world volume
% ($i,j= 0,1,2$ and $A,B$ are tangent-space 11d indices) 
 satisfying %\foot{{\bf Notation?  $\mu$ ?! how  consistent with rest?}}
$
\langle \re_{i}, \re_{j}\rangle = G_{MN}\,\re_{i}^{M}\re_{j}^{N} =
 \eta_{ij}
$   (cf. \rf{a9}).
Using the fact that the induced metric \rf{02}  is diagonal with constant coefficients, it is enough to take $\re_{i}^{M}$
to be proportional to $\partial_{i}X^{M}$, i.e.
%Fixing the proportionality constants gives
\be \label{317}
\re_{0} = \frac{2}{\mu}(\kappa\partial_{t}+\kappa\partial_{\beta}+\nu\partial_{\eta}), \qquad\qquad  \re_{1} =\frac{2}{\mu}\partial_{\rho}, \qquad\qquad  \re_{2}= k\partial_{\varphi}.
\ee
An orthonormal basis in the normal bundle can be chosen as  %$\{\rn_{p}\}_{p=1,\dots, 8}$
\ba \la{318}
\rn_{1} =& \frac{2}{\sinh \rho }\partial_{\alpha}, \qquad
\rn_{2} = 2(\tanh \rho \ \partial_{\rho}+\coth \rho \ \partial_{\beta}), \qquad
\rn_{3} =\frac{2}{\mu}(\nu\partial_{\rho}+\nu\partial\beta+\kappa\partial_{\eta}), \nonumber \quad
 \\
\rn_{4}=&\partial_\gamma, \qquad \rn_{5} = 2\sqrt{2}\partial_{\theta_{1}},\qquad
\rn_{6} = 2\sqrt{2}\partial_{\theta_{2}},\qquad
\rn_{7} = 2\sqrt{2}\partial_{\phi_{1}},\qquad
\rn_{8} = 2\sqrt{2}\partial_{\phi_{2}}.
\ea
One can introduce a dual basis  satisfying $
\re^{i}(\re_{j}) = \delta^{i}_{j}, \ \ \rn^{p}(\rn_{q}) = \delta^{p}_{q}, \ \  \rn^{p}(\re_{i}) = \re^{i}(\rn_{p}) = 0.
$
We define the set of 11d gamma matrices with respect to the orthonormal frame \eqref{317}, \eqref{318}, \ie we introduce $\Gamma_{A}=(\rho_{i}, \gamma_{p})$ \foot{Here   $A$ is the 11d tangent space index; for the explicit  form 
of the matrices see  Appendix \ref{app:basis}.} %such that
\ba\la{311}
 &\{\rho_{i}, \rho_{j}\} = 2\eta_{ij}{\rm I}_{32}\, , \qquad i = 0, 1, 2; \qquad  \qquad \qquad 
 \{\gamma_{p}, \gamma_{q}\} = 2\delta_{pq}\, {\rm I}_{32}\,, \qquad p=1, \dots, 8.
\ea
Assuming  the $\kappa$-symmetry gauge   $(1+\Gamma)\theta=0$
the  quadratic term in  the fermionic part of the M2 brane   action   may be written as ($\rho^i=  \eta^{ij} \r_j$)
\ba\la{312} %SK1
&  S_{\rm F} = T_{2}\int d^{3}\xi \sqrt{-\bag }\,\bar\theta(1-\Gamma)\rho^{i}D_{e_{i}}\theta,\qquad \ \ \  \qquad \Gamma = \tfrac{1}{3!}\eps^{ijk}\rho_{i}\rho_{j}\rho_{k} \ . 
\ea
The   corresponding Dirac operator is  % appearing can be written as
\ba
& \slashed{D} = \rho^{i}D_{e_{i}} = \rho^{i}\Big[\nabla_{e_{i}}+\tfrac{1}{12}\big(\rho_{i}{\rm F}_{4}-3 {\rm F}_{4,i}\big)\Big], \qquad \qquad \nabla_{e_{i}}=\partial_{e_{i}}
+\tfrac{1}{4}\Omega^{AB}_{i}\Gamma_{AB}, \la{313} \\
& {\rm F}_{4}=\tfrac{1}{4!}F_{ABCD}\, \Gamma^{ABCD}, \qquad \ \ \ {\rm F}_{4,i}=\tfrac{1}{3!}F_{iBCD}\, \Gamma^{BCD}, 
\ea
where  $\Omega^{AB}(e_{i})$ is the spin connection on $\ads_{4}\times S^{7}/\mathbb Z_{k}$. 
Using the frame \eqref{317}, \eqref{318}, we get  %an explicit form for $\slashed{D}$:
\be\la{315}
 \slashed{D} = \rho^{i}\nabla^{\perp}_{e_{i}} +\frac{3\kappa}{2\mu}\rho^{1}\rho^{2}\gamma^{1}\gamma^{2},\qquad \ \ \ 
 \qquad \nabla^{\perp}_{e_{i}} = \partial_{e_{i}}+\tfrac{1}{4}\Omega^{pq}_{i}\gamma_{pq},
 \ee
 with the  normal bundle connection $\Omega^{pq}_ i = \langle \rn^{p}, \nabla_{e_{i}}\rn^{q}\rangle$ given  in the Appendix \ref{app:basis}.\footnote{As in the cases discussed in Appendix A of \cite{Giombi:2024itd}, 
 the membrane metric \rf{02} is flat,  implying  $\Omega^{ij}_{k} = 0$.  Also,  since the classical solution  represents 
  a minimal surface (WZ term does not contribute at the classical level)  
% In addition, since the interaction with the RR-form does not contribute at the level of the equations of motion, the %membrane is a minimal surface, and t
 the term involving $\rho^{i} \rho^{j} \Omega_{jp ,  i}$ vanishes.
}

The resulting  determinant of the fermionic operator   computed   for a single Fourier mode 
 as in \rf{3.14} 
 has the  following factorized structure
\ba
\mc D_{\rm F}(p_{0},p_{1},p_{2})& = P_{4}(p_{0},p_{1},p_{2})\ P_{4}(-p_{0},p_{1},p_{2})\ \tilde P_{4}(p_{0},p_{1},p_{2})\ \tilde P_{4}(p_{0},p_{1},-p_{2}),\la{3155}\\
P_{4}(p_{0},p_{1},p_{2}) =& 
p_0^4
-2 \nu  p_0^3
+p_0^2 (-2 \kappa ^2+\nu ^2-2 p_1^2-\tfrac{1}{2} k^2 \mu^2  p_2^2)
+\tfrac{1}{2} \nu  p_0 (4 \kappa ^2+4 p_1^2+k^2 \mu^2  p_2^2)\lp
+p_1^4
+\tfrac{1}{2} p_1^2 (4 \kappa ^2-2 \nu ^2+k^2 \mu^2  p_2^2)
+\tfrac{1}{16} \mu^2  (16 \kappa ^2-8 k^2 \kappa ^2 p_2^2+k^4 \mu^2  p_2^4), \\
%%%
\tilde P_{4}(p_{0},p_{1},p_{2}) =& 
p_0^4 - p_0^2( 2p_1^2+\kappa^2+\tfrac{1}{2}\mu^2kp_2(kp_2-2)) 
+(p_1^2+\tfrac{1}{4}\mu^2k^2p_2^2)(p_1^2+\tfrac{1}{4}\mu^2(kp_2-2)^2).\no
\ea
The characteristic frequencies $p_0(p_1,p_2)$ are found by solving $\mc D_{\rm F}(p_{0},p_{1},p_{2})=0$. 
In the string theory  limit $p_{2}=0$ they reduce to the ones given in  \eqref{2.13}, \eqref{2.14}. 

\subsection*{1-loop correction to energy } %Membrane correction to the energy and large $k$ expansion}

The 1-loop correction to the energy generalizing \rf{2.15}  to the M2 brane case 
 can be represented  in terms of the  determinants  in \rf{3.14}  and \rf{315} as (see, e.g.,  \cite{Beccaria:2012xm,Giombi:2024itd}) 
\ba
\la{3.30}
{\rm E}_{1} =& \frac{1}{2\kappa}\sum_{p_{1},p_{2}\in\mathbb Z}
\int_{-\infty}^{\infty}{dw\ov 2\pi}\, \log\frac{\mc D_{\rm B}(-iw, p_{1}, p_{2})}{\mc D_{\rm F}(-iw, p_{1}, p_{2})}
= E_1 + E^{\rm M2}_{1} \ .
\ea
Here the  $p_{2}=0$ term in the sum  represents the string  mode contribution $E_1$ in \rf{2.15},\rf{1.22},
 while the terms with $p_{2}\neq 0$  give the  additional M2 brane  mode contribution   denoted as $E^{\rm M2}_{1} $
as  in  \rf{26}.
%SK
We can evaluate this part in the large $k$ limit by converting the sum over $p_{1}$ into an integral in the long string limit,
by  rescaling $w$ and $p_{1}$ by  $k$
 and  then expanding in  large $k$. This yields
\ba
\no E_{1}^{\rm M2}(x, \J, k) =& \frac{k^{2}}{2\pi\kappa}\sum_{p_{2}=1}^{\infty} 
\int_{-\infty}^{\infty}dw'\, \int^\infty_{-\infty} dp'_{1}\, \log\frac{\mc D_{\rm B}(-ikw', kp'_{1}, p_{2})}{\mc D_{\rm F}(-ikw', kp'_{1}, p_{2})} \\
=& \sum_{p_{2}=1}^{\infty}\Big[
\mc C_{2} \frac{1}{(kp_{2})^{2}}+\mc C_{4}\frac{1}{(kp_{2})^{4}}+\dots\Big] 
= \mc C_{2}\frac{\zeta(2)}{k^{2}}+\mc C_{4}\frac{\zeta(4)}{k^{4}}+\dots, \la{3.31}
\ea
where 
\be\la{334}
\mc C_{2} = \frac{2(2\kappa^{2}+3\nu^{2})}{\kappa}, \qquad\qquad \qquad 
\mc C_{4} = \frac{8\kappa^{2}+76\kappa^{2}\nu^{2}-3\nu^{4}}{2\kappa(\kappa^{2}-\nu^{2})}.
\ee
 For  some  higher order $\mc C_{2n}$ terms see  Appendix \ref{app:higher}.
Expressing these  coefficients  in terms of $\J=\nu$ and $x$, we get  %a linear dependence on $\J$
\ba
&\mc C_{2n} = \J\, C_{2n}(x), 
\qquad \ \ 
C_{2}(x) = \frac{2(5+2x^{2})}{\sqrt{1+x^{2}}} = 4\sqrt{1+x^{2}}+\frac{6}{\sqrt{1+x^{2}}},  \la{336}\\
&C_{4}(x) = \frac{81+92x^{2}+8x^{4}}{2x^{2}\sqrt{1+x^{2}}}  = 4\frac{(1+x^{2})^{3/2}}{x^{2}}+38\frac{\sqrt{1+x^{2}}}{x^{2}}
-\frac{3}{2}\frac{1}{x^{2}\sqrt{1+x^{2}}}.\la{337}
\ea
%where we have separated the factor  present in classical energy
%so that the remaining terms are subleading at  small   $\J$   when $x$ is  large.
Since $\J x = \frac{\log\S}{\pi}$ (see  (\ref{1.20}))
the expression   (\ref{3.31}) gives the $1/k$ corrections to  the   generalized  cusp anomaly 
$q_{0}(x,k)$ in \rf{1.9} (\ref{1.24}) already quoted  in  \rf{1.24a} %SK4
\be\la{340}
\bar q_{0}(x,k) = \frac{C_{2}(x)}{6 x}\frac{\pi}{k^{2}}+\frac{C_{4}(x)}{90 x}\frac{\pi^{3}}{k^{4}}+\dots \ . 
\ee
The  small/large  $x$   expansions  of \rf{336},\rf{337} are %we have 
\ba
\la{3.36}
x \ll1: \ \ \ C_{2}(x) =& \te 10-x^{2}+\frac{7}{4}x^{4}+\dots, \qquad\qquad 
C_{4}(x) = \frac{81}{2x^{2}}+\frac{103}{4}-\frac{61}{16}x^{2}+\dots, 
\\
x \gg 1: \ \ \ C_{2}(x) =&\te 4x+\frac{8}{x}+\dots, \qquad\qquad \qquad \ \ \ \ 
C_{4}(x) = 4x+\frac{44}{x}+\dots\ . \la{339}
\ea

 %%%%%%%%%%%%%%%%%%%%%%%%%%%%
 \section{Circular $(S,J)$  string   in $\ads_{4}\times \cp$ \la{sec3}}

%\section{Circular $(S,J)$ string}
Let us now  consider  the   circular   string solution with  spins $S$ and $J$  in $\ads_{4}\times \cp$ 
which was  discussed in  \cite{McLoughlin:2008he} (for the analogous solution in $\ads_{5}\times S^{5}$ see \cite{Arutyunov:2003za,Park:2005ji}).
Using the coordinates in \rf{A.2},\rf{A.5}   it  is described by 
%This configuration resides in $\ads_{3}\times S^{1}\subset \ads_{4}\times \cp$. Its projection onto $\ads_{3}$ forms a  constant radius circle with winding number k, 
%while its projection onto $\cp$ traces a big circle with winding number m.
%The relevant 10d metric is
%\be\la{4.1}ds^{2}_{\ads_{4}\times S^{1}} = -\cosh^{2}\rho\, dt^{2}+d\rho^{2}+\sinh^{2}\rho\, (d\alpha^{2}+\cos^{2}\alpha\, d\beta^{2})+d\eta^{2},\ee
%where $\eta$ is a $2\pi$-periodic angle in $\cp$ (sum of two $\pi$-periodic angles). Using a conformal gauge with flat worldsheet metric, 
% the classical solution is 
\ba
\la{4.2}
t = \kappa\, \tau, \qquad \rho=\rho_{*}=\const, \qquad \alpha =0, \qquad \beta = w\, \tau+{\rmk}\, \sigma, \qquad 
\eta =& \omega\, \tau+{\rm m}\, \sigma,
\ea
with   other angles  being  trivial and $\rmk$ and $\rm m$  being   integer winding numbers. 
The conformal gauge (Virasoro) constraints read ($r_{0}=\cosh\rho_{*}$, $r_{1}=\sinh\rho_{*}, \  r_0 = \sqrt{ r^2_1+1}$)
\ba
\la{4.4}
w^{2}-(\kappa^{2}+{\rmk}^{2}) = 0, \qquad r_{1}^{2}w\,{\rmk}+\omega\, {\rm m} = 0, \qquad -r_{0}^{2}\kappa^{2}+r_{1}^{2}(w^{2}+{\rmk}^{2})+\omega^{2}+
{\rm m}^{2}=0.
\ea
The three conserved  charges  are  as in \rf{34} given by 
\be
\la{4.5}(E_0,S,J) =\sbl\,  (\E_0,\S,\J) \ , \qquad \qquad  
\E_0 = r_{0}^{2}\,\kappa, \qquad \S = r_{1}^{2}\,w, \qquad \J = \omega.
\ee
The second constraint in (\ref{4.4})  implies that 
\be
{\rmk}\,S+{\rm m}\, J = 0.\la{44}
\ee
In the following we will  set 
\be
{\rmk}=1, \qquad \ \  \qquad  u \equiv  \frac{\S}{\J}  = -{\rm m} \ .\la{45}
\ee
The dependence on $\rmk$ can be restored by rescaling of $\J$   and we will  formally treat $u$ as a continuous parameter. 
The solution with minimal energy for given values of spins corresponds to $u=-{\rm m}=1$. 
The conditions \rf{4.4}  can be solved in terms  of   independent   parameters $u, r_{1}$ as 
\be
\la{4.9}
\omega = r_{1}^{2}\frac{\sqrt{1+2r_{1}^{2}+u^{2}}}{\sqrt{u^{2}-r_{1}^{4}}}, \qquad
w = \frac{u\omega}{r_{1}^{2}} = u\frac{\sqrt{1+2r_{1}^{2}+u^{2}}}{\sqrt{u^{2}-r_{1}^{4}}}, \qquad \kappa = \frac{r_{1}^{2}+u^{2}}{\sqrt{u^{2}-r_{1}^{4}}}, 
\ee
%together with the obvious $r_{0}=\sqrt{1+r_{1}^{2}}$. 
so that 
%From (\ref{4.9}) we can express the classical energy as  
\be
\la{4.10}
\E_0 = \frac{(1+r_{1}^{2})(r_{1}^{2}+u^{2})}{\sqrt{u^{2}-r_{1}^{4}}}.
\ee
In the case of $S=J$ or  $u=1$   one finds  explicitly $r_{1}^{2} = \tfrac{1}{4}\J(\sqrt{8+\J^{2}}-\J)$   and thus 
 %we can provide exact expressions, instead of series expansions. From Virasoro constraints, one gets 
\be\la{48}
u=1: \qquad  \qquad \ \ \ 
\E_0(\J) = \sqrt{1+\tfrac{5}{2}\J^{2}-\tfrac{1}{8}\J^{4}+\tfrac{1}{8}\J \sqrt{(8+\J^{2})^{3}}}
%\J = \sqrt{2}\frac{r_{1}^{2}}{\sqrt{1-r_{1}^{2}}}, \qquad
 \ . 
\ee
For general $u$ one   can expand $\E_0$  at large or small $\J$, corresponding to $r_{1}\to \sqrt{u}$ or $r_{1}\to 0$:
\ba
\la{4.11}
\J\gg1: \ \ \ \   &r_{1}^{2}  = u-\frac{1}{2}u(1+u)^{2}\frac{1}{\J^{2}}+\frac{1}{8}u(1+u)^{2}(3+u)(1+3u)\frac{1}{\J^{4}}+\dots,\\
\la{4.12}
&\E_0(\J,u) %\stackrel{\J\gg 1}
{=} (1+u)\,\J+\frac{1}{2\J}u(1+u)-\frac{1}{8\J^{3}}u(1+u)(1+3u+u^{2})+\dots,\\
\J\ll 1: \ \ \ \   &r_{1}^{2} = \frac{u}{\sqrt{1+u^{2}}}\J-\frac{u^{2}}{(1+u^{2})^{2}}\J^{2}-\frac{u(1-3u^{2}+u^{4})}{2(1+u^{2})^{7/2}}\J^{3}+\frac{2u^{2}(-1+u^{2})^{2}}{(1+u^{2})^{5}}
\J^{4}+\dots,\la{4.13} \\
&\E_0(\J, u) {=} u+\sqrt{1+u^{2}}\, \J+\frac{u}{2(1+u^{2})}\J^{2}-\frac{u^{2}}{2(1+u^{2})^{5/2}}\J^{3}+\dots\ . \la{4.14}
%\lp
%+\frac{1}{16\J^{5}}u(1+u)(1+7u+13u^{2}+7u^{3}+u^{4})+\mc O(1/\J^{7}).
\ea
The expression  in \rf{4.12} starts with the familiar  ``fast-string'' term $\E_0 = (1+u) \J + \dots = \S + \J + \dots$, while 
 the first term in \rf{4.14} $\E_0(\J, u) {=} u+... = - {\rm m} + ...$   is  the   contribution to the energy due to  the string winding the circle.
A comparison of  (\ref{4.12}) and (\ref{4.14})  for $u=1$ with the exact  expression for $\E_0(\J)$ in \rf{48} 
  is  illustrated  in Fig. \ref{fig:a}.
\begin{figure}[H]
\centering
\includegraphics[width=0.7\textwidth]{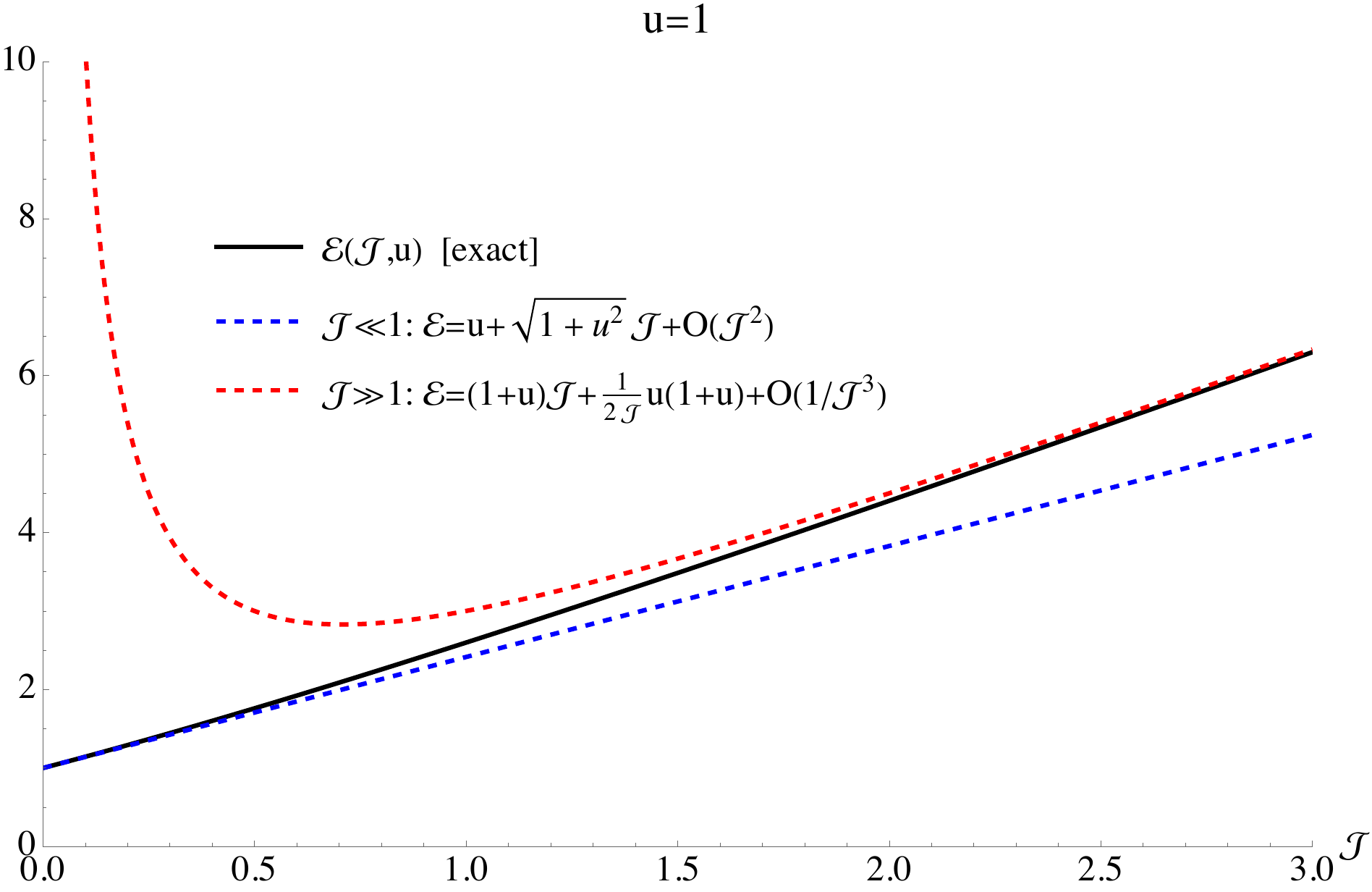}
\caption{Comparison of the exact expression $\E_0(\J,u=1)$ with its  small and large $\J$ expansions.}
% (we kept the terms in legend). All curves are drawn for $u=1$.}
\la{fig:a}
\end{figure}
To compute  the 1-loop correction to the  energy one needs to combine the fluctuation frequencies $p_0=p_0(p_1)$  as in \rf{2.15}.
Fixing  the static gauge  on $t$ and $\eta$  one finds as in \cite{McLoughlin:2008he} that\footnote{As in the folded string case   discussed above, we  note  
 that different parametrizations of fluctuations may introduce constant shifts of $p_0$ but these  
do not affect the 1-loop correction to the energy.}  
\ba
&\cp: \ \ \ \ \ (p_{0})_{1,2,3,4} = \sqrt{p_{1}^{2}+\tfrac{1}{4}(\J^{2}-u^{2})},\qquad \qquad \la{413}
(p_{0})_{5} = \sqrt{p_{1}^{2}+\J^{2}-u^{2}} \ , \\
&\ads_4: \ \ \ \ \ 
p_{0} = \sqrt{p_{1}^{2}+\kappa^{2}}, \qquad 
\la{4.19}
(p_{0}^{2}-p_{1}^{2})^{2}+4r_{1}^{2}\kappa^{2}p_{0}^{2}-4(1+r_{1}^{2})\big(p_{0}\sqrt{\kappa^{2}+1}-p_{1}\big)^{2}=0,  % \ \ p_{0}=(p_{0})^{\ads}_{2,3}.
\ea
where  two  out   of the  total three     $\ads_4$  frequencies are given by  the positive  solutions  of the quartic  equation in \rf{4.19}. 

The fermionic spectrum contains four different frequencies, each being doubly-degenerate.  Two  are given by  \cite{McLoughlin:2008he}
\be
\la{4.20}
{\rm F}: \ \ \ \ p_{0} = \pm \frac{r_{0}^{2}\kappa u}{2(u^{2}+r_{1}^{2})}+\sqrt{(p_{1}\pm b)^{2}+\J^{2}+r_{1}^{2}}, \ \ \ \ 
\qquad b \equiv  -\frac{\kappa u}{w}\frac{w^{2}-\J^{2}}{2(u^{2}+r_{1}^{2})},\ \  %\ \ \ \ \ p_{0}=(p_{0})^{F}_{1,2}.
\ee
and two   are the solutions of the equation which is very similar to the one in \rf{4.19} 
\be
\la{4.21}
{\rm F}: \ \ \ \ (p_{0}^{2}-p_{1}^{2})^{2}+r_{1}^{2}\kappa^{2}p_{0}^{2}-(1+r_{1}^{2})\big(p_{0}\sqrt{\kappa^{2}+1}-p_{1}\big)^{2}=0 \ . %,\ \ \ p_{0}=(p_{0})^{F}_{3,4}.
\ee
The resulting 1-loop correction $E_{1}$  is given  by the sum (\ref{2.15}). It  
can be  computed explicitly    in the large $\J$  expansion  (cf.  \rf{1.33},\rf{1.34}),  where  the ``odd'' term part   is   related  \rf{1.19} 
to the corresponding \adss   expression \rf{1.33}. 
 The small $\J$ expansion of $E_1$  was not studied previously   and we present  it in  Appendix \ref{app:circular-short-oneloop}.

%\section{M2 brane generalization of circular $(S,J)$ solution}
\section{1-loop energy of circular  $(S,J)$ M2 brane  in $\ads_{4}\times S^{7}/\mathbb{Z}_{k} $}
\la{sec:fermions-circular}

Let us now consider the   direct  uplift of the circular string solution of the previous section 
 to the M2 brane solution  in $\ads_{4}\times S^{7}/\mathbb{Z}_{k}$   where in addition to  \rf{4.2}
 we set   $\vp = \s'$ as in section 3.  Using \rf{4.9}  the  resulting induced metric  may be written as    (cf. \rf{02}). 
\be
\bag_{ij} = \frac{\L^{2}}{4}
\begin{pmatrix}
-(r_{1}^{2}+u^{2}) & 0& 0 \\
0 & r_{1}^{2}+u^{2} & 0  \\
0 & 0 & \frac{4}{k^{2}}
\end{pmatrix}, \qquad \qquad 
\sqrt{-\bag } = \frac{\L^{3}}{4k}(r_{1}^{2}+u^{2}). \la{51}
\ee
The classical   conserved  charges are the same as in \rf{4.5} (WZ term  in \rf{a10}  does not contribute at the classical level). 
Choosing the static gauge where $t$, $\beta$  and $\vp$  do not fluctuate  we have  ($\rmk=1$)
\ba
& t = \kappa\tau, \qquad
\rho = \rho_{*}+\hat\rho(\xi), \qquad
\alpha =\hat\alpha(\xi), \qquad
\beta = w\, \tau+\sigma, \\
& \eta = \omega \tau+\rm m\,  \sigma+\hat\eta(\xi), \quad
\gamma = \tfrac{\pi}{4}+\hat\gamma(\xi), \quad
\theta_{1,2} = \tfrac{\pi}{2}+\hat\theta_{1,2}(\xi), \quad
\phi_{1,2} = \hat\phi_{1,2}(\xi), \qquad \vp=\s' \ . 
\ea
 
 % background, wrapping around the 11-dimensional
%circle, is done by the same construction as  in the long folded string case. At  the classical level, the solution remains again 
%unchanged, except for the addition of the $\phi$ coordinate, and the action is same.  

\subsection*{Bosonic fluctuations}

The expansion of the volume part of the M2 brane action gives the  quadratic fluctuation  Lagrangian of the same form as in (\ref{3.9}) while  the analog of \rf{3.11} is 
%The WZ term contributes at this order and reads
\ba %SK2
S^{(2)}_{\rm WZ} =& -T_{2} (-\tfrac{3}{8}\L^{3})\int \cosh\rho_{*}\sinh^2\rho_{*}\, \hat\alpha\, d(\kappa\, \tau)\ww d\hat\rho\ww 
d(w\, \tau+\sigma)
= \tfrac{3}{8}\T_{2}r_{1}^{2}r_{0}\kappa\int d\tau d\sigma d\s'\ \hat\alpha\partial_{\s'}\hat\rho,\no \\
&\qquad \qquad L^{(2)}_{\rm WZ} = \frac{3}{2} \frac{k r_{1}^{2}r_{0}\kappa}{r_{1}^{2}+u^{2}}\, \hat\alpha\partial_{\s'}\hat\rho. \qquad 
\ea
The bosonic fluctuation frequencies are determined by solving the analog of 
eq. (\ref{3.14}).
  In the string theory  limit ($p_{2}=0$)  one  finds   for the characteristic polynomial 
% (up to irrelevant numerical constants) %the factorized product
\ba
\la{5.7}
\mc D_{\rm B}(p_{0},p_{1},0) =& (p_{0}^{2}-p_{1}^{2}-\kappa^{2})(p_{0}^{2}-p_{1}^{2}-\omega^{2}+u^{2}) \lp
\times [(p_{0}+\tfrac{1}{2}\omega)^{2}-(p_{1}-\tfrac{1}{2}u)^{2}-\tfrac{1}{4}(\omega^{2}-u^{2})]^{2}\, 
[(p_{0}-\tfrac{1}{2}\omega)^{2}-(p_{1}+\tfrac{1}{2}u)^{2}-\tfrac{1}{4}(\omega^{2}-u^{2})]^{2}\lp
\times [(p_{0}^{2}-p_{1}^{2})^{2}+4r_{1}^{2}\kappa^{2}p_{0}^{2}-4(1+r_{1}^{2})(\sqrt{1+\kappa^{2}}\,p_{0}-p_{1})^{2}].
\ea
Then the solutions   of $  \mc D_{\rm B}(p_{0},p_{1},0) = 0$   are equivalent   to the frequencies  in \rf{2.11},\rf{2.12} (up to constant shifts related  to the  choice of parametrization
of fluctuations).  For generic $p_{2}$ we  get
\be
\la{5.8}
\mc D_{\rm B}(p_{0},p_{1},p_{2}) = P_{4}(p_{0},p_{1},p_{2})\,  P_{4}(-p_{0},-p_{1},+p_{2})\, P_{8}(p_{0},p_{1},p_{2}),
\ee
where $P_{r}$ stands for a polynomial of degree $r$ in $p_{0}$. From the  zeroes of the two $P_{4}$ factors   we get % the 8 exact frequencies
\ba
\la{5.9}
p_{0} = \tfrac{1}{2}\Big[s_{1}\omega\pm\sqrt{4p_{1}^{2}+4s_{2}u p_{1}+\omega^{2}+(r_{1}^{2}+u^{2})kp_{2}(kp_{2}+2s_{3})}\ \Big],
\ea
with $s_{1,2,3} = \pm 1$ (8 frequencies in total). The zeros  of  $P_{8}$ 
 cannot be easily written in a closed form  but can be found in an  expansion in  large $\J$ (see  Appendix \ref{app:M2-largeJ-bosfreqs}) 
   or small $\J$.

\subsection*{Fermionic fluctuations}

The determination of fermionic frequencies is analogous to the   case  of the  long folded  M2   solution in section  3. 
The orthonormal basis in the tangent  and normal bundles may be chosen as  (cf. \rf{317}) 
\ba
\la{5.10}
\re_{0} =& \frac{2}{\vv}(\kappa\,\partial_{t}+w\,\partial_{\beta}+\omega\,\partial_{\eta}), \qquad
\re_{1} = \frac{2}{\vv}(\partial_{\beta}+\mathrm {m}\,\partial_{\eta}), \qquad 
\re_{2} =  k\, \partial_{\vp},\qquad \ \ \  \vv\equiv  \sqrt{r_{1}^2+u^2} \ , \\  \label{5.11}
\rn_{1} =& 2\partial_{\rho}, \qquad \rn_{2} = 
\frac{2\omega}{r_{0}r_{1}}\partial_{t}+\frac{2\kappa u r_{0}}{\vv^{2}r_{1}}\partial_{\beta}+\frac{2\kappa r_{1}r_{0}}{\vv^{2}}
\partial_{\eta}, \qquad
\rn_{3} = \frac{2}{r_{1}}\partial_{\alpha}, \qquad \rn_{4}=\partial_{\gamma}, \nonumber \\
\rn_{5} =& 2\sqrt{2}\partial_{\theta_{1}}, \qquad
\rn_{6} = 2\sqrt{2}\partial_{\theta_{2}}, \qquad \rn_{7}=2\sqrt{2}\partial_{\phi_{1}}, \qquad \rn_{8}=2\sqrt{2}\partial_{\phi_{2}}.
\ea
The  analog of the Dirac operator in the $\kappa$-fixed  quadratic fermionic action reads (cf. \rf{313},\rf{315})
 \be
 \la{5.13}
 \slashed{D} = \rho^{i}\nabla^{\perp}_{e_{i}} +\frac{3}{2}\frac{\kappa r_{0}r_{1}}{\vv^{2}}
 \rho^{0}\rho^{1}\gamma^{1}\gamma^{3}, \qquad \ \ \qquad \nabla^{\perp}_{e_{i}} = \partial_{e_{i}}+\frac{1}{4}\Omega^{pq}_{i}\gamma_{pq},
 \ee
%SK
with 
 the spin connection   presented in Appendix \ref{app:basis} (see \rf{b3}). 
 The zeros of the   corresponding  determinant   for one Fourier mode $\mc D_{\rm F}(p_{0},p_{1},p_{2})$ 
 give $p_0=p_0(p_1,p_2)$.  
 In the string limit  $p_{2}=0$   we get ($s_{1,2}=\pm 1 $) %SK2 %, the determinant vanishes at the  frequencies
\be
p_{0} = s_{1} \frac{r_{0}^{2}\kappa u}{2(u^{2}+r_{1}^{2})}+\ha s_{2}{\omega}+\sqrt{\big(p_{1}+s_{1} b+\ha s_{2}{u}\big)^{2}+(\omega^{2}+r_{1}^{2})}, 
\qquad b = -\frac{\kappa u}{w}\frac{w^{2}-\omega^{2}}{2(u^{2}+r_{1}^{2})}, %\qquad p_{0}\equiv (p_{0})^{F}_{1,2},
\ee
which are the shifted versions of those in (\ref{4.20}), 
 and also  find  the fermionic frequencies  in  (\ref{4.21}) (with no shifts).
 
\subsection*{1-loop  correction to energy} % and large $k$ expansion}

The 1-loop correction to the M2 brane energy is given again by  (\ref{3.30}),\rf{3.31}. 
The expressions  for the few leading $1/k^2$ coefficients in \rf{3.31}  are % found to be 
% constants $\mc C_{2n}$ has a rapidly increasing complexity. The first three cases are 
\ba
\la{5.15}
\mc C_{2} =& \frac{2(5r_{1}^{2}+8r_{1}^{4}-3u^{2})}{\sqrt{u^{2}-r_{1}^{4}}}, \qquad 
\mc C_{4} = \frac{232r_{1}^{6}+148r_{1}^{8}-70r_{1}^{2}u^{2}-3u^{4}+r_{1}^{4}(81-64u^{2})}{2(u^{2}-r_{1}^{4})^{3/2}}, \\
\mc C_{6} =& \frac{2}{15 (u^2-r_1^4){}^{5/2}}\Big[
 -175 u^6+u^4 r_1^2 (1192+1717 r_1^2)-u^2 r_1^4 (3751+9886 
r_1^2+6660 r_1^4)\lp\qquad \qquad \ \ \ \qquad \qquad 
+r_1^6 (1926+9529 r_1^2+14472 r_1^4+7044 r_1^6)\Big].
\ea
%We can now examine the form of these coefficients at small and large $\J$. It is straightforward to expand them in series of large or small $\J$ at fixed $u$.
Expanding  in large/small $\J$  for fixed $u$  we find (making use of (\ref{4.11}),(\ref{4.13}))
\ba
\J \gg1: \ \ \ \mc C_{2} =& 10\J-\frac{u(6+11u)}{\J}+\frac{u(12+68u+88u^{2}+27u^{3})}{4\J^{3}}+\dots, \no \\
\mc C_{4} =& \frac{81}{2u(1+u)}\J^{3}+\frac{81+103u-221u^{2}}{4u(1+u)}\J-\frac{81+729u+1077u^{2}+173u^{3}-499u^{4}}{16u(1+u)}\frac{1}{\J}+\dots, \no \\
\mc C_{6} =& \frac{1284}{5u^{2}(1+u)^{2}}\J^{5}-\frac{2(-1926-2027u+4714u^{2})}{15u^{2}(1+u)^{2}}\J^{3}+\dots,  \la{5.17}
\\
\J \ll1: \ \ \  \mc C_{2} =& -6u+\frac{10}{\sqrt{1+u^{2}}}\J+\frac{u(3+13u^{2})}{(1+u^{2})^{2}}\J^{2}+\dots,\no  \\
\mc C_{4} =& -\frac{3}{2}u-\frac{35}{\sqrt{1+u^{2}}}\J+\frac{162+165u^{2}-137u^{4}}{4u(1+u^{2})^{2}}\J^{2}+\dots,\no  \\
\mc C_{6} =& -\frac{70}{3}u+\frac{2384}{15\sqrt{1+u^{2}}}\J+\frac{-7502-7327u^{2}+2559u^{4}}{15u(1+u^{2})^{2}}\J^{2}+\dots. \la{5.18}
\ea
In Appendix \ref{app:M2-short} we present 
 the generating function for  the coefficients of the  leading $\sim u\J^{0}$  terms in the small $\J$ expansion
 generalizing those given  in \rf{5.18}. 
 
In the special case of   $\S=\J$, \ie $u=1$  we  get explicitly 
\ba
u=1:\ \ \ \ \ \ \ E_{1}^{\rm M2} = &( -6+5\sqrt 2\, \J+4\J^{2}+\dots)\frac{\zeta(2)}{k^{2}}+
(-\tfrac{3}{2}-\tfrac{35}{\sqrt 2}\J+\tfrac{95}{8}\J^{2}+\dots)\frac{\zeta(4)}{k^{4}}+\dots\lp
= 
\Big[-\Big(\frac{\pi}{k}\Big)^{2}-\tfrac{1}{60}\Big(\frac{\pi}{k}\Big)^{4}+\dots\Big]
+\Big[\tfrac{5}{3\sqrt 2}\Big(\frac{\pi}{k}\Big)^{2}-\tfrac{7}{18\sqrt 2}\Big(\frac{\pi}{k}\Big)^{4}+\dots\Big]\, \J
\lp  \ \ \ \ 
+\Big[\tfrac{2}{3}\Big(\frac{\pi}{k}\Big)^{2}+\tfrac{19}{244}\Big(\frac{\pi}{k}\Big)^{4}+\dots\Big]\, \J^{2}+\mc O(\J^{3}).\la{516}
\ea
%%%%%%%%%%%%%%%%%%%%%%%%%%%%%%%%%%%%%%%%%%%%%%%%%

 \section{1-loop  correction to semiclassical M2 brane energy in flat space} 
 % correction Flat space}

As discussed in the Introduction, similar semiclassical 1-loop  computations  can be done for M2 branes
  in flat 11d  background. 

Expanding M2 brane action \rf{a9}  in flat  background 
near a classical  solution,  the  action for quadratic fluctuations for 8  ``transverse'' bosonic 
and fermionic  fluctuations 
may be written as (see, e.g., Appendix A in  \cite{Giombi:2024itd})\foot{We denote  by $X$ the classical solution and by 
 $\mathrm{X}$ the corresponding 3d  fluctuations.}
%($d^3V = d^3\xi \sqrt{-\bag }$) is given by
\begin{align}\label{610}
    &S_{\rm B} = %\frac{1}{2} \int d^{3}\xi \sqrt{-\bag} \, \mathrm{X} L \mathrm{X} = 
     \frac{1}{2} \int  d^{3}\xi \sqrt{-\bag}   \, \bag^{ij} \Big( \langle\nabla^{\perp}_{i}\mathrm{X} , \nabla^{\perp}_{j}\mathrm{X}\rangle - \langle \KK^{ij}, \mathrm{X}\rangle \langle \KK_{ij}, \mathrm{X}\rangle \Big),\\
      & S_{\rm F} = \int d^{3}\xi \sqrt{-\bag}  \,\bar\theta(1-\Gamma)   \slashed{\nabla}  \theta, \qquad \Gamma = \tfrac{1}{3!}\epsilon^{ijk}\rho_{i}\rho_{j}\rho_{k}, \ \ \ 
     \slashed{\nabla} = \rho^{i}\nabla_{\re_{i}}, \quad \nabla_{\re_{i}}=\partial_{\re_{i}} +\tfrac{1}{4}\Omega^{AB}_{i}\Gamma_{AB}.
\la{666}\end{align}
   Here $\bag_{ij}$ is the classical induced metric,  $\nabla^{\perp}$ is the connection on the normal bundle,   $\re_{i}$ denotes a local orthonormal frame tangent to the M2 brane
   surface   and $\KK_{ij}$ is the extrinsic curvature. The latter 
is given by     ${\KK}_{ij} = (\text{I} - \text{P})(\del_i \del_j X)$ 
where   $\text{P}$ is  the projector on the tangent bundle of the  classical M2 brane surface.

%%%%%%%%%%%%%%%%%%%%%%%%%%
%\subsection{Circular  $J_1=J_2$  brane  }
For example, the   M2    brane solution  generalizing 
the    circular string rotating in two  planes   and wrapped on the  11d circle is given by ($\xi^i=(\tau,\sigma,\sigma')$; cf. \rf{008})
\ba\la{601}
  &   X^{0} = \kappa\, \tau , \qquad \ \ X^{1}+iX^{2} = \tfrac{1}{2}\kappa\,  e^{i(\tau+\sigma)} , \qquad  \ \ X^{3}+iX^{4} = \tfrac{1}{2}\kappa\,   e^{i(\tau-\sigma)} , \qquad  \ \  X^{10}= R_{11}\sigma', \\
  &  \qquad \qquad  \qquad \qquad  \bag_{ij} = \text{diag} \big{(}-\tfrac{1}{2}\kappa^2, \ \tfrac{1}{2}\kappa^2, \ R_{11}^2 \big{)} \ ,\la{602}
\ea
where $\sigma'\in (0, 2 \pi)$  and $\bag _{ij} $ is the   induced 3d metric. 
The    M2 brane energy and spins  are  given by % flat and becomes
\ba
  &   E_0 = \frac{\kappa R_{11}}{\ell_{P}^3}= \alpha'^{-1}\kappa,\ \ \   \ \ \ J_{1} = J_{2}\equiv J = \frac{\kappa^2R_{11}}{4\, \ell_{P}^3}=\tfrac{1}{4} {\alpha'^{-1}\kappa^2}, \ \qquad \ \  \ E_0 = 2\sqrt{\alpha'^{-1}J} \ , \la{63}
\ea
where we used  \rf{007}.
Considering quadratic  fluctuations near  this solution (for  details  see Appendix \ref{flat})
one finds that  the 1-loop correction to the   energy    can be  expressed   in terms of the  determinants of the bosonic and fermionic fluctuation operators as in \rf{3.30}   where ($p_1,p_2$  are again the  integer mode  numbers) 
\ba
    &  \mathcal{D}_{\mathrm{B}} (-iw, p_1, p_2) = \left(\Lambda ^2 p_2^2+p_1^2+w^2\right)^5 \Big[ \Lambda ^6 p_2^6+3 \Lambda ^4 p_2^4 p_1^2-16 \Lambda ^2 p_2^2+3 \Lambda ^2 p_2^2 p_1^4-8 \Lambda ^2 p_2^2 p_1^2 \nonumber \\
    & +w^4 \left(3 \Lambda ^2 p_2^2+3 p_1^2+8\right)-w^2 \left(-3 \Lambda ^4 p_2^4-8 \Lambda ^2 p_2^2-6 \Lambda ^2 p_2^2 p_1^2-3 p_1^4-16\right)+p_1^6-8 p_1^4+16 p_1^2+w^6 \Big] , \no \\
    &  \mathcal{D}_{\mathrm{F}} (-iw, p_1, p_2) = \Big[\Lambda ^4 p_2^4+2 p_1^2 \left(\Lambda ^2 p_2^2+w^2-1\right)+2 w^2 \left(\Lambda ^2 p_2^2+1\right)+p_1^4+w^4+1\Big]^4 ,\la{66} \\
&\qquad \qquad  \ \ \ \Lambda^2\equiv \frac{\kappa^2}{2R_{11}^2} = \frac{2J}{\gs^2}.  \la{67}
\ea
One  finds that the  string-mode ($p_2=0$) 
 contribution  vanishes.   Using  the integral approximation for  the sum over $p_1$ for $p_2\neq 0$
 we then get for the M2  brane mode  contribution (cf. \rf{003}; $\kappa= \a' E_0$) 
\begin{equation}
 \rE_1= \frac{1}{\kappa}\bae\ , \ \ \ \ \ \qquad   \bae=     \sum_{p_2=1}^\infty\int_{-\infty}^\infty dp_1\int_{-\infty}^\infty \frac{dw}{2\pi} \log \frac{\mathcal{D}_{\mathrm{B}}(-iw, p_1, p_2)}{\mathcal{D}_{\mathrm{F}}(-iw, p_1, p_2)} \ . \la{68}
\end{equation}
This   can be  expanded in  the parameter $\Lambda \gg 1$ after  rescaling  $w \rightarrow \Lambda w$, $p_1 \rightarrow \Lambda p_1$.  As a result,
\be
\bae(\Lambda) = \sum_{p_2=1}^{\infty}F(p_2\Lambda)  =  
\sum_{p_2=1}^{\infty}\Big(-\frac{4}{p_2^{2}\Lambda^{2}}-\frac{152}{15p_2^{6}\Lambda^{6}}+\dots\Big)
=  -\frac{4\, \zeta(2)}{\Lambda^{2}}-\frac{152\, \zeta(6)}{15\Lambda^{6}}+\dots \ , \la{69} \ee
where  the function  $F(x)$  has the following  integral representation 
\ba
\la{8.2}
F(x) &= \tfrac{1}{2}x^{2}\int_{0}^{\infty}dt\, \Big[-3+\frac{6}{1+t}+\frac{4x^{4}(t^{2}-1)-4}{\sqrt{x^{8}(1+t)^{4}+2x^{4}(1+2t-t^{2})+1}}\lp\qquad \qquad \qquad \qquad 
+\frac{x^{4}(1-t)(1+t)^{3}-16(1+2t-t^{2})}{\sqrt{
x^{8}(1+t)^{8}-32x^{4}(1+t)^{4}(1+t^{2})+256(1-t^{2})^{2}}}\Big] \ . 
\ea
Its  asymptotic expansion for $x\to \infty$ is given by (cf. \rf{004})
\ba
\la{8.3}
F(x)  &=\sum_{n=1}^\infty { c_n\ov x^{2n-4}}=
  -\frac{4}{x^{2}} - \frac{152}{15}\frac{1}{x^{6}} - \frac{2496}{35}\frac{1}{ x^{10}} -\frac{9439168}{15015}\frac{1}{ x^{14}} 
- \frac{4871031808}{765765}\frac{1}{ x^{18}} %\lp\qquad \qquad\te 
%- \frac{3578963968}{51051}\frac{1}{ x^{22}} 
+...\ , 
%- \frac{1370845977591808}{1673196525}\frac{1}{ x^{26}}+\dots.
\ea
with the $\bae$ in \rf{69}  determining  all terms in  \rf{001},\rf{004}
(with $\sum_{p_2=1}^\infty { p_2^{-2n+4}}= \zeta(2n-4)$).
We shall   present the exact  formula for the coefficients $c_n$ in \rf{8.3} 
in Appendix \ref{flat}.

One  can   evaluate the sum in \rf{69} numerically 
  getting  the  values  of the function $\bae(\Lambda) = \sum_{p_2=1}^{\infty}F(p_2\Lambda)$  
   that    turn out to be real for $\Lambda\ge 2$.
   The results for several  values  of $\Lambda$ are represented  by  blue circles in Fig.~\ref{fig:b},
%MB   
by summing up to $p_{2, \rm max}=2000$.
    They are  in good agreement  with the plot of just  the first two terms  in \rf{69}
    (orange curve). More precisely, the relative error is
below 0.1 percent already at $\Lambda=2.5$ and fully negligible at $\Lambda = 3.0$. A more careful 
evaluation at the convergence radius $\Lambda=2$ (see Appendix \ref{flat}) is beyond our aims
and would require a larger maximum value of $p_{2}$.
    %     Integrating each term at a fixed value of $p_2$ and estimating the sum, one finds 
%the blue circles in Fig.~(\ref{fig:b}) that we compare with the first two terms in (\ref{8.5}).
\begin{figure}[H]
\centering
\includegraphics[width=0.6\textwidth]{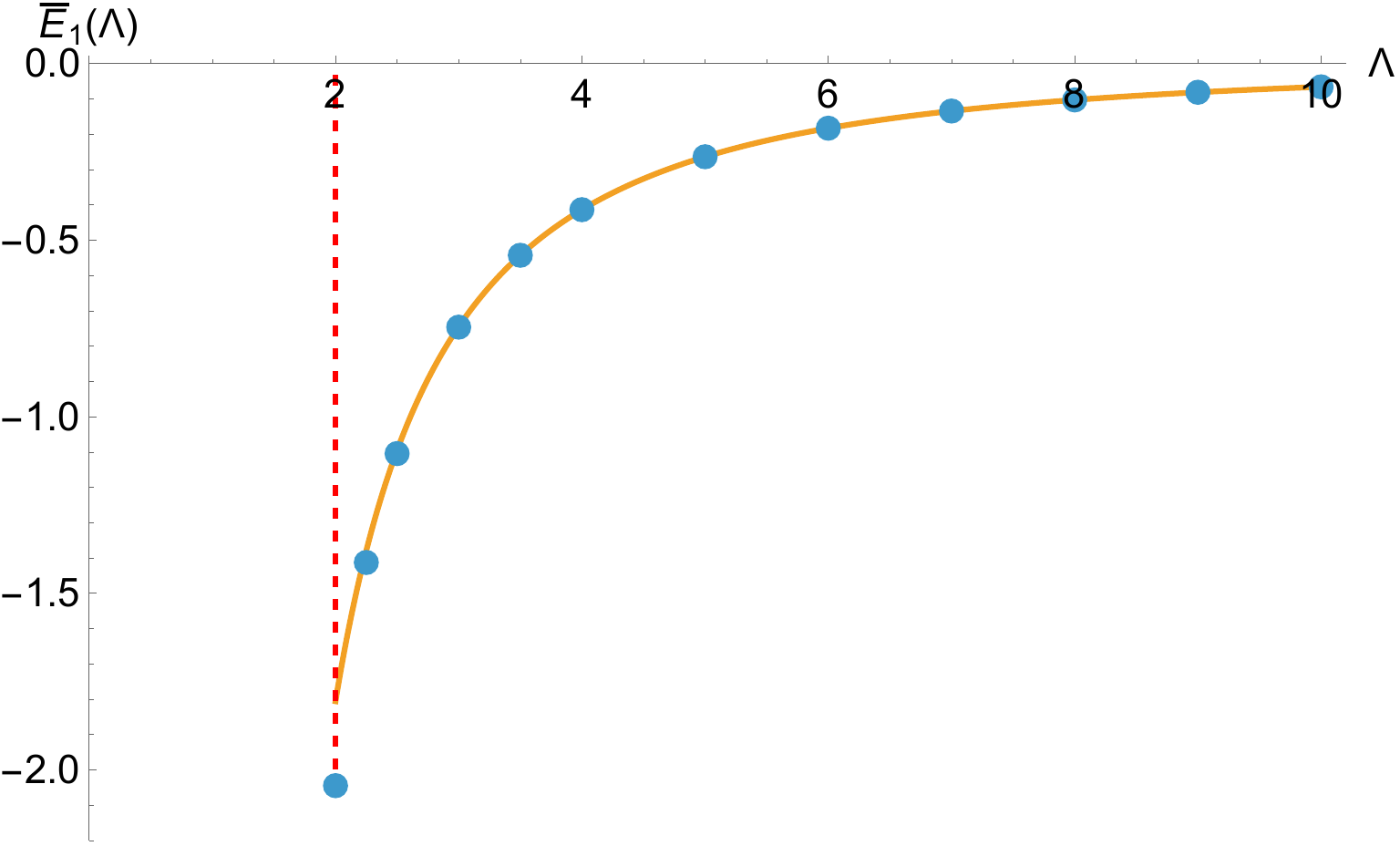}
\caption{The values of $\bae(\Lambda)$ for $\Lambda\ge 2$  found by numerical evaluation of the sum in \rf{69}   (blue circles). 
The orange line represents  the sum of the first two terms of the large $\Lambda$ expansion of 
$\bae(\Lambda)$.   }
\la{fig:b}
\end{figure}

%\subsection*{Folded  spinning brane}

One  may   repeat  a similar  analysis   of  the 1-loop M2  brane correction in the  case of the solution  generalizing 
 the folded string   rotating in one  plane in flat space (cf. \rf{601},\rf{602},\rf{63})
 %given by
\ba \label{611}
  &  X^{0} = \kappa\, \tau\, , \qquad X^{1} = \kappa\, \sin \sigma\ \cos \tau\,  , \qquad X^{2} =  \kappa\,  \sin \sigma\ \sin \tau\,  , \qquad X^{10} = R_{11} \s' \ , \\
  & 
   \qquad \qquad  \qquad \qquad  \bag_{ij} = \text{diag}\big{(}-\kappa^2\cos^{2}\sigma, \ \kappa^2\cos^{2}\sigma, \ R_{11}^2 \big{)}\ , \la{79}\\
&
    E_0= \frac{\kappa R_{11}}{\ell_{P}^3}=\alpha'^{-1} \kappa\, , \qquad\qquad  J  = \frac{\kappa^2R_{11}}{2\ell_{P}^3}=\tfrac{1}{2}\alpha'^{-1}\kappa^2\,, \qquad \qquad E_0 = \sqrt{2\alpha'^{-1}J}\ . \la{78}
\ea
In contrast to the circular solution in \rf{601}  here the induced metric is no longer  flat  which complicates the derivation of the fluctuation spectrum.  After the Fourier 
expansion   in extra $\s'$ direction   one finds that the 7+1   bosonic and 8  fermionic fluctuation operators 
are given by  (see Appendix \ref{flat} for details)
\ba
& %\rK_1=
\Delta =  -\frac{1}{\sqrt{-\bag }}\partial_{i}(\sqrt{-\bag}\,\bag^{ij}\partial_{j}) \ \to \   \partial^2_{\tau}-\partial^2_{\sigma} 
+ \Lambda^2 
\, p_{2}^2\, \cos^2 \sigma \ , \ \ \ \ \ \ \ \  \   \Lambda^2\equiv  \frac{\kappa^2}{R_{11}^2} = { 2 J\ov \gs^2} \ , \label{6.33}
\\
&\qquad \qquad  %\rK_{2} =
 \Delta + R^{(2)}\ \to \     \partial^2_{\tau}-\partial^2_{\sigma} +  \Lambda^2  \, p_{2}^2\, \cos^2 \sigma  + 
\frac{2}{\cos^2\sigma} \ , 
\label{6.34}\\
&  \qquad \qquad \ \  \slashed{\nabla} = \rho^{\alpha}\partial_{\alpha} + i\Lambda\, p_{2}\,\cos \sigma\ . \la{635}
\ea
%The simplicity of such a solution is related to the fact that it can be considered as a hypersurface in $(X^{0}, X^{1}, X^{2}, X^{10})$. To understand the structure of the fluctuations that appear, let us first consider a general case where the M2 brane can be viewed as a hypersurface.
While  finding their   eigenvalues for general $p_2$ appears  to be non-trivial, the fact that  
they depend  on $J$ and $\gs$ only  through   the  parameter $\Lambda$  implies that 
the  M2 brane correction to the energy should have  a similar   structure \rf{004},\rf{69} 
 as in the  above circular  case.

 %%%%%%%%%%%%%%%%%%%%%%%%%%%%%%%%%%%%%%%%%%%%%%%
 \section*{Acknowledgments}
We would like to thank  J. Russo  for useful comments.
 MB is supported by the INFN grant GAST.
SAK  acknowledges support of   the President's  PhD Scholarship of Imperial College London. 
AAT is  supported by the STFC grant ST/T000791/1.  

%%%%%%%%%%%%%%%%%%%%%%%%%%%%%%%%%%%
\appendix

\section{Definitions and notation}
\la{apa}

The $U(N)_{k}\times U(N)_{-k}$ ABJM theory in the large $N$ limit with fixed $k$ is dual  %\ci{Aharony:2008ug} 
 to the M-theory on $\ads_{4}\times S^{7}/\mathbb Z_{k}$
 with the following metric and 3-form background
\ba
\la{A.1}
& ds^{2}_{11} = \L^{2}(\tfrac{1}{4}ds^{2}_{\ads_{4}}+ds^{2}_{S^{7}/\mathbb Z_{k}}), \qquad\qquad  \L = (2^{5}\pi^{2}Nk)^{1/6}\ell_{P}, \\
\la{A.2}
& ds^{2}_{\ads_{4}} = -\cosh^{2}\rho\, dt^{2}+d\rho^{2}+\sinh^{2}\rho\, (d\alpha^{2}+\cos^{2}\alpha\, d\beta^{2}), \\
\la{A.3}
& ds^{2}_{S^{7}/\mathbb Z_{k}} = ds^{2}_{\cp}+\frac{1}{k^{2}}(d\vp+k\, {\rm A})^{2}, \qquad\qquad  \vp\equiv\vp+2\pi, \\
\la{A.4}
& C_{3} = -\tfrac{3}{8}\, \L^{3}\, \cosh\rho\, \sinh^{2}\rho\, \sin\alpha\, dt\ww d\rho\ww d\beta.
\ea
%Our preferred choice for metric in $\cp$ (and 1-form) will be 
We shall use the  following  parametrization of $ds^{2}_{\cp}$ and ${\rm A}$
\ba
\la{A.5}
ds_{\cp}^{2} =&{d}\gamma ^{2}+\cos ^{2}\gamma \sin
^{2}\gamma \big( {d}\eta +\tfrac{1}{2}\cos \theta _{1}
{d}\phi_{1}-\tfrac{1}{2}\cos \theta _{2}{d}\phi _{2}\big) ^{2} \notag 
 \\
&+\tfrac{1}{4}\cos ^{2}\gamma \left( {d}\theta _{1}^{2}+\sin ^{2}\theta
_{1}{d}\phi _{1}^{2}\right) +\tfrac{1}{4}\sin ^{2}\gamma \left(
{d}\theta _{2}^{2}+\sin ^{2}\theta _{2}{d}\phi
_{2}^{2}\right) , \\
\la{A.6}
{\rm A} =& \tfrac{1}{2}\big[\cos(2\gamma) d\eta+\cos^{2}\gamma\cos\theta_{1}d\phi_{1}+\sin^{2}\gamma\cos\theta_{2}\, {d}\phi_{2}\big]\ . 
\ea
The limit of  large $k$    with   fixed  $\l \equiv \frac{N}{k}$
corresponds to the  't Hooft 
 expansion in  the 3d gauge theory which  is dual to the perturbative type IIA string theory  on 
$\ads_{4}\times \cp$    background     with the metric 
\ba\la{A.7}
& ds^{2}_{10} = L^{2}(\tfrac{1}{4}ds^{2}_{\ads_{4}}+ds^{2}_{\cp}), \qquad\qquad  L = \gs^{1/3}\, \L  \ . 
\ea
The  string  coupling $\gs$ and the effective dimensionless string tension $\T$ (defined with respect 
to the radius $\ha L$  of the $\ads_{4}$)   are given by 
(we set $\ell_{P} =\ell_{\rm s}= \sqrt{\alpha'}$)
\ba
\la{A.8}
& \gs = (\frac{\L}{k\,\ell_{P}})^{3/2} = \frac{\sqrt\pi(2\l)^{5/4}}{N}\ , \\
   &\T = \frac{\frac{1}{4}L^{2}}{2\pi \alpha'} = \sqrt\frac{\l}{2} =\frac{\sqrt{\bar\l}}{2\pi}\ , 
\qquad\qquad  \bar\l \equiv  2\pi^{2}\l\ , \la{a8} \\
& \frac{1}{k^{2}} = \frac{\l^{2}}{N^{2}} = \frac{\gs^{2}}{8\pi\T}\ .  \la{A.9}
\ea
%SK3
The bosonic part of the M2  brane  action is given by \eqref{a10}. 
The M-theory expansion corresponding  to the  large $N$  with  fixed $k$   expansion on the  gauge theory side 
  is   for $\L\gg \ell_{P}$.  This is  the  expansion in the 
large effective dimensionless M2 brane tension
\be\la{a14}
\T_{2} \equiv  \L^{3}\, T_{2} = \frac{1}{\pi}\sqrt{2Nk} \ .  %, \qquad T_{2} = \frac{1}{(2\pi)^{2}\ell_{P}^{3}},
\ee
It is related to the effective dimensionless type IIA  string tension in \rf{A.8} as 
\be\la{a15}
\T = \frac{\pi}{2k}\, \T_{2} \ , 
\ee
which follows also  upon   double dimensional   reduction   \ci{Duff:1987bx}  of the membrane action in $\ads_{4}\times S^{7}/\mathbb Z_{k}$  
 to the  type IIA   GS string action in $\ads_{4}\times \cp$.

\section{Gamma matrices and spin connection}
%Details on the quadratic fluctuations for fermions}
\la{app:basis}

The explicit choice for the gamma  matrices $(\rho^{i}, \gamma^{p})$ used in this paper is
\ba
&\rho^{0} = (i\sigma^{2})\otimes \sigma^{2}\otimes \sigma^{2}\otimes \sigma^{2}\otimes \sigma^{2}, \quad 
\rho^{1} = \sigma^{1}\otimes \sigma^{2}\otimes \sigma^{2}\otimes \sigma^{2}\otimes \sigma^{2}, \quad
\rho^{2} = \sigma^{3}\otimes \sigma^{2}\otimes \sigma^{2}\otimes \sigma^{2}\otimes \sigma^{2},\no 
\\
&\gamma^{1} = \pauli{0}{2}{0}{1}{2}, \quad 
\gamma^{2} = \pauli{0}{3}{0}{0}{0}, \quad
\gamma^{3} = \pauli{0}{2}{1}{2}{0}, \quad\no \\
&
\gamma^{4} = \pauli{0}{2}{3}{2}{0}, \quad
\gamma^{5} = \pauli{0}{2}{2}{0}{1}, \quad
\gamma^{6} = \pauli{0}{2}{0}{3}{2}, \quad\no \\
&\gamma^{7} = \pauli{0}{2}{2}{0}{3}, \qquad
\gamma^{8} = \pauli{0}{1}{0}{0}{0}, \la{b1}
\ea
where $\s^i$ are Pauli matrices and $\s^0$ is a unit $2\times 2$  matrix.

The spin  connection in the normal bundle  in the basis \rf{317},\eqref{318}
  corresponding to the  long folded M2 brane solution  %for the long folded membrane 
can be written as
\ba
\Omega_{pq}(\re_{i}) = \langle \rn_{p} , \nabla_{\re_{i}}\rn_{q}  \rangle , \qquad 
\Omega = E_{2,3}\frac{2\nu }{\mu}\re^{1} + E_{5,7} \big(\frac{\nu}{\mu}\re^{0}+\re^{2} \big) + E_{6,8} \big( -\frac{\nu}{\mu}\re^{0}+\re^{2} \big) - E_{3,4} \frac{\kappa}{\mu} \re^{2} ,\la{b2}
\ea
where $E_{p, q}$ is a matrix  that  has $+1$ in $(p, q)$  place and $-1$ in $(q, p)$  place  with  other  entries  being 0.

Similarly, the connection in the normal bundle in the basis \eqref{5.10},\eqref{5.11} for the circular $(S, J)$ M2 brane  solution is given by %takes form :
\ba
%&\Omega_{pq}(\re_{i}) = \langle \rn_{p} , \nabla_{\re_{i}}\rn_{q} \rangle , \\
\Omega = &E_{1,2}\big( -\frac{2 \omega \aA}{\kappa r_{1}^2}\re^{0} - \frac{2\kappa u r_{0}^2}{\aA^{3}}\re^{1}\big) + E_{5,7}\big(\frac{\omega}{\aA}\re^{0}-\frac{u}{\aA}\re^{1}+\re^{2} \big) \no \\ 
&+ E_{6,8} \big( -\frac{\omega}{\aA}\re^{0}+\frac{u}{\aA}\re^{1}+\re^{2} \big) + E_{2,4} \big( -\frac{\kappa r_{0} r_{1}}{\aA^2} \re^{2}\big),
\qquad \ \ \  \aA\equiv  \sqrt{r_{1}^2+u^2}\ . \la{b3}\ea

\section{Higher order $1/k^2$  corrections  in \rf{3.31}}  % membrane corrections to long folded string}
\la{app:higher}

The explicit expressions for the next two   coefficients in the large $k$ expansion of the folded membrane 
1-loop energy in \rf{3.31}  are found to be 
%We collect here a few higher order contributions to the large $k$ expansion of the membrane contributions
%to the energy of the long folded string, \cf  (\ref{3.31}).
\ba
\mc C_{6} =&\te  -\J\, \frac{2 (808 \kappa ^6-1892 \kappa ^4 \nu ^2-667 \kappa ^2 \nu 
^4-175 \nu ^6)}{15 \kappa  (\kappa^{2} -\nu^{2} )^2}=  -\J\, \frac{2 (-1926-2027 x ^2+532 x ^4+808 x ^6)}{15 x ^4 \sqrt{1+x ^2}}, \no\\
\mc C_{8} =&\te -\J\, \frac{467328 \kappa ^8-250560 \kappa ^6 \nu ^2-729664 \kappa ^4 \nu^4
-231824 \kappa ^2 \nu ^6+13685 \nu ^8}{420 \kappa  (\kappa^{2}-\nu^{2} )^3}\no\\
= &\te -\J\, \frac{-731035-573520 x ^2+1322624 x ^4+1618752 x ^6+467328 x^8}{420 x ^6 \sqrt{1+x ^2}} . \la{c1}%, \ \ \ \ \ 
%\mc C_{2n} =\J C_{2n}\la{c2}
\ea
Their small/large $x$ expansions are (cf. \rf{3.36},\rf{339})  %Small $x$ expansion:
\ba
x\ll 1: \ \ \ \ \ \ \ \mc C_{6} =& \te \J\,\big(\frac{1284}{5}\frac{1}{x^{4}}+\frac{2128}{15}\frac{1}{x^{2}}-\frac{3293}{30}+\dots\big), \no \ \\
\mc C_{8} =&\te \J\,\big(\frac{146207}{84}\frac{1}{x^{6}}+\frac{83201}{168}\frac{1}{x^{4}}-\frac{10681967}{3360}\frac{1}{x^{2}}-\frac{3106619}{1344}+\dots\big), \la{C.5}\\
x \gg1: \ \ \ \ \ \  \mc C_{6} =&\te  \J\, \big(-\frac{1616}{15}x-\frac{256}{15}\frac{1}{x}+\dots \big), \qquad 
\mc C_{8} =\J\, \big(-\frac{38944}{35}x-\frac{115424}{35}\frac{1}{x}+\dots\big).\la{c4}
\ea
Comparing the small $x$ expansions  of  $\mc C_{2n}$  in \rf{3.36},\rf{C.5}  in the folded   membrane  case % (\ref{3.36}, \ref{C.5}) 
with the large $\J$ expansions  of the coefficients in the large $k$ expansion of 
the  1-loop energy of the  circular $(S,J)$ membrane in (\ref{5.17})  suggests a conjecture  that for any $n$ 
\ba
\mc C_{2n}^{\rm circular} \stackrel{\J\gg 1}{=} {\rm c}_{2n}\, \frac{\J^{2n-1}}{[u(1+u)]^{n-1}}+\mc O(\J^{2n-3}), \qquad \qquad 
\mc C_{2n}^{\rm folded} \stackrel{x\ll 1}= {\rm c}_{2n}\, \frac{\J}{x^{2n-2}}+\mc O\big({1\ov x^{2n-4}}\big), \la{c44}
\ea
where ${\rm c}_{2n}$ are the  {\it same} numerical coefficients in both cases.

%%%%%%%%%%%%%%%%%%%%%%%%%%%%%%%%%%%%%%%%%%%%%%%%
\section{Expansion of 1-loop  energy of circular $(S,J)$ string  for  small $\J$}
\la{app:circular-short-oneloop}

In this Appendix  we study  the small $\J$ expansion of the 1-loop string energy for the circular $(S,J)$ solution
discussed in section \ref{sec3}.
The small $\J$    expansion of $\cp$ fluctuation frequencies in \rf{413} reads (here we set $p_1=n$)\footnote{For simplicity, we continue branches of solutions of algebraic equations  from large $n$ to small values of $n$. For a more careful discussion
see \cite{Beccaria:2012xm}.}
\ba
\cp: \ \ \ \ (p_{0})_{1,2,3,4} = \tfrac{1}{2}\sqrt{4n^{2}-u^{2}}+\mc O(\J^{2}),  \qquad \ \ \ \ 
(p_{0})_{5}  = \sqrt{n^{2}-u^{2}}+\mc O(\J^{2}).
\ea
For $n=0$ the first  four degenerate frequencies are imaginary, indicating    an instability.\foot{This is a reflection of the fact that the  
 string wrapped on a circle in the  internal space is   unstable, unless  the  contraction due to its tension  is balanced  by 
 its  rotation. For a discussion of a similar instability 
 in \adss  context see \cite{Beisert:2005mq}.}
%This means that the considered solution is not 
%dual to a lowest state with given $S,J$ (as the folded string has lower energy, while winding further increases the energy).
%Consequently, negative modes  should be rotated.
% \footnote{This instability is an artefact of expanding at  strong coupling
% -- at weak coupling anomalous  dimensions are real.} 
%Similar issues appeared in the past, see for instance page 6 in 
For the $\ads_4$   frequencies in \rf{4.19}   we get  % sector we have
\ba
\ads_4: \ \ \ \  & (p_{0})_{1} 
= \sqrt{n^{2}+u^{2}}+\frac{u}{\sqrt{1+u^{2}}\sqrt{n^{2}+u^{2}}}\J+\mc O(\J^{2}), \\
&
(p_{0})_{2,3} = 
\mp
\frac{(n\pm 1)^{2}+u^{2}+(-n\mp 2)\sqrt{(1+u^{2})\big[(n\pm 1)^{2}+u^{2}\big]}}{(1+u^{2})\big[(n\pm 1)^{2}+u^{2}\big]}\,\J+\mc O(\J^{2}).
\ea
The fermionic frequencies \rf{4.20},\rf{4.21}  have the following  expansions
\ba
& %{\rm F}:\ \ \ \ \  
(p_{0})_{1} = \begin{cases}
\frac{1}{2}(1+2n-\sqrt{1+u^{2}})+\big[\frac{u}{2\sqrt{1+u^{2}}}-\frac{u(-3-3u^{2}+2n\sqrt{1+u^{2}}}{2(1+u^{2})^{3/2}(2n-\sqrt{1+u^{2}})}\big]\J+\mc O(\J^{2}), & n>0 \\
\frac{1}{2}(1-2n+\sqrt{1+u^{2}})+\big[\frac{u}{2\sqrt{1+u^{2}}}-\frac{u(-3-3u^{2}+2n\sqrt{1+u^{2}}}{2(1+u^{2})^{3/2}(-2n+\sqrt{1+u^{2}})}\big]\J+\mc O(\J^{2}), & n<0,
\end{cases}\  \no 
\\
&(p_{0})_{2} = \begin{cases}
\frac{1}{2}(-1+2n+\sqrt{1+u^{2}})+\big[-\frac{u}{2\sqrt{1+u^{2}}}+\frac{u(3+3u^{2}+2n\sqrt{1+u^{2}}}{2(1+u^{2})^{3/2}(2n+\sqrt{1+u^{2}})}\big]\J+\mc O(\J^{2}) ,& n>0 \\
\frac{1}{2}(-1-2n-\sqrt{1+u^{2}})+\big[-\frac{u}{2\sqrt{1+u^{2}}}-\frac{u(3+3u^{2}+2n\sqrt{1+u^{2}}}{2(1+u^{2})^{3/2}(2n+\sqrt{1+u^{2}})}\big]\J+\mc O(\J^{2}), & n<0 ,
\end{cases}\   
\\
&\te (p_{0})_{3,4} =\tfrac{1}{2}(\sqrt{(2n\pm 1)^{2}+u^{2}}\mp\sqrt{1+u^{2}}) %\lp\qquad 
\mp\frac{u((2n\pm 1)^{2}+u^{2}\mp 2(1\pm n)\sqrt{(1+u^{2})[(2n\pm 1)^{2}+u^{2})}]}
{2(1+u^{2})[(2n\pm 1)^{2}+u^{2}]}\J+\mc O(\J^{2}).\no 
\ea
The  1-loop correction to the energy \rf{2.15} can  be written  as 
\ba
\la{D.9}
&E_{1}=\frac{1}{2\kappa}\sum_{n\in\mathbb Z}  \hat \omega_{n}(\J)= \frac{1}{2\kappa}\ho_0(\J)+\frac{1}{\kappa}\sum_{n=1}^{\infty}\bar \omega_{n}(\J), 
\qquad\qquad \qquad 
 \bar \omega_{n} = \frac{1}{2}(\ho_{n}+\ho_{-n}) \ , \\
\la{D.8}
& \hat \omega_{n}(\J) \equiv \sum_{\{p_{0}\}}(-1)^{F}p_{0} =  \ho_{n}^{(0)}+\ho_{n}^{(1)}\,\J+\mc O(\J^{2}),
\\
&\te \ho_0 (\J) = (1+3i)\,u+\frac{1}{\sqrt{1+u^{2}}}\J+\mc O(\J^{2}),\la{D.10}
\\ 
&\bar \om_{n}^{(0)} = \te  -4n+\sqrt{n^{2}-u^{2}}+2\sqrt{4n^{2}-u^{2}}+\sqrt{n^{2}+u^{2}}+\sqrt{(n+1)^{2}+u^{2}}+\sqrt{(n-1)^{2}+u^{2}} \lp\te \qquad \qquad \qquad 
 -\sqrt{(2n+1)^{2}+u^{2}}-\sqrt{(2n-1)^{2}+u^{2}}, \la{D10}
\\
&\bar \om_{n}^{(1)} =\te  \frac{8u n}{\sqrt{1+u^{2}}(1-4n^{2}+u^{2})}+\frac{u}{\sqrt{1+u^{2}}}\frac{1}{\sqrt{n^{2}+u^{2}}}
+\frac{u(2-n)}{\sqrt{1+u^{2}}\sqrt{(n-1)^{2}+u^{2}}}\lp\te \qquad  \ \ \ \ +\frac{u(2+n)}{\sqrt{1+u^{2}}\sqrt{(n+1)^{2}+u^{2}}}
+\frac{2u(n-1)}{\sqrt{1+u^{2}}\sqrt{(2n-1)^{2}+u^{2}}}
-\frac{2u(n+1)}{\sqrt{1+u^{2}}\sqrt{(2n+1)^{2}+u^{2}}}.\no 
\ea
The sums over $n$ are convergent since 
$
\bar \om _{n}^{(0)} = \frac{7u^{2}-4u^{4}}{8}\frac{1}{n^{3}}+\mc O(n^{-4}), \quad %\qquad \qquad
\bar \om_{n}^{(1)} = \frac{10u-u^{3}}{8\sqrt{1+u^{2}}}\frac{1}{n^{3}}+\mc O(n^{-4}), 
$
and  can be computed numerically at  fixed $u$. For example,  for  $\S=\J$, \ie $u=1$, we find 
\be
u=1: \ \ \ \ \ \ \ \ \ \ \ \ \sum_{n=1}^{\infty}\bar \om _{n}^{(0)} = -0.40652, \qquad\qquad \qquad 
\sum_{n=1}^{\infty}\bar \om_{n}^{(1)} = -1.5001,
\ee
and as a result  the  real and imaginary parts of $E_1$ are given by 
\be\la{d10}
E_{1} = 0.093-1.212\, \J+\mc O(\J^{2})  + i \big[ 1.5-1.06\, \J+\mc O(\J^{2})\big] \ .
\ee
One may also  consider the expansion of \rf{D.9}  in small $u$.
%\paragraph{Small $u$ expansion}
Splitting the sum into $n=1$ and $n\geq 2$ parts we get  % ({\bf check first line notation?!})
%plit the term $n=1$ 
%and sum over $n\ge 2$ the expansion over $u$. For $\bar S_{n}^{(0)}$ this gives
\ba
&\bar\om^{(0)}_1 = -4+u+\sqrt{1-u^{2}}+2\sqrt{4-u^{2}}+\sqrt{4+u^{2}}-\sqrt{9+u^{2}} \te = u-\frac{11}{12}u^{2}-\frac{289}{1728}u^{4}+\dots,\\
&
\sum_{n=2}^{\infty}\bar \om_{n}^{(0)} = \sum_{n=2}^{\infty}\big[\te \frac{7n^{2}-1}{2n(1-5n^{2}+4n^{4})}u^{2}+
\frac{-9+135 n^2-807 n^4+2005 n^6-2028 n^8+864 n^{10}-1024 n^{12}}{32 (n-5 n^3+4 n^5)^3}u^{4}+\dots\big]\lp
\qquad \qquad \te = (\frac{11}{12}-\log 2)u^{2}+\big[\frac{289}{1728}-\frac{5}{16}\zeta(3)\big]u^{4}+\dots,
\\  & \qquad \qquad 
\sum_{n=1}^{\infty}\bar \om_{n}^{(0)} =\te  u-\log 2 \ u^{2}-\frac{5}{16}\zeta(3) u^{4}+\dots.\la{d12}
\ea
Similarly, we find that 
\be
\sum_{n=1}^{\infty}\bar \om_{n}^{(1)} = \te 1+3(1-2\log 2)u-\frac{1}{2}u^{2}+\big[-\frac{5}{2}+3\log 2-\frac{5}{8}\zeta(3)\big]u^{3}+\frac{3}{8}u^{4}+\dots\ .
\ee
Using that (cf. \rf{4.9}) 
$
\frac{1}{\kappa}= \frac{1}{u}-\frac{1}{u^{2}\sqrt{1+u^{2}}}\J+\mc O(\J^{2}),
$   we  obtain the following expression  for  the real part of  $E_1$ 
(the imaginary part  comes from  $3i\,u$ term in (\ref{D.10}))\foot{Note that here the term $-u \log 2$ can  be again be interpreted as coming from the
 redefinition $\sqrt{\lb}\to 2\bar h(\lb)$ in  the 
classical part of  the energy  (cf. \rf{1.19}). 
Other $\log 2$ terms do not have this  origin.}
\ba
\la{D.21}
\text{Re}\,E_{1} =& \te \big[\frac{3}{2}+(3-5\log 2)\J+\mc O(\J^{2})\big]-\big[\log 2+\mc O(\J^{2})\big]\, u\lp\te 
+\big[(-\frac{5}{2}+\frac{5}{2}\log 2-\frac{5}{16}\zeta(3))\, \J+\mc O(\J^{2})\big]\,u^{2}-\big[\frac{5}{16}\zeta(3) +\mc O(\J^{2})\big]\,u^{3}+\mc O(u^{4}).
\ea

\subsubsection*{Comparison with the $\ads_{5}\times S^5$ case}
%%%%%%%%%%%%%%%%%%%%%%%%%%%%%%%%%
It is of interest  to compare  the  above expressions with  $E_1$  for a similar circular $(S,J)$ solution in \adss 
%Let us compare with the circular string with spin $S$ in $\ads_{5}$ wrapped around a big circle of $S^{5}$ and carrying also momentum $J$
(see  \cite{Park:2005ji}). Using the same notation 
%Notation matches the one we adopted in $\ads_{4}$. In particular,  the classical solution is identical.
here  one has $4+2 +2 $ bosonic  fluctuation frequencies (cf. \rf{4.19}) 
  and $4+4$   fermionic frequencies   given by 
\ba
& {\rm B}: \ \  \ \ \ \ (p_{0})_{1,2,3,4} = \sqrt{n^{2}+\J^{2}-u^{2}},\qquad
(p_{0})_{5,6} = \sqrt{n^{2}+\kappa^{2}},
\\
&\qquad \ \ \ \ (p_{0}^{2}-n^{2})^{2}+4 r_{1}^{2}\kappa^{2}p_{0}^{2}-4(1+r_{1}^{2})(\sqrt{1+\kappa^{2}}p_{0}-p_{1})^{2}=0, \qquad p_{0}=(p_{0})_{7,8}\ , 
\\
&
 {\rm F}: \ \ \ \ \ \ (p_{0})_{1,2} = \sqrt{(n\pm c)^{2}+a^{2}},
\\
&\qquad \qquad a^{2}=\te \frac{1}{2}(\kappa^{2}+\J^{2}-u^{2}), \qquad c=\frac{1}{2}\kappa\big[1+\frac{2(1+r_{1}^{2})}{\kappa^{2}-\J^{2}+u^{2}}\big]
\sqrt{\frac{\kappa^{2}-\J^{2}+u^{2}-2r_{1}^{2}}{2(\kappa^{2}+1)}}.
\ea
Their  small $\J$ expansion  reads %of bosonic frequencies can be taken from previous expressions
\ba
& {\rm B}: \ \ (p_{0})_{1,2,3,4} = \sqrt{n^{2}-u^{2}}+\mc O(\J^{2}), \te \qquad 
(p_{0})_{5,6} = \sqrt{n^{2}+u^{2}}+\frac{u}{\sqrt{1+u^{2}}\sqrt{n^{2}+u^{2}}}\J+\mc O(\J^{2}),\no \\
&(p_{0})_{7,8} = \sqrt{(n\pm 1)^{2}+u^{2}}\mp \sqrt{1+u^{2}}\te
\mp
\frac{(n\pm 1)^{2}+u^{2}+(-n\mp 2)\sqrt{(1+u^{2})((n\pm 1)^{2}+u^{2}}}{(1+u^{2})((n\pm 1)^{2}+u^{2})}\,\J+\mc O(\J^{2}), \\
& 
 {\rm F}: \ \   (p_{0})_{1,2} = \tfrac{1}{2}\sqrt{1+u^{2}+4n(n\pm \sqrt{1+u^{2}})}\te
+\frac{u(4n^{2}\sqrt{1+u^{2}}\pm 8n(1+u^{2})+3(1+u^{2})^{3/2})}
{2(1+u^{2})^{3/2}(\pm 2n+\sqrt{1+u^{2}})\sqrt{1+u^{2}+4n(n\pm\sqrt{1+u^{2}})}}\J+\mc O(\J^{2}).\no
\ea
The corresponding 1-loop  energy  is  given by (\ref{D.8}) and (\ref{D.9})  where now 
\ba
&\ho_{0}(\J) = 2(1+2i)u-2\sqrt{1+u^{2}}+\frac{2(1+u^{2}-4u\sqrt{1+u^{2}})}{(1+u^{2})^{3/2}}\J+\mc O(\J^{2})\ , \no \\
&
\bar \om_{n}^{(0)} = 4\sqrt{n^{2}-u^{2}}+2\sqrt{n^{2}+u^{2}}+\sqrt{(n-1)^{2}+u^{2}}+\sqrt{(n+1)^{2}+u^{2}}\lp\ \ \ \ \ \ \ \
-2\sqrt{4n^{2}-4n\sqrt{1+u^{2}}+1+u^{2}}-2\sqrt{4n^{2}+4n\sqrt{1+u^{2}}+1+u^{2}},\\
&
\bar \om_{n}^{(1)} =\te \frac{2u}{\sqrt{1+u^{2}}\sqrt{n^{2}+u^{2}}}+\frac{u(2-n)}{\sqrt{1+u^{2}}\sqrt{(n-1)^{2}+u^{2}}}
+\frac{u(2+n)}{\sqrt{1+u^{2}}\sqrt{(n+1)^{2}+u^{2}}}\lp
\te\ \ \ \ \ \ \ \   +2u\big(\frac{2n}{\sqrt{1+u^{2}}}-3\big)\frac{1}{\sqrt{1+u^{2}}\sqrt{4n^{2}-4n\sqrt{1+u^{2}}+1+u^{2}}} \te 
-2u\big(\frac{2n}{\sqrt{1+u^{2}}}+3\big)\frac{1}{\sqrt{1+u^{2}}\sqrt{4n^{2}+4n\sqrt{1+u^{2}}+1+u^{2}}}.\no 
\ea
The sums are again convergent and can be computed numerically at fixed $u$. At the special point $\S=\J$, \ie $u=1$, we find 
$
\sum_{n=1}^{\infty}\bar \om_{n}^{(0)} = -1.9621, \quad
\sum_{n=1}^{\infty}\bar \om_{n}^{(1)} = -3.1034,
$
and thus (cf. \rf{d10})
\be\la{dd10}
E_{1} = -2.376-2.716\, \J+\mc O(\J^{2})  + i \big[  2-1.414\, \J+\mc O(\J^{2})\big].
\ee
Expanding in  small $u$ we get  (cf. \rf{D.21})
\ba
\text{Re}\,E_{1} =&\te  -\big[{u}^{-1}+\mc O(\J^{2})\big]+\big[2+(\frac{1}{2}-8\log 2)\J+\mc O(\J^{2})\big]-\big[1+\mc O(\J^{2})\big]\, u\no \\ & \qquad \te 
+\big[(\frac{5}{8}+4\log 2-\zeta(3))\, \J+\mc O(\J^{2})\big]\,u^{2}
+\big[\frac{1}{4}-\zeta(3) +\mc O(\J^{2})\big]\,u^{3}
+\mc O(u^{4}).
\ea
The $u^{-1}$ term (that was   absent in   (\ref{D.10})) 
comes from the mode $n=0$ that has a $\mc O(\J^{0})$ part which is not vanishing for $u=0$.

To make the comparison  between  the $\ads_4\times \cp$   and \adss cases more transparent, we may  remove the contribution of  $\ho_{0}(\J)$  in both cases 
to  get 
\ba
E_{1}^{\ads_{4}\times \cp}\big|_{n\neq 0} =& \big[1+(3-5\log 2)\J+\mc O(\J^{2})\big]-\big[\log 2+\mc O(\J^{2})\big]\, u\lp\te 
+\big[(-\frac{5}{2}+\frac{5}{2}\log 2-\frac{5}{16}\zeta(3))\, \J+\mc O(\J^{2})\big]\,u^{2}+\big[-\frac{5}{16}\zeta(3) +\mc O(\J^{2})\big]\,u^{3}+\mc O(u^{4}),\no  \\
%%%
E_{1}^{\ads_{5}\times S^{5}}\big|_{n\neq 0} =& \te \big[1+(\frac{9}{2}-8\log 2)\J+\mc O(\J^{2})\big]-\big[\frac{1}{2}+\mc O(\J^{2})\big]\, u\lp\te 
+\big[(-\frac{27}{8}+4\log 2-\zeta(3))\, \J+\mc O(\J^{2})\big]\,u^{2}+\big[\frac{1}{8}-\zeta(3) +\mc O(\J^{2})\big]\,u^{3}+\mc O(u^{4}).
\ea
%and now similarity is better, apart from numerical values of coefficients.
%As we mentioned, the term $-u \log 2$ in $\ads_{4}$ case comes from redefinition $\sqrt{\lb}\to 2\bar h(\lb)$. 
%It is then natural that such specific contribution is absent in $\ads_{5}$.

\section{Large $\J$ expansion of  circular M2  bosonic fluctuation  frequencies}
\la{app:M2-largeJ-bosfreqs}

The frequencies corresponding to zeros of $P_4$ polynomials \rf{5.8}  given  in (\ref{5.9}) have the following 
 large $\J$ expansion
\ba
 {\rm B}: \ \  \qquad (p_{0})_{1, 2} =& \J+\big[p_{1}^{2}+up_{1}+\tfrac{1}{4}u(1+u)kp_{2}(\pm 2+kp_{2})\big]\,\frac{1}{\J}+\dots,\no  \\
(p_{0})_{3,4} =& -\J-\big[p_{1}^{2}-up_{1}+\tfrac{1}{4}u(1+u)kp_{2}(\pm 2+kp_{2})\big]\,\frac{1}{\J}+\dots,\no  \\
(p_{0})_{5,6} =& \big[p_{1}^{2}-up_{1}+\tfrac{1}{4}u(1+u)kp_{2}(\pm 2+kp_{2})\big]\,\frac{1}{\J}+\dots, \no \\
(p_{0})_{7,8} =& -\big[p_{1}^{2}+up_{1}+\tfrac{1}{4}u(1+u)kp_{2}(\pm 2+kp_{2})\big]\,\frac{1}{\J}+\dots.
\ea
For the  frequencies   corresponding to the zeros of  the   polynomial $P_{8}$ in (\ref{5.8})  one finds 
\ba
\qquad \qquad (p_{0})_{9} =& \phantom{-} 2\J+\big[p_{1}^{2}-2(1+u)p_{1}+2(1+3u+u^{2})+\tfrac{1}{2}u(1+u)k^{2}p_{2}^{2}\big]\frac{1}{2\J}+\dots, \no \\
(p_{0})_{10} =& -2\J-\big[p_{1}^{2}+2(1+u)p_{1}+2(1+3u+u^{2})+\tfrac{1}{2}u(1+u)k^{2}p_{2}^{2}\big]\frac{1}{2\J}+\dots,\no  \\
(p_{0})_{11} =& \J+\big[p_{1}^{2}-u^{2}+\tfrac{1}{2}u(1+u)k^{2}p_{2}^{2}\big]\frac{1}{2\J}+\dots, \quad 
\no\\
(p_{0})_{12} = &-\J-\big[p_{1}^{2}-u^{2}+\tfrac{1}{2}u(1+u)k^{2}p_{2}^{2}\big]\frac{1}{2\J}+\dots,\no  \\
(p_{0})_{13} =& \J+\big[p_{1}^{2}+u(2+u)-\tfrac{1}{2}u(1+u)k^{2}p_{2}^{2}\big]\frac{1}{2\J}+\dots, \\
(p_{0})_{14} =& -\J-\big[p_{1}^{2}+u(2+u)-\tfrac{1}{2}u(1+u)k^{2}p_{2}^{2}\big]\frac{1}{2\J}+\dots, \no\\
(p_{0})_{15,16} = \big[2(1+u)p_{1}&\pm\sqrt{p_{1}^{4}+4u(1+u)p_{1}^{2}(1-\tfrac{1}{4}k^{2}p_{2}^{2})-u^{2}(1+u^{2})k^{2}p_{2}^{2}(1-\tfrac{1}{4}k^{2}p_{2}^{2})}\ \big]
\frac{1}{2\J}+\dots.\no 
\ea
%
%
%\

\section{1-loop  energy of circular M2 brane in  small $\J$ limit} %Leading small $\J$  corrections to the circular string}
\la{app:M2-short}

One  can obtain  the leading  $\sim u\,\J^{0}$ terms  in the coefficients (\ref{5.18}) 
of the $1/k^2$  corrections 
from the small $\J$ expansion of the  corresponding membrane fluctuation  frequencies
%At leading order we find for bosons as well as fermions 
%the following expressions for $p_{0}$ 
\ba
 {\rm B}: \ \   (p_{0})_{1,2,3,4} =& \tfrac{1}{2} \sqrt{4 n^2+4 s_{1} n u+k u^2 \ell  (2 s_{2}+k \ell )}+\mc O(\J),\ \ \qquad s_{1}, s_{2} = \pm 1 \no \\
(p_{0})_{5,6} =& \tfrac{1}{2} \sqrt{4 n^2+u^2 (\pm 4+k^2 \ell ^2)}+\mc O(\J), \no  \\
(p_{0})_{7,8} =& \pm \sqrt{1+u^2}+\tfrac{1}{2} \sqrt{4 (n\mp 1)^2+u^2 (4+k^2 \ell ^2)}+\mc O(\J), \no 
\\
 {\rm F}: \ \   (p_{0})_{1,2,3,4} =& \tfrac{1}{2} (s_{1}\sqrt{1+u^2}+\sqrt{(s_{2}-2 s_{1}s_{2}n+u)^2+k^2 u^2 \ell ^2})+\mc O(\J), \qquad \ \ \ s_{1},s_{2}=\pm 1,\no  \\
(p_{0})_{5,6,7,8} =& \tfrac{1}{2} (s_{1}\sqrt{1+u^2}+\sqrt{1+4 (-s_{1}+n) n+u^2 (s_{2}+k \ell )^2})+\mc O(\J)\ , 
\ea
where we renamed the integer mode numbers as  $p_{1}\equiv  n,  \ p_{2}\equiv \ell$. 
Replacing the sum over $n$ with an integral we get 
\ba
E_1^{\rm M2} =  \frac{1}{2\kappa}\sum_{\ell\in\mathbb{Z}-\{0\}}\sum_{n\in \mathbb{Z}}\sum(-1)^{F}p_{0} \to \ 
-6u\frac{\zeta(2)}{k^{2}}-\frac{3}{2}u\frac{\zeta(4)}{k^{4}}+\dots + \mc O(\J)\ , \la{f2}
\ea
reproducing the leading  terms in (\ref{5.18}). One can similarly  find  the order $\J^2$  terms  in 
  (\ref{5.18}).
  
  One can   obtain  the  generating function for all  $1/k^{2r}$  coefficients  at leading order in small $\J$ 
  as\foot{It is interesting to note a similarity between (\ref{f4})
and Eq.~(2.16)  appearing  in the expression for the cusp anomalous dimension in \cite{Giombi:2024itd}.}
%The generating function at leading order in $\J\to 0$ can also be obtained in closed form and the result is 
\ba
&\qquad \qquad  \sum_{n=-\infty}^{\infty}\sum(-1)^{F}p_{0} \stackrel{k\ell\gg 1}{=} u^{2}F(k\ell)+\mc O(\J),
\la{F.7}\\
& F(y) = \te \frac{1}{2}(y-1)^{2}\log(y-1)+\frac{1}{2}(y+1)^{2}\log(y+1)+y^{2}\log y
+\frac{1}{8}(4-y^{2})\log(y^{2}-4)\lp   \quad \te -\frac{3}{8}(4+y^{2})\log(y^{2}+4)
+\frac{1}{4}(1+2y-y^{2})\log(-1-2y+y^{2})+\frac{1}{4}(1-2y-y^{2})\log(-1+2y+y^{2}),\no\\
&\qquad  \stackrel{y\gg 1}= \te -\frac{6}{y^{2}}-\frac{3}{2}\frac{1}{y^{4}}-\frac{70}{3}\frac{1}{y^{6}}-\frac{391}{12}\frac{1}{y^{8}}+\dots.\la{f4}
\ea
The first  two terms in \rf{f4}   are seen   to be 
in agreement with the  corresponding coefficients   in (\ref{5.18})  after using that 
$1/\kappa = 1/u+\mc O(\J)$.  
%We  note a  similarity between (\ref{F.4})
%and Eq.~(2.16) in \cite{Giombi:2024itd}.

%%%%%%%%%%%%%%%%%%%%%%%%%%%
\section{Details of 1-loop M2 brane  computations in   flat space \la{flat}} % \blue{REMOVE ?}}

%\subsubsection*{Semiclassical M2 brane  expansion in flat background}

%SK3

\def \M  {{\cal M}}  \def \La{\Lambda} 

\subsubsection*{Circular $J_1=J_2$  brane}

In the case of the solution in \rf{601}   we may choose 
%Since the solution rotates in two planes, the M2 brane can be viewed as a 3d surface in 6d. We choose 
the frame in the normal bundle  so that  $ \rn_{r} = \partial_{X^r}$  for $r=5,...,9$    and  ($\xi^i=(\tau,\s,\s')$)
\begin{align}
    &\rn_{1} = -\partial_{X^{0}}+\sin(\tau + \s)\partial_{X^{1}}-\cos(\tau + \s)\partial_{X^{2}} + \sin(\tau - \s)\partial_{X^{3}}-\cos(\tau - \s)\partial_{X^{4}},  \\ 
    &\rn_{2} = \cos(\tau + \s)\partial_{X^{1}} + \sin(\tau + \s)\partial_{X^{2}} , \qquad 
    \rn_{3} = \cos(\tau - \s)\partial_{X^{3}}+\sin(\tau - \s)\partial_{X^{4}} \ .
    % \qquad     n_{4,5,6,7,8} = \partial_{X^{5,6,7,8,9}} .
\end{align}
Then  the  fluctuation operator  in $r=5,...,9$  directions   is given by the massless  Laplacian for the metric $\bag$ in \rf{602}.
The ``mass matrix'' in the fluctuation operator  \rf{610}  in $\rn_{1,2,3}$  directions is  %given by 
\begin{equation}\label{15}
 \small    \M_{pq}= - (\KK_{ij})_{p}(\KK^{ij})_{q} = \begin{pmatrix}
        0 & 0 & 0 \\
        0 & 0 &  -\frac{4}{\kappa^2}  \\ 
        0 & -\frac{4}{\kappa^2} & 0 
    \end{pmatrix}_{pq} , \qquad \ \ \ \  p, q=1,2,3.
\end{equation}
The  covariant  derivative part  of the operator in \rf{610} 
$ \Delta^{\perp}= \frac{1}{\sqrt{-\bag}}(\nabla_{i}^{\perp})^{\dagger}(\sqrt{-\bag}\bag^{ij}\nabla^{\perp}_{j}) $ 
can be found using  that the connection on the normal bundle is  given by 
\begin{equation}
\small     \Omega_{pq} = \langle n_{p}, \partial n_{q}\rangle  = \begin{pmatrix}
        0 & -1 & -1 \\
        1 & 0 & 0 \\
        1 & 0 & 0 
    \end{pmatrix}d\tau + \begin{pmatrix}
        0 & -1 & 1 \\
        1 & 0 & 0 \\
        -1 & 0 & 0 
    \end{pmatrix}d\sigma \ .
\end{equation}
As a result,\foot{We ignore the overall constant factor $2\kappa^{-2}$ that can be  absorbed into the fluctuation fields.}
\begin{equation}
 \small    \Delta^{\perp}+\M = % \frac{2}{\kappa^2} 
 \begin{pmatrix}
        \partial^2_{0}-\partial^2_{1}-\La^2 \partial^{2}_{2} & -2(\partial_{0}-\partial_{1}) & -2(\partial_{0}+\partial_{1}) \\
        2(\partial_{0}-\partial_{1}) & \partial^{2}_{0}-\partial^{2}_{1}-\La^2 \partial^{2}_{2} & -4 \\
        2(\partial_{0}+\partial_{1}) & -4 & \partial^2_{0}
-\partial^{2}_{1}-\La^2 \partial_{2}^2 
\end{pmatrix} \ , \qquad\qquad  \La^2 \equiv {\kappa^2\ov 2 R^2_{11}} \ . \la{g10}
\end{equation}
Expanding in Fourier modes $(p_0,\, p_1,\  p_2)$ (cf. \rf{3.14}), the  characteristic frequencies  for $q=1,2,3$  directions
 can be determined from the  polynomial equation\foot{In the string case $p_{2}=0$ 
 one gets   3+3  modes    with $    p_0^2(p_1, 0) = \{(p_1+2)^2 , \ p_1^2 , \ (p_1-2)^2\}$.}
$
   \det ( \Delta^{\perp} + \M) = 0.
$
In addition, there are 5 ``massless''  modes 
$
  p_0^2=p_1^2+\La^2 p_2^2   \ .
$
In total, we get the agreement with the expression for ${\cal D}_{\rm B}$ in \rf{66}. 

Since the  connection along the membrane surface is flat, the  square of the Dirac operator may be written as (cf. \rf{b1}) 
%$    \slashed \nabla =\slashed \partial - \frac{1}{2} \gamma_{1}\gamma_{2} d (\tau+\s) -\frac{1}{2} \gamma_{1}\gamma_{3}d(\tau-\s) ,$   with its square being 
\begin{equation}
    \slashed{\nabla}^{2}=- \partial_{{0}}^{2}+\partial_{{1}}^2+\La^2 \partial_{{2}}^{2}+\gamma_{1}\gamma_{2}(\partial_{{0}}-\partial_{{1}})+\gamma_{1}\gamma_{3}(\partial_{{0}}+\partial_{{1}}) +\rho^{0}\rho^{1}\gamma_{2}\gamma_{3} \ .
\end{equation}
One can  choose a $\kappa$-symmetry gauge and the  $\gamma$-matrices basis such that 11d Majorana fermion can be split as 
$
    \theta = \chi \otimes \eta ,   \ \rho^{0}\rho^{1} \eta = \eta \ ,
$
where $\chi$ is a 2d Majorana fermion  so that  %and $\gamma_{1, 2, 3}$ can be chosen to be $\sigma_{3,1,2}\otimes \gamma_{9}$ so that:
\begin{equation}
\small     \slashed{\nabla}^{2} \chi = \begin{pmatrix}
        -\partial_{{0}}^{2}+\partial_{{1}}^2+\La^2 \partial_{{2}}^{2} + i & (\partial_{{0}}-\partial_{{1}})-i(\partial_{{0}}+\partial_{{1}}) \\
       -(\partial_{{0}}-\partial_{{1}})-i(\partial_{{0}}+\partial_{{1}})& -\partial_{{0}}^{2}+\partial_{{1}}^2+\La^2\partial_{{2}}^{2} -i   \end{pmatrix} \chi \ .
\end{equation}
As a result, the 4+4 fermionic frequencies are found to be  the same as the zeroes of ${\cal D}_{\rm F}$  in \rf{66} 
%so that the fermionic modes are (same was obtained before in polar coordinates):
\begin{equation}
    p_0^2(p_1,p_2)=  p_1^2 +\La^2 p_2^2 +1 \pm \sqrt{4p_1^2 +    \La^2 p_2^2    }  \ .
\end{equation}

%%%%%%%%%%%%%%%%%%%%%%%%%%%%%%%%%%%%%%%%%%%%%%
\subsubsection*{Exact formula for the coefficients $c_n$  in (\ref{8.3})}

One may represent  $c_n$ with $n>1$  as   ($c_1=-4$)  
\be \la{g1} 
c_n=\ha( \alpha_{n} + \beta_n) \ , \qquad\qquad \ \ \   %\text{with}\ \ \ 
\alpha_{n} \equiv  (-1)^{n}\frac{4^{n}\Gamma(n-\frac{1}{2})\Gamma(n+\frac{1}{2})}{\sqrt\pi\, n\, \Gamma(2n+\frac{1}{2})}\ ,  
\ee
where   $\beta_{n}$ are given by the solution of the second order difference equation
\ba
& 64 n^2 (1+n) (-1+2 n) (1+2 
n) (4+3 n) (31+42 n+14 
n^2) \beta_{n}\lp
-8 (1+n) (918+9705 
n+41860 n^2+94960 n^3+122292 
n^4+89510 n^5+34580 n^6+5460 
n^7) \beta_{n+1}\lp
+(2+n) (3+2 
n)^2 (1+3 n) (5+4 n) (7+4 
n) (3+14 n+14 n^2) \beta_{n+2}=0\ , 
\ea
with  the initial values
$
\beta_{2} = -\frac{2176}{105}, \ \  \beta_{3} = -\frac{493568}{3465}. 
$
Explicitly, we  get
\ba
& \alpha_{n} =\te  \frac{16}{35},-\frac{128}{693},\frac{128}{1287},-\frac{14336}{230945},\dots,
\qquad \ \ \ \ 
\beta_{n} =\te -\frac{2176}{105}, -\frac{493568}{3465}, -\frac{56639488}{45045}, -\frac{185098305536}{14549535},
% -\frac{3128015454208}{22309287},
\dots \no \\
& c_n= \ha( \alpha_{n} + \beta_n) 
\te =-  \frac{152}{15},\ -\frac{2496}{35},\ -\frac{9439168}{15015},\dots \ , \qquad  \qquad \ \ \ n=2,3, ... \ . \la{g2}
\ea
The %relation for  $\alpha_{n}$ and the
 recursion relation 
 for $\beta_{n}$ allows one  to compute the  coefficients  $c_n$  to very high order.
 In particular, one can check that  the series \rf{8.3} for $F(x)$ has the  radius of convergence 
$
\frac{1}{x^{4}} < \frac{1}{16} $, i.e.   $x>2$, 
as presented  in Fig.~\ref{fig:c}. 
This is also the radius of convergence of $\bae (\Lambda)$ in \rf{69} 
 since $\zeta(n)\to 1$ for large $n$, consistently with the  data  presented  in Fig.~\ref{fig:b}.
\begin{figure}[H]
\centering
\includegraphics[width=0.5\textwidth]{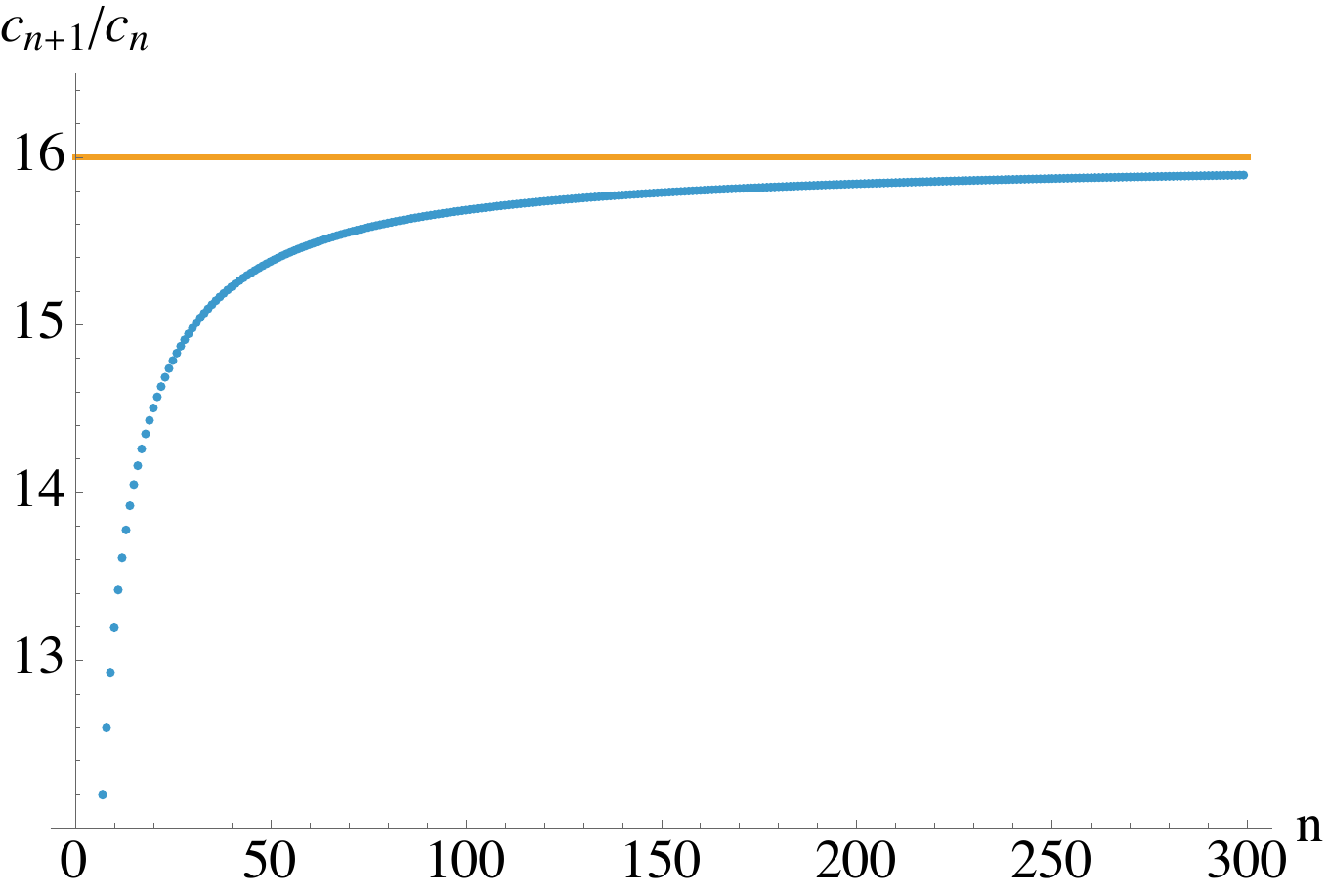}
\caption{Ratio test for the coefficients of the function in (\ref{8.3}). 
%The coefficients are $c_{n}=\frac{\alpha_{n}+\beta_{n}}{2}$ and the ratio $c_{n+1}/c_{n}$ tends to 16,
% implying the convergence radius in (\ref{8.14}).
}
\la{fig:c}
\end{figure}

%%%%%%%%%%%%%%%%%%%%
\subsubsection*{Folded  spinning brane}

In the case of the solution in \rf{611}, we get  7  modes  with the ``massless''  Laplace  operator $\Delta$  in \rf{6.33}  and one mode  with 
the operator $
    \Delta + R^{(2)} $ in \rf{6.34}. To get the latter one may use
  the Gauss equation to  express the second term in the operator in \rf{610} 
in terms of the scalar curvature of the metric in \rf{79} as 
$  -{\KK}^{ij}\ {\KK}_{ij} =   R^{(3)}=R^{(2)}$.

  The   fermionic operator  in \rf{666} simplifies to
$
    \slashed{\nabla} = \rho^{i}\big{(}\partial_{\re_i} + \frac{1}{4}\Omega^{jk}_{i} \rho_{jk}\big{)}, \  \Omega_{jk;i} = \langle \re_{j}, \nabla_{\re_{i}}\re_{k} \rangle.
$
A convenient choice of the gamma matrices basis is such that $\Gamma = \rho_{0}\rho_{1}\rho_{2}= \sigma^{0}\otimes \gamma_{9}$. Imposing the $\kappa$-symmetry gauge $(1+\Gamma)\theta=0$, the 11d Majorana fermion can be written in terms of the 3d and 8d Majorana fermions
$
    \theta = \psi \otimes \eta, \ \  (1+\gamma_{9})\eta = 0.
$
Representing  the metric in \rf{79} as  $ds^2= \kappa^2 \, e^{2\l(\s)} ( - d\tau^2 + d \s^2) + R^2_{11} d\s'^2$
where $e^\l=% \k 
%AT
\cos \s$  we get for the Dirac operator %SK2
(after expanding in modes in $\xi^2=\s'$)
%\foot{Here $\Omega^{0}_{\ 1} = e^{-\lambda}(\partial_{\sigma}\lambda) \re^{0}+e^{-\lambda}(\partial_{\tau}\lambda)\re^{1}.$}
\begin{align}\la{g14}
    &\slashed{\nabla} = -e^{-\lambda}\rho_{0}\partial_{\tau}+e^{-\lambda}\rho_{1}\partial_{\sigma}+ip_{2} \La\rho_{2} +\ha {e^{-\lambda}}\big{(} -\partial_{\tau}\lambda\rho_{0}+\partial_{\sigma}\lambda \rho_{1}\big{)}.
 %   &\Omega^{0}_{\ 1} = e^{-\lambda}(\partial_{\sigma}\lambda) e^{0}+e^{-\lambda}(\partial_{\tau}\lambda)e^{1}.
\end{align}
After the   conformal transformation of the     fermions $\theta \rightarrow e^{-\lambda/2}\theta$ it may be written as 
%\begin{equation}
 $   \slashed{\nabla}(e^{-\lambda/2}\theta) =  e^{-3\lambda/2}\big{(}\rho^{\alpha}\partial_{\alpha}+
    ip_{2}\La\, e^{\lambda}\rho_{2}\big{)}\theta, $
%\end{equation}
 with the non-trivial part of the  determinant equivalent to the one of the operator in  \rf{635}. 

In the string theory   limit, i.e.   $p_{2}=0$, 
 the rescaled  fermionic operator becomes the  flat space one
  and thus has the same spectrum as the  massless bosonic operator  $ \partial^2_\tau-\del_\s^2 $  in \rf{6.33}.
  The ``massive''  operator in \rf{6.34} 
  may be written as $\Delta + R^{(2)} =  \del_\tau^2 -\del_\s^2  - 2\partial^2_\s(\log \cos\s )$.
   A peculiar property of $e^{\lambda} =\cos \s$ 
    is that it is the same as  the ground state wave function of the operator $-\partial_{\sigma}^2$ in the case of the Dirichlet boundary conditions at $\sigma = \pm\ha \pi$. This implies that the  massless and ``massive'' operators 
      are related by the Darboux transformation and have the same spectrum except for the ground state (see, e.g., \cite{Cooper:1994eh}, Section 7.1). 
      Explicitly, their  eigenfrequencies  are, respectively, 
$  \omega_{n} = n+1, \ \  n=1,2,... $ and $  \omega_{n} = n+1, \ \  n=0,1,2,...
$.
This implies the expected triviality of the 1-loop   correction to the energy  coming   from the string theory modes only. 

%SK2
%AT
In the case when $p_{2}\neq  0$, we note that both operators $\Delta$ in \rf{6.33}  and $\Delta+R^{(2)}$  in \rf{6.34} 
acting on $\rm X(\tau,\s)$ can be written in the form that appears in  the angular oblate spheroidal equation\footnote{We thank Gerald  Dunne for pointing this out.}
\begin{align}
    & \frac{d}{dz}\Big{[}(1-z^2)\frac{d}{dz}f(z)\Big{]}+\Big{[}\omega^2-\tfrac{1}{4}-\gamma^2(1-z^2)-\frac{\mu^2}{1-z^2}\Big{]}f(z) = 0\, , \la{g15} \\
    &\qquad  \mathrm{X} = e^{i\omega \tau}(1-z^2)^{1/4}f(z) \, , \ \ \ \qquad z = \sin \sigma,\no 
\end{align}
where $\gamma^2 = \Lambda^2p_{2}^2$ and $\mu^2 = \four$  in the case  of 
$\Delta$ while   $ \mu^2 = 2+\four $   in the case  of 
$\Delta+R^{(2)}$. For a recent discussion on the eigenvalues of the spheroidal equation, see \cite{Meynig:2025lnk} and references therein. The Fredholm determinant of a more general Heun differential operator was also recently studied in \cite{Arnaudo:2024rhv}.

%This  can be used for  determining their spectrum. 
 %operator respectively. 

%Asymptotic properties of the eigenvalues for $\gamma \gg 1$ and integer $\mu$ are well-known.

%Let us consider a fixed nonzero value of $p_{2}$. In the semiclassical limit, $\Lambda = p_{2}\kappa/R_{11} \gg 1$. Consider $L_{1}$. For $\Lambda \gg 1$, there are two minima around which one can expand:
%\begin{equation}
%    \sigma_{0} = \pm\frac{\pi}{2} \mp \frac{2^{1/4}}{\Lambda^{1/2}}+\dots.
%\end{equation}

\small
\bibliography{BT-Biblio}

\providecommand{\href}[2]{#2}\begingroup\raggedright\begin{thebibliography}{10}

\bibitem{Aharony:2008ug}
O.~Aharony, O.~Bergman, D.~L. Jafferis and J.~Maldacena, \emph{{N=6 superconformal Chern-Simons-matter theories, M2-branes and their gravity duals}}, \href{https://doi.org/10.1088/1126-6708/2008/10/091}{\emph{JHEP} {\bfseries 10} (2008) 091} [\href{https://arxiv.org/abs/0806.1218}{{\ttfamily 0806.1218}}].

\bibitem{Beisert:2010jr}
N.~Beisert et~al., \emph{{Review of AdS/CFT Integrability: An Overview}}, \href{https://doi.org/10.1007/s11005-011-0529-2}{\emph{Lett. Math. Phys.} {\bfseries 99} (2012) 3} [\href{https://arxiv.org/abs/1012.3982}{{\ttfamily 1012.3982}}].

\bibitem{Giombi:2024itd}
S.~Giombi, S.~A. Kurlyand and A.~A. Tseytlin, \emph{{Non-Planar Corrections in ABJM Theory from Quantum M2 Branes}}, \href{https://doi.org/10.1007/JHEP11(2024)056}{\emph{JHEP} {\bfseries 11} (2024) 056} [\href{https://arxiv.org/abs/2408.10070}{{\ttfamily 2408.10070}}].

\bibitem{Giombi:2023vzu}
S.~Giombi and A.~A. Tseytlin, \emph{{Wilson Loops at Large N and the Quantum M2-Brane}}, \href{https://doi.org/10.1103/PhysRevLett.130.201601}{\emph{Phys. Rev. Lett.} {\bfseries 130} (2023) 201601} [\href{https://arxiv.org/abs/2303.15207}{{\ttfamily 2303.15207}}].

\bibitem{Beccaria:2023ujc}
M.~Beccaria, S.~Giombi and A.~A. Tseytlin, \emph{{Instanton contributions to the ABJM free energy from quantum M2 branes}}, \href{https://doi.org/10.1007/JHEP10(2023)029}{\emph{JHEP} {\bfseries 10} (2023) 029} [\href{https://arxiv.org/abs/2307.14112}{{\ttfamily 2307.14112}}].

\bibitem{Beccaria:2025vdj}
M.~Beccaria and A.~A. Tseytlin, \emph{{Non-Planar Corrections to ABJM Bremsstrahlung Function from Quantum M2 Brane}},  \href{https://arxiv.org/abs/2501.06858}{{\ttfamily 2501.06858}}.

\bibitem{Bergshoeff:1987cm}
E.~Bergshoeff, E.~Sezgin and P.~K. Townsend, \emph{{Supermembranes and eleven-dimensional supergravity}}, \href{https://doi.org/10.1016/0370-2693(87)91272-X}{\emph{Phys. Lett.} {\bfseries B189} (1987) 75}.

\bibitem{deWit:1998yu}
B.~de~Wit, K.~Peeters, J.~Plefka and A.~Sevrin, \emph{{The M theory two-brane in AdS$_4\times S^7$ and AdS$_7\times S^4$}}, \href{https://doi.org/10.1016/S0370-2693(98)01340-9}{\emph{Phys. Lett.} {\bfseries B443} (1998) 153} [\href{https://arxiv.org/abs/hep-th/9808052}{{\ttfamily hep-th/9808052}}].

\bibitem{McLoughlin:2008ms}
T.~McLoughlin and R.~Roiban, \emph{{Spinning Strings at One-Loop in Ad$S^4$ $\times$ $ CP^3$}}, \href{https://doi.org/10.1088/1126-6708/2008/12/101}{\emph{JHEP} {\bfseries 12} (2008) 101} [\href{https://arxiv.org/abs/0807.3965}{{\ttfamily 0807.3965}}].

\bibitem{Alday:2008ut}
L.~F. Alday, G.~Arutyunov and D.~Bykov, \emph{{Semiclassical Quantization of Spinning Strings in Ad$S^4$ $\times$ $ CP^3$}}, \href{https://doi.org/10.1088/1126-6708/2008/11/089}{\emph{JHEP} {\bfseries 11} (2008) 089} [\href{https://arxiv.org/abs/0807.4400}{{\ttfamily 0807.4400}}].

\bibitem{McLoughlin:2008he}
T.~McLoughlin, R.~Roiban and A.~A. Tseytlin, \emph{{Quantum Spinning Strings in Ad$S^4$ $\times$ $ CP^3$: Testing the Bethe Ansatz Proposal}}, \href{https://doi.org/10.1088/1126-6708/2008/11/069}{\emph{JHEP} {\bfseries 11} (2008) 069} [\href{https://arxiv.org/abs/0809.4038}{{\ttfamily 0809.4038}}].

\bibitem{Bandres:2009kw}
M.~A. Bandres and A.~E. Lipstein, \emph{{One-Loop Corrections to Type IIA String Theory in AdS$_{4}\times\mathbb{CP}^{3}$}}, \href{https://doi.org/10.1007/JHEP04(2010)059}{\emph{JHEP} {\bfseries 04} (2010) 059} [\href{https://arxiv.org/abs/0911.4061}{{\ttfamily 0911.4061}}].

\bibitem{Beccaria:2012qd}
M.~Beccaria, G.~Macorini, C.~Ratti and S.~Valatka, \emph{{Semiclassical folded string in $AdS_5 \times S^5$}}, \href{https://doi.org/10.1007/JHEP05(2012)137}{\emph{JHEP} {\bfseries 05} (2012) 030} [\href{https://arxiv.org/abs/1203.3852}{{\ttfamily 1203.3852}}].

\bibitem{Gubser:2002tv}
S.~S. Gubser, I.~R. Klebanov and A.~M. Polyakov, \emph{{A Semiclassical Limit of the Gauge / String Correspondence}}, \href{https://doi.org/10.1016/S0550-3213(02)00373-5}{\emph{Nucl. Phys.} {\bfseries B636} (2002) 99} [\href{https://arxiv.org/abs/hep-th/0204051}{{\ttfamily hep-th/0204051}}].

\bibitem{Frolov:2002av}
S.~Frolov and A.~A. Tseytlin, \emph{{Semiclassical Quantization of Rotating Superstring in $\mathrm{AdS}_5$ $\times$ $ S^5$}}, \href{https://doi.org/10.1088/1126-6708/2002/06/007}{\emph{JHEP} {\bfseries 06} (2002) 007} [\href{https://arxiv.org/abs/hep-th/0204226}{{\ttfamily hep-th/0204226}}].

\bibitem{Krishnan:2008zs}
C.~Krishnan, \emph{{Ad$S^4$/CFT(3) at One Loop}}, \href{https://doi.org/10.1088/1126-6708/2008/09/092}{\emph{JHEP} {\bfseries 09} (2008) 092} [\href{https://arxiv.org/abs/0807.4561}{{\ttfamily 0807.4561}}].

\bibitem{Bianchi:2014ada}
L.~Bianchi, M.~S. Bianchi, A.~Bres, V.~Forini and E.~Vescovi, \emph{{Two-Loop Cusp Anomaly in ABJM at Strong Coupling}}, \href{https://doi.org/10.1007/JHEP10(2014)013}{\emph{JHEP} {\bfseries 10} (2014) 013} [\href{https://arxiv.org/abs/1407.4788}{{\ttfamily 1407.4788}}].

\bibitem{Gromov:2014eha}
N.~Gromov and G.~Sizov, \emph{{Exact Slope and Interpolating Functions in ${\mathcal{N}}\!=6$ Supersymmetric Chern-Simons Theory}}, \href{https://doi.org/10.1103/PhysRevLett.113.121601}{\emph{Phys. Rev. Lett.} {\bfseries 113} (2014) 121601} [\href{https://arxiv.org/abs/1403.1894}{{\ttfamily 1403.1894}}].

\bibitem{Gromov:2008qe}
N.~Gromov and P.~Vieira, \emph{{The All Loop AdS4/CFT3 Bethe Ansatz}}, \href{https://doi.org/10.1088/1126-6708/2009/01/016}{\emph{JHEP} {\bfseries 01} (2009) 016} [\href{https://arxiv.org/abs/0807.0777}{{\ttfamily 0807.0777}}].

\bibitem{Frolov:2003qc}
S.~Frolov and A.~A. Tseytlin, \emph{{Multispin String Solutions in $\mathrm{AdS}_5$ $\times$ $ S^5$}}, \href{https://doi.org/10.1016/S0550-3213(03)00580-7}{\emph{Nucl. Phys. B} {\bfseries 668} (2003) 77} [\href{https://arxiv.org/abs/hep-th/0304255}{{\ttfamily hep-th/0304255}}].

\bibitem{Belitsky:2006en}
A.~V. Belitsky, A.~S. Gorsky and G.~P. Korchemsky, \emph{{Logarithmic Scaling in Gauge/String Correspondence}}, \href{https://doi.org/10.1016/j.nuclphysb.2006.04.030}{\emph{Nucl. Phys. B} {\bfseries 748} (2006) 24} [\href{https://arxiv.org/abs/hep-th/0601112}{{\ttfamily hep-th/0601112}}].

\bibitem{Frolov:2006qe}
S.~Frolov, A.~Tirziu and A.~A. Tseytlin, \emph{{Logarithmic Corrections to Higher Twist Scaling at Strong Coupling from AdS/CFT}}, \href{https://doi.org/10.1016/j.nuclphysb.2006.12.013}{\emph{Nucl. Phys. B} {\bfseries 766} (2007) 232} [\href{https://arxiv.org/abs/hep-th/0611269}{{\ttfamily hep-th/0611269}}].

\bibitem{Giombi:2010fa}
S.~Giombi, R.~Ricci, R.~Roiban, A.~A. Tseytlin and C.~Vergu, \emph{{Generalized scaling function from light-cone gauge $AdS_{5}\times S^{5}$ superstring}}, \href{https://doi.org/10.1007/JHEP06(2010)060}{\emph{JHEP} {\bfseries 06} (2010) 060} [\href{https://arxiv.org/abs/1002.0018}{{\ttfamily 1002.0018}}].

\bibitem{Arutyunov:2003za}
G.~Arutyunov, J.~Russo and A.~A. Tseytlin, \emph{{Spinning Strings in $\mathrm{AdS}_5$ $\times$ $ S^5$: New Integrable System Relations}}, \href{https://doi.org/10.1103/PhysRevD.69.086009}{\emph{Phys. Rev. D} {\bfseries 69} (2004) 086009} [\href{https://arxiv.org/abs/hep-th/0311004}{{\ttfamily hep-th/0311004}}].

\bibitem{Park:2005ji}
I.~Y. Park, A.~Tirziu and A.~A. Tseytlin, \emph{{Spinning Strings in $\mathrm{AdS}_5$ $\times$ $ S^5$: One-Loop Correction to Energy in $SL(2)$ Sector}}, \href{https://doi.org/10.1088/1126-6708/2005/03/013}{\emph{JHEP} {\bfseries 03} (2005) 013} [\href{https://arxiv.org/abs/hep-th/0501203}{{\ttfamily hep-th/0501203}}].

\bibitem{Beisert:2005mq}
N.~Beisert, A.~A. Tseytlin and K.~Zarembo, \emph{{Matching Quantum Strings to Quantum Spins: One-Loop Versus Finite-Size Corrections}}, \href{https://doi.org/10.1016/j.nuclphysb.2005.03.030}{\emph{Nucl. Phys. B} {\bfseries 715} (2005) 190} [\href{https://arxiv.org/abs/hep-th/0502173}{{\ttfamily hep-th/0502173}}].

\bibitem{Beisert:2005cw}
N.~Beisert and A.~A. Tseytlin, \emph{{On Quantum Corrections to Spinning Strings and Bethe Equations}}, \href{https://doi.org/10.1016/j.physletb.2005.09.054}{\emph{Phys. Lett. B} {\bfseries 629} (2005) 102} [\href{https://arxiv.org/abs/hep-th/0509084}{{\ttfamily hep-th/0509084}}].

\bibitem{Sundborg:1988ai}
B.~Sundborg, \emph{{Self-energies of Massive Strings}}, \href{https://doi.org/10.1016/0550-3213(89)90084-9}{\emph{Nucl. Phys. B} {\bfseries 319} (1989) 415}.

\bibitem{Amano:1988ht}
K.~Amano and A.~Tsuchiya, \emph{{Mass Splittings and the Finiteness Problem of Mass Shifts in the Type {II} Superstring at One Loop}}, \href{https://doi.org/10.1103/PhysRevD.39.565}{\emph{Phys. Rev. D} {\bfseries 39} (1989) 565}.

\bibitem{Chialva:2003hg}
D.~Chialva, R.~Iengo and J.~G. Russo, \emph{{Decay of long-lived massive closed superstring states: Exact results}}, \href{https://doi.org/10.1088/1126-6708/2003/12/014}{\emph{JHEP} {\bfseries 12} (2003) 014} [\href{https://arxiv.org/abs/hep-th/0310283}{{\ttfamily hep-th/0310283}}].

\bibitem{Chialva:2004xm}
D.~Chialva, R.~Iengo and J.~G. Russo, \emph{{Search for the most stable massive state in superstring theory}}, \href{https://doi.org/10.1088/1126-6708/2005/01/001}{\emph{JHEP} {\bfseries 01} (2005) 001} [\href{https://arxiv.org/abs/hep-th/0410152}{{\ttfamily hep-th/0410152}}].

\bibitem{Sen:2016gqt}
A.~Sen, \emph{{One Loop Mass Renormalization of Unstable Particles in Superstring Theory}}, \href{https://doi.org/10.1007/JHEP11(2016)050}{\emph{JHEP} {\bfseries 11} (2016) 050} [\href{https://arxiv.org/abs/1607.06500}{{\ttfamily 1607.06500}}].

\bibitem{Mezincescu:1987kj}
L.~Mezincescu, R.~I. Nepomechie and P.~van Nieuwenhuizen, \emph{{Do supermembranes contain massless particles?}}, \href{https://doi.org/10.1016/0550-3213(88)90085-5}{\emph{Nucl. Phys. B} {\bfseries 309} (1988) 317}.

\bibitem{Iengo:2006gm}
R.~Iengo and J.~G. Russo, \emph{{Handbook on string decay}}, \href{https://doi.org/10.1088/1126-6708/2006/02/041}{\emph{JHEP} {\bfseries 02} (2006) 041} [\href{https://arxiv.org/abs/hep-th/0601072}{{\ttfamily hep-th/0601072}}].

\bibitem{Chialva:2009pg}
D.~Chialva, \emph{{String Mass Shifts}}, \href{https://doi.org/10.1016/j.nuclphysb.2009.04.023}{\emph{Nucl. Phys. B} {\bfseries 819} (2009) 225} [\href{https://arxiv.org/abs/0903.3979}{{\ttfamily 0903.3979}}].

\bibitem{Roiban:2007ju}
R.~Roiban and A.~A. Tseytlin, \emph{{Spinning Superstrings at Two Loops: Strong-Coupling Corrections to Dimensions of Large-Twist SYM Operators}}, \href{https://doi.org/10.1103/PhysRevD.77.066006}{\emph{Phys. Rev. D} {\bfseries 77} (2008) 066006} [\href{https://arxiv.org/abs/0712.2479}{{\ttfamily 0712.2479}}].

\bibitem{Beccaria:2012xm}
M.~Beccaria, S.~Giombi, G.~Macorini, R.~Roiban and A.~A. Tseytlin, \emph{{'Short' spinning strings and structure of quantum $AdS_5 \times S^5$ spectrum}}, \href{https://doi.org/10.1103/PhysRevD.86.066006}{\emph{Phys. Rev.} {\bfseries D86} (2012) 066006} [\href{https://arxiv.org/abs/1203.5710}{{\ttfamily 1203.5710}}].

\bibitem{Duff:1987bx}
M.~J. Duff, P.~S. Howe, T.~Inami and K.~S. Stelle, \emph{{Superstrings in D=10 from Supermembranes in D=11}}, \href{https://doi.org/10.1016/0370-2693(87)91323-2}{\emph{Phys. Lett. B} {\bfseries 191} (1987) 70}.

\bibitem{Cooper:1994eh}
F.~Cooper, A.~Khare and U.~Sukhatme, \emph{{Supersymmetry and quantum mechanics}}, \href{https://doi.org/10.1016/0370-1573(94)00080-M}{\emph{Phys. Rept.} {\bfseries 251} (1995) 267} [\href{https://arxiv.org/abs/hep-th/9405029}{{\ttfamily hep-th/9405029}}].

\bibitem{Meynig:2025lnk}
M.~Meynig, \emph{{Non-perturbative asymptotics of the eigenvalues of the spheroidal equation}},  \href{https://arxiv.org/abs/2503.06780}{{\ttfamily 2503.06780}}.

\bibitem{Arnaudo:2024rhv}
P.~Arnaudo, G.~Bonelli and A.~Tanzini, \emph{{One loop effective actions in Kerr-(A)dS black holes}}, \href{https://doi.org/10.1103/PhysRevD.110.106006}{\emph{Phys. Rev. D} {\bfseries 110} (2024) 106006} [\href{https://arxiv.org/abs/2405.13830}{{\ttfamily 2405.13830}}].

\end{thebibliography}\endgroup
\bibliographystyle{JHEP-v2.9}
\end{document}